\documentclass[a4paper,12pt]{article}
\usepackage[latin1]{inputenc}
\usepackage{harvard}
\usepackage{amssymb}
\usepackage{amsmath}
\usepackage{amsthm}
\usepackage{graphicx}
\usepackage{lscape}
\usepackage{times}
\usepackage{url}

\newcommand{\be}{\begin{equation}}
\newcommand{\ee}{\end{equation}}
\newcommand{\bea}{\begin{eqnarray}}
\newcommand{\eea}{\end{eqnarray}}
\newcommand{\bi}{\begin{itemize}}
\newcommand{\ei}{\end{itemize}}
\newcommand{\Cov}{{\rm Cov}}
\newcommand{\Var}{{\rm Var}}

\newcommand{\can}{\citeasnoun}
\newcommand{\e}{{\rm e}}

\renewcommand{\epsilon}{\varepsilon}
\renewcommand{\phi}{\varphi}
\renewcommand{\[}{\left[}
\renewcommand{\]}{\right]}
\renewcommand{\(}{\left(}
\renewcommand{\)}{\right)}

\def\ds{\displaystyle}

\def\E{\textrm{E}}
\def\Var{\textrm{Var}}

\theoremstyle{plain}
\newtheorem{proposition}{Proposition}

\theoremstyle{definition}

\title{Heterogeneous expectations and long range correlation of the volatility of asset returns
\thanks{The authors acknowledge helpful discussions and
exchanges with the participants at the $32^{\rm nd}$ joined actuarial
seminar at the Universities of Lausanne and Lyon, at ETH Zurich, at the $13^{\rm th}$ International Conference on Forecasting Financial Markets and at the 
$23^{\rm rd}$ International Conference of the French Finance Association.}
}
\author{J. Coulon$^{1}$ and Y. Malevergne$^{2,3}$\\
\\
\small $^1$ \it ISFA Graduate School of Actuarial Science - University of Lyon 1\\
\small \it 50 avenue Tony Garnier, 69366 Lyon Cedex 07, France\\
\small $^2$ \it University of Saint-Etienne -- Institute of Business Administration\\
\small \it 2 rue Trefilerie, 42023 Saint-Etienne, France\\
\small $^3$ \it EM-Lyon Business School\\
\small \it 23 avenue Guy de Collongue, 69134 Ecully Cedex, France\\
\\
{\small e-mails : \url{jerome.coulon@univ-lyon1.fr} and \url{ymalevergne@ethz.ch}}}
\date{}

\begin{document}
\maketitle

\begin{abstract}

Inspired by the recent literature on aggregation theory, we aim at relating the long range correlation 
of the stocks return  volatility to the heterogeneity of the investors' expectations about the level 
of the future volatility.
Based on a semi-parametric model of investors' anticipations, we make the connection between 
the distributional properties of the heterogeneity parameters and the auto-covariance/auto-correlation functions 
of the realized volatility. We report different behaviors, or change of convention, whose observation 
depends on the market phase under consideration. In particular, we report and justify the fact that 
the volatility exhibits significantly longer memory during the phases of speculative bubble than during 
the phase of recovery following the collapse of a speculative bubble.
\end{abstract}

JEL classification:  G10, G14, D84, C43, C53

{\it Keywords}: Realized volatility, aggregation model, long memory, bounded rationality


\section*{Introduction}

The slow hyperbolic decay of the auto-correlation function characterizing the behavior of many 
economic and financial time series has been the topic of active debates for more than two decades. 
Several interpretations and models have been provided in an attempt to explain the origin of this phenomenon, 
also known as the long-memory phenomenon. Among the most relevant explanations that have been proposed up to now, 
one can focus on three major mechanisms acting separately or in conjunction: (i) the aggregation approach 
suggested by \can{granger80} who have shown that the time series resulting from the aggregation 
of micro-variables exhibiting short-memory often yields long-memory, (ii) the presence of infrequent 
structural breaks, which allows mimicking long term non-stationarity of the economic and financial activity 
\cite[among others]{diebold01,gourieroux01,granger04,gadea04}, and (iii) the presence of  non-linearities 
in economic an financial systems (see \can{davidson05} for a survey).

Our aim, in this article, is to provide a model that relates the long memory of the realized volatility 
of assets returns, which is a pervasive feature of financial time series 
\cite[for the pioneering works]{Taylor86,ding83,Dacorogna93}, to the heterogeneous behavior of the economic agents. 
Based on the fact that the market participants perform heterogeneous anticipations about the future level 
of the realized volatility and that they act as  bounded rationality agents, we propose an explanation 
of the long memory phenomenon that relies on the aggregation by the market of the heterogeneous beliefs 
of the investors, revealed through the market pricing process.

The heterogeneity of market investors is now an well-recognized fact -- in particular amongst the supporters 
of behavioral finance (see \citeasnoun{BT2003} for a survey) -- that can take several forms. Heterogeneity 
can first be considered as resulting from the diversity of the nature, the size and strategies of the economic agents: 
individual investors who invest their money to finance the education of their children, traders who manage money 
for their own account or for the account of their clients, institutional investors who manage pension funds, 
and so on... do not have the same financial resources (depending on their size), the same purposes (short term 
or long term profits and allocation frequency for example), or the same skills. All these differences make 
them focusing on different pieces of information and therefore anticipating differently the future value of the firms.

In this respect, \can{Dacorogna05} have recently provided evidence for the existence of a relation between 
the degree of heterogeneity amongst the market participants and the stage of development of financial markets. 
Indeed, relating the stage of maturity of a financial market to the speed of the hyperbolic decay of 
the auto-correlation function of the volatility of the assets traded on the  market under consideration, 
Di Matteo {\it et al.} suggest that the more mature and efficient the market is, the larger is the number 
of different classes of agents and strategies and the smaller is the effect of long-memory.
In addition, other phenomena such as mass psychology and contagion must be taken into account.
Indeed, clear evidence has shown that rumors \cite{banerjee93}, mimetism \cite{orlean95}, herding 
\cite{banerjee92,froot92,kirman93,cont00}, fashions \cite{shiller89} and so on, affect the agents' behavior.

In order to account for these various sources of heterogeneity but still keep a parsimonious representation, 
we provide a model that relates the investors' behavior to few heterogeneity parameters that allow accounting 
for their individual tendency to perform optimal anticipations on the basis of the flow of incoming news 
they receive or, on the contrary, on the basis of a self-referential approach which leads them to mainly 
focus on their past anticipations and on their past observations of the market volatility. This approach 
permits us to focus on the fact that, in addition to the heterogeneity of sizes and strategies, another 
main source of heterogeneity comes from the way the agents actually anticipate the many factors which impact 
future earnings of the firm and their volatility. These factors are captured, in our model, by help of a flow 
of incoming public and private information which can be considered as embedding several macroeconomic variables 
affecting stocks volatility such as the business cycles (see \can{schwert89} who has found a higher volatility 
of many key economic variables during the Great Recession), oil price whose volatility is important in explaining 
technology stock return volatility \cite{sadorsky03}, or inflation and interest rates which have large impacts 
on the stock market volatility \cite{kearney98}.

Then, we make the connection between the distributional properties of the heterogeneity parameters and 
the auto-covariance/auto-correlation functions of the realized-volatility. It allows us to discuss the kind 
of economic behaviors that yields the long memory of the realized-volatility time series.
Finally, the calibration of our model over the last decade, on a database of 24 US stocks of large and 
middle capitalizations, allows us reconstructing the distribution of the heterogeneity parameters and then 
to have access to the overall behavior of the investors. Notably different behaviors are observed, 
depending on the market phase under consideration with (i) a strong tendency to self-referential anticipations 
before the crash of the Internet bubble, and (ii) a redistribution in favor of the investors performing 
their forecasts on the basis the incoming piece of information after the crash.

The paper is organized as follows. The next section briefly recalls some basic stylized facts about the 
so-called realized-volatility, a measure of the volatility introduced by \can{andersen02} and 
\can{barndorff01}, among others. Then in section 2, we present our model of bounded rationality investors
with heterogeneous beliefs in order to investigate the impact of both the agents' bounded rationality
and their heterogeneous beliefs in a market that is assumed to perform an aggregation of the individual 
anticipations. In section 3, we discuss the calibration issues of the model and derive the asymptotic law 
of the estimator of the parameters of the model. The fourth section present our empirical results while 
the fifth section concludes.


\section{Stylized facts about the realized volatility of asset prices}
\label{sec1}

Many stylized facts about the volatility of financial asset prices have been reported in several studies 
over the recent years \cite[among many others]{ding83,lo91}. In particular, people now agree on the fact 
that (i) returns display, at any time scale, a high degree of variability which is revealed by the presence 
of irregular bursts of volatility and (ii) the volatility displays a positive auto-correlation over large 
time lags, which quantifies the fact that high (resp. low) volatility events tends to cluster in time, 
as already reported by \can{mandelbrot63} and \can{fama65} who first mentioned evidence that large changes 
in the price of assets are often followed by other large changes, and small changes are often followed 
by small changes. This behavior has also been noticed by several other studies, such as \can{baillie96}, 
\can{chou88} or \can{schwert89} for instance.

In this section, after we have recalled the definition of the notion of realized volatility, we document 
some of the common features of the process of the asset price realized volatility and relate them to the 
relevant literature.

\subsection{Definition of the realized volatility}

Based upon the quadratic variation theory within a standard frictionless arbitrage-free pricing environment, 
\can{andersen02} have suggested  a general framework for the use of high frequency data in the measurement, 
the modeling and the prediction of the daily volatility of asset returns. In fact, as also recalled by 
\citename{barndorff01} \citeyear{barndorff01,barndorff02}, when the underlying asset price process is a 
semi-martingale, the realized variance (the squared realized volatility) provides a consistent estimator of 
the quadratic variation, in the limit of large samples. Indeed, denoting by $P_{i,t}$ the $i^{th}$ observation 
of the asset price during the trading day $t$ and by $r_{i,t} = \ln(P_{i,t} )- \ln(P_{i-1,t})$ the continuously 
compounded return on the asset under consideration over the period $i - 1$ to $i$, the realized variance, at day $t$, defined by
\begin{equation}
\label{eq:SSR}
\hat \sigma_t^2 = \sum_{i=2}^{n_t} r_{i,t}^2,
\end{equation}
with $n_t$  the number of observations during this day, is a consistent estimator of the integrated variance of the price 
process. In addition, as shown by \can{barndorff01}, the asymptotic properties of this estimator are such that 
\be
\frac{\hat \sigma_t^2 - \int_{t_1}^t \sigma^2(s)~ ds}{\sqrt{\frac{2}{3} \ds \sum_{i=1}^{n_t} r_{i,t}^4}} 
\stackrel{{\cal L}}{\longrightarrow} {\cal N}(0,1), \qquad \text{and} \qquad \frac{\ln \hat \sigma_t^2 - 
\ln \int_{t_1}^t \sigma^2(s)~ ds}{\sqrt{\frac{2}{3} \ds 
\frac{\sum_{i=1}^{n_t} r_{i,t}^4}{\[\sum_{i=1}^{n_t} r_{i,t}^2 \]^2}}} \stackrel{{\cal L}}{\longrightarrow} 
{\cal N}(0,1),
\ee
where $\sigma(t)$ is the instantaneous (or spot) volatility of the log price process.

If the estimator (\ref{eq:SSR}) does not require the time series of asset returns to be homoscedastic during 
day $t$, it is however assumed that the returns are uncorrelated. Therefore, in order to account for the market 
microstructure effects, which may produce spurious correlations \cite[for instance]{roll84}, it is often necessary either 
to filter the raw series of intraday returns to remove these correlations\footnote{See the comparative study 
in \can{bollen02} for instance.} or to focus on sufficiently large time scales -- 5 minutes instead of 1 minute 
or tick by tick quotations, for instance -- in order to smooth out the microstructure effects and correlations. 
The immediate drawback of this later approach is to decrease the number $n_t$ of available observations, which 
may bias the estimates.

\begin{landscape}
\begin{figure}
\begin{center}
\includegraphics[width=11cm]{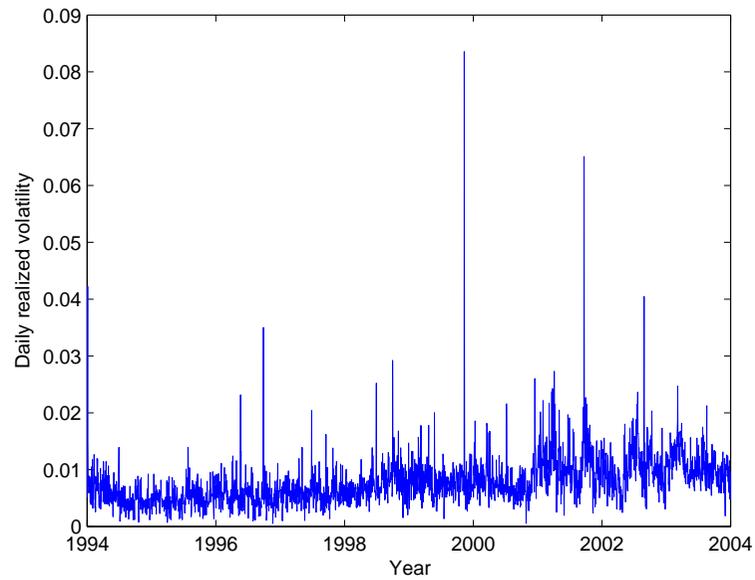} 
\includegraphics[width=11cm]{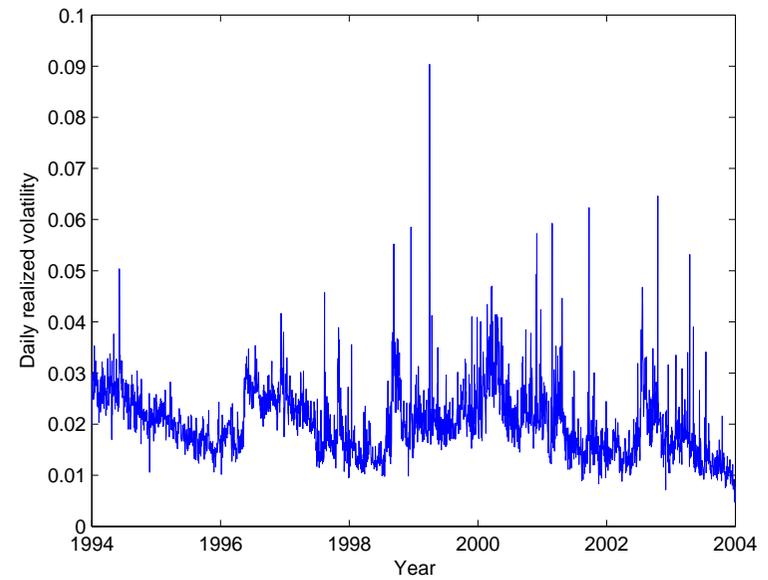}
\end{center}
\caption{\label{fig:VR} Daily realized volatility of two assets from 01/01/1994 to 12/31/2003. 
On the left a middle capitalization : The Washington Post, and on the right a large capitalization : Coca Cola.}
\end{figure}
\end{landscape}

\vspace{1cm}
\centerline{\it [Insert figure~\ref{fig:VR} about here]}
\vspace{1cm}

As an illustration, we present, on the figure \ref{fig:VR}, the realized volatility for two time series drawn from our intraday database of 24 US stocks prices (see section 4 for details on the database). On the left panel, we can observe the realized volatility of the daily returns 
of a middle capitalization stock (The Washington Post) during the time period from 01/01/1994 to 12/31/2003 while, 
on the right panel, is depicted the realized volatility of a large capitalization stock (Coca-Cola) over the 
same time period. Since our investigation of the time dependence between the intraday {\em returns} of these asset
prices has not revealed the presence of a significant correlation beyond the one 
minute time scale, the market microstucture effect are negligible, which allows us to 
directly estimate the realized volatility from the one-minute raw returns. 

\begin{landscape}
\begin{figure}
\begin{center}
\includegraphics[width=11cm]{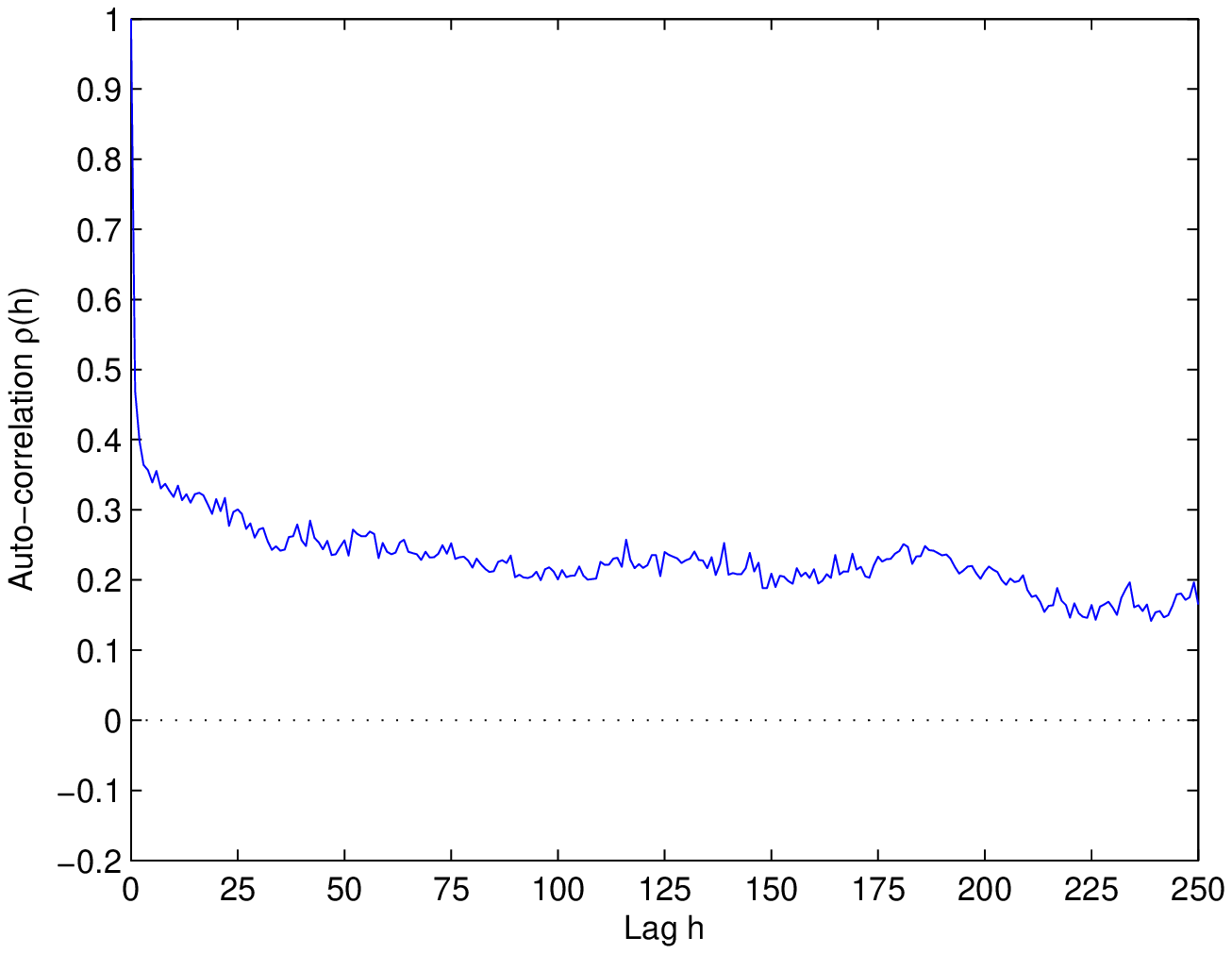} 
\includegraphics[width=11cm]{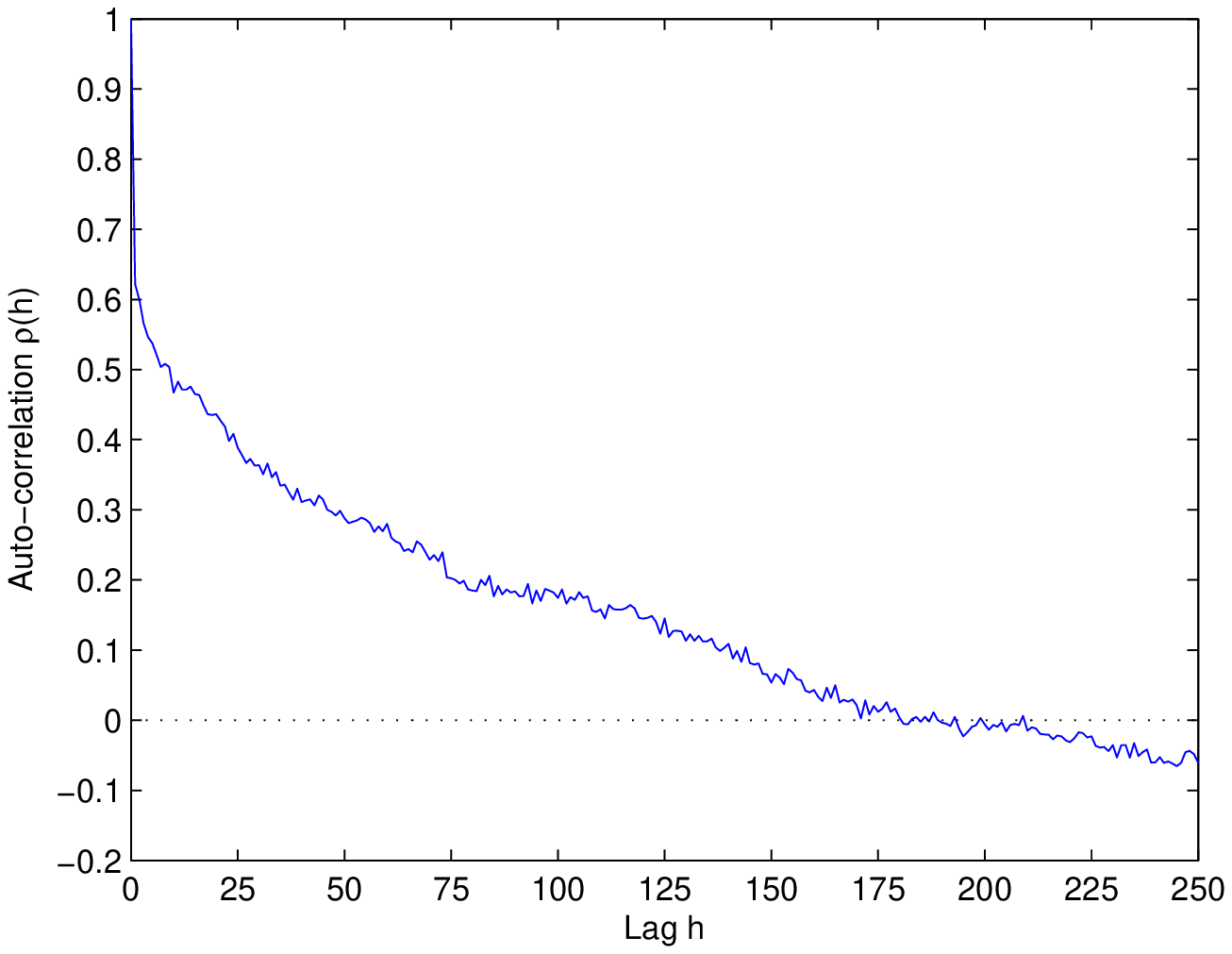}
\end{center}
\caption{\label{fig:VR_acfs}Auto-correlation function of the realized volatility of two assets from 01/01/1994 to 12/31/2003.
On the left a middle capitalization : The Washington Post, and on the right a large capitalization : Coca Cola.} 
\end{figure}
\end{landscape}

\vspace{1cm}
\centerline{\it [Insert figure~\ref{fig:VR_acfs} about here]}
\vspace{1cm}

On the figure~\ref{fig:VR_acfs}, we have drawn the auto-correlation function of the previous realized volatility 
(still The Washington post on the left and Coca-Cola on the right) from lag 0 to lag 250 days, which corresponds 
to a one year period or so. The slow decay is characteristic of the long memory. However, the two graphs seem 
different. For The Washington Post the auto-correlation falls down to 0.4 quickly and then decreases very slowly, 
whereas for Coca-Cola, it decreases more regularly and the auto-correlation becomes not significantly different 
from zero beyond lag 180, or so. This general feature could mean that the correlation function of the volatility 
of large capitalization stocks would exhibit shorter memory than middle capitalization stocks. This remark, 
that will be confirmed later, is in line with the observation reported by \can{Dacorogna05} and according 
to which the more efficient a stock (or a market), the faster the decay of the correlation function of its volatility.


\subsection{Normality of the log-volatility}
\label{subsec:LogVolNorm}

We now turn to the distributional properties of the realized volatility. As suggested by many previous studies 
\cite[among others]{andersen01,andersen01b,barndorff02}, the log-normal distribution\footnote{Notice that \can{barndorff02} 
also show that the inverse Gaussian law provides an accurate fit of the distribution of the log-volatility. 
In fact, the log-normal and the inverse Gaussian  distributions are indistinguishably close to each other over 
the entire range of interest.} provides an adequate description of the distribution of the volatility, at least 
in the bulk of the distribution. These observations clearly support the modeling of the realized volatility 
in terms of log-normal models, which goes back to \can{clark73}  (see also \can{Taylor86}).

\begin{landscape}
\begin{figure}
\begin{center}
\includegraphics[width=11cm]{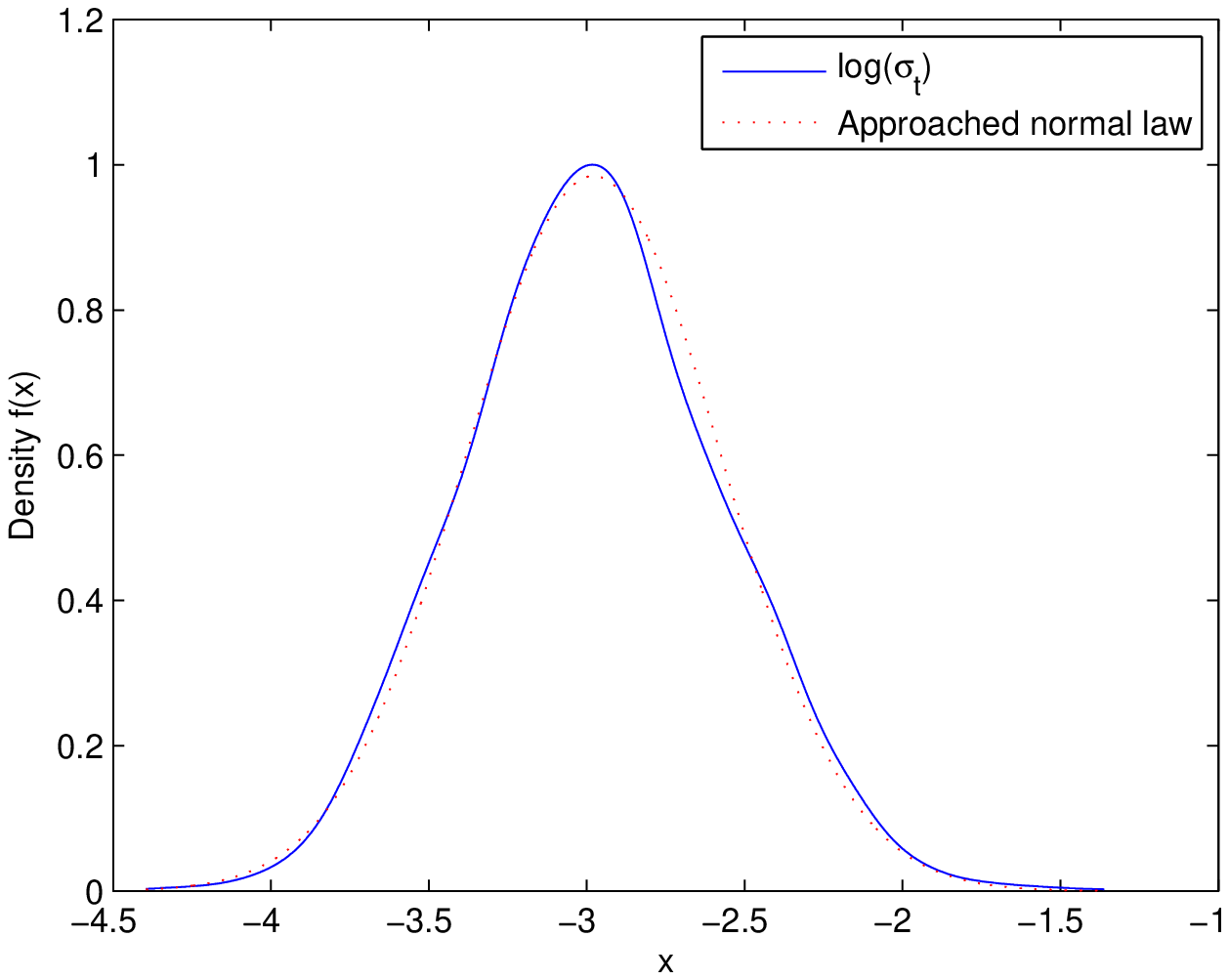} 
\includegraphics[width=11cm]{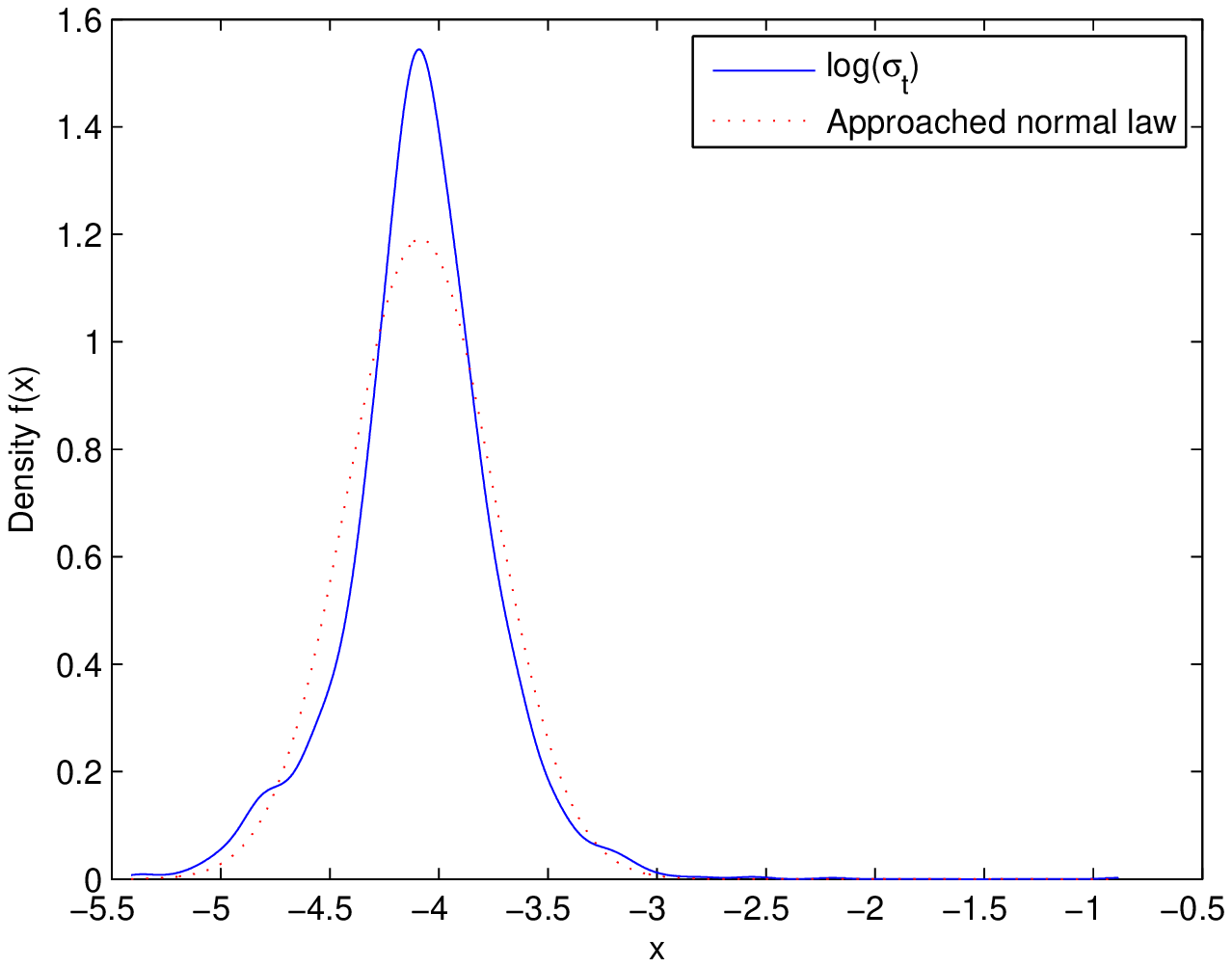}
\end{center}
\caption{\label{fig:LVR_densities} Density of $\log(\hat \sigma_t)$ and their approached normal law for different assets.
On the left a middle capitalization : Microchip Technology inc., and on the right a large capitalization : Procter \& Gamble co.} 
\end{figure}
\end{landscape}

Starting from these observations, let us illustrate the relevance of the log-normal model for the realized volatility 
of the returns on the prices of the assets in our database. To this aim, let us denote by $\{\omega_t\}_{t\ge 1}$ 
the logarithm of the series of the realized volatility. Figure \ref{fig:LVR_densities} shows the kernel estimate 
\cite{pagan99} of the density of the log-volatility of a  mid-cap (Microchip Technology) on the left panel and 
of a large cap (Coca Cola) on the right one, over the whole time period. The density of log-volatility of the price returns 
of Microchip Technology seems very close to a normal density at the naked eye. On the overall, it is the case 
for all the mid-caps. On the contrary, the densities of the large caps present sharp peaks and fat tails so that 
they significantly depart from the normal density. These visual impressions will be formalized later on, 
in section~\ref{sec:computations}.

\vspace{1cm}
\centerline{\it [Insert figure~\ref{fig:LVR_densities} about here]}
\vspace{1cm}


\section{\label{sec:heterogeneity}Heterogeneity model}

As recalled in introduction, several explanations have been proposed, in the now large body of literature 
about the long memory of financial time series, to point out the various origins of this phenomenon. 
The first one, addressed by \can{granger80}, concerns the nature of financial time series. Indeed, they 
consider that financial series results from the aggregation of micro-variables and that this aggregation is 
responsible for the long memory. The second one is the presence of infrequent structural breaks (also called 
structural changes) in financial time series \cite[among others]{diebold01,gourieroux01,granger04,gadea04,davidson05}. 

Our model is based upon \can{granger80}'s proposition; it explains the long memory of the log-volatility 
of asset prices by the aggregation of micro-variables intended to represent the  heterogeneous 
expectations of each market participant. For, we suppose that the market aggregates the agents' anticipations 
of the log-volatility and that this aggregation drives the realized log-volatility. We will formalize this assumption later on.

\subsection{General framework}

The role of the log-volatility as the central object underpinning our model comes from the remark that, 
as recalled in the previous section, it is reasonable, in a first approximation, to consider that the realized 
volatility follows a log-normal distribution. As an additional hypothesis, we will assume that 
\begin{quote}
{\bf (H1)}\hspace{1cm} {\em the log-volatility process $\{ \omega_t\}_{t \in {\mathbb Z}}$ follows a Gaussian stochastic process,}
\end{quote}
which obviously ensures that the volatility itself has a log-normal stationary distribution.

Now, the standard economic theory tells us that any rational agent facing a decision problem aims at optimizing 
the output of her actions, based upon the entire set of information she has at her disposal. In the present context, 
it means that any rational investor strives for the best prediction of the future realized volatility 
$\hat \sigma_{t+\tau}$, $\tau \ge 1$, based on the set $\{ \sigma_1, \dots, \sigma_t\}$ of her past observations. 
In the sense of the minimum mean squared error, the best predictor is given by
\bea
\hat \sigma_{t+\tau} &=& \E \[ \sigma_{t+\tau} |  \sigma_1, \dots, \sigma_t \],\\
&=& \E \[\left. \e^{\omega_{t+\tau}} \right| \omega_1, \dots, \omega_t\],\\
&=& {\rm e}^{2 \sigma_\omega(\tau)^2} \cdot {\rm e}^{\hat \omega_{t+\tau}},
\eea
where $\hat \omega_{t+\tau}$ denotes the best predictor of the log-volatility, based on the same set of 
observations and $\sigma_\omega(\tau)^2$ is the ($\tau$-step ahead) prediction error, {\it i.e.} the mean 
squared error $\E \[ \(\hat \omega_{t+\tau} - \omega_{t+\tau} \)^2 \]$. Since the log-volatility is assumed 
to follow a Gaussian process, the best predictor $\hat \omega_{t+\tau}$ is given by a linear combination of 
the past observations and the past anticipations. In particular (see \can[pp. 162-168]{brockwell90})
\be
\label{eq:pred}
\hat \omega_{t+1} = \bar \omega + \sum_{i=0}^{t-1} \phi_{t,i} \cdot \( \hat \omega_{t-i} - \omega_{t-i}\),
\ee
where the sequence of coefficients $\{\phi_{t,i}\}_{i\ge 0}$ and the long term mean $\bar \omega$ depend on the specific 
model each investor relies on and on the particular calibration method she uses to estimate the structural 
parameters of her model. Thus, one expects that the $\phi_i$'s are specific to each market actor so that each 
investor performs different expectations about the level of the future realized volatility. In addition, each 
agent can incorporate some exogenous economic variables or some piece of (private) information she thinks to 
improve her prediction. 

Besides, considering that the market carries out an aggregation of the agents' anticipations, we postulate that 
\begin{quote}
{\bf (H2)}\hspace{1cm} {\em the realized log-volatility is the average of all the individual anticipations.}
\end{quote}
 Thus, denoting by 
$\hat \omega_{i,t}$ the agent $i$'s forecast, the realized log-volatility is given by
\be
\label{eq:aggregation}
\omega_t(n) = \frac{1}{n} \sum_{i=1}^{n} \hat \omega_{i,t},
\ee
if the market in made of $n$ participants. Under the assumption of an infinitely large number of investors, 
the realized log-volatility writes
\be
\omega_t = \lim_{n \rightarrow \infty} \omega_t(n).
\ee

Our approach can appear utterly simplistic insofar as the real process yielding the realized volatility 
certainly involves non-linear transforms of the individual anticipations before they are actually aggregated 
through the price formation process, which is well-known to rely on various positive or negative feedback 
mechanisms \cite{shiller00,sornette03}. However, in this first attempt to capture the impact of the 
heterogeneity of the investors' anticipations on the dynamics of the realized volatility, and in the 
absence of arguments allowing us to model these non-linear interactions, it is necessary to restrict 
ourselves to this assumption of linearity.

Moreover, in order to be able to make tractable calculations, we need some other simplifying assumptions. 
Focusing on agent $i$, we denote by $\hat X_{i,t}$ its expectation about the future level of the {\it excess}
log-volatility over its long term mean
\be
\hat X_{i,t}=\hat \omega_{i,t} - \bar \omega
\ee
on day $t$, and by 
\be
\bar X_{n,t}=\omega_{t}(n) - \bar \omega = \frac{1}{n} \sum_{i=1}^n \hat X_{i,t}
\ee
the excess of the realized log-volatility over its long term mean. We assume that the anticipation $\hat X_{i,t}$ 
depends on the anticipation, $\hat X_{i,t-1}$, the agent made the day before and on the excess realized log-volatility 
$\bar X_{n,t-1}$, but also on a public piece of information $\varepsilon_t$ as well as on a private piece of 
information $\eta_{i,t}$ according to the recursion equation
\be
\label{eq:ARDyn}
\hat X_{i,t} = \varphi_i \cdot \hat X_{i,t-1} + \psi_i \cdot \bar X_{n,t-1}+ c_i\varepsilon_t + \eta_{i,t} 
\qquad t\geq 0, \qquad i=1,2,\dots,n.
\ee
From equation~(\ref{eq:pred}), one should expect that $\psi_i = - \phi_i$ if we consider rational agents who strive for the best prediction (in the minimum mean-squared sense) of the volatility. However, 
in order to generalize our model to the case where the market participants can be {\em bounded} rationality agents, we allows for $\psi_i \neq -\phi_i$. Nonetheless, we will assume that these coefficients remains constant from time to time.

Let us stress that the first order dynamics (\ref{eq:ARDyn}) can seem too simple, 
and it probably is. However, our approach amounts to postulate that the agents assume that the (log)-volatility follows some kind of ARCH process, which is quite reasonable. In addition, even if real agents use more sophisticated prediction schemes to perform their expectations about the future level of the realized (log)-volatility, and thus use higher order dynamics, we can still make the assumption that these higher order dynamics can be reduced to an aggregation of first order dynamics. 
Therefore, $\hat X_{i,t}$ does not exactly characterize the expectation of individual agent but more precisely of a 
class of agents.

In this basic setting, the $n$-dimensional vector $\hat X_t = \(\hat X_{1,t}, \dots, \hat X_{n,t}\)'$  follows a first 
order (vectorial) auto-regressive process. The Gaussian noises $\{\varepsilon_t\}$, $\{\eta_{1,t}\}$, $\{\eta_{2,t}\}$, 
\dots, are independent, centered, and of variance $\Var(\varepsilon_t) = \sigma_\varepsilon^2$ and $\Var(\eta_{i,t}) = 
\sigma_{\eta_i}^2$, $\forall i$, respectively. Moreover the structural coefficients $\(\varphi_i, \psi_i\)$ and $c_i$, which 
characterize the investors' behavior, can be considered as the result of independent draws from the same law 
(called the heterogeneity distribution). These draws are independent of the values of the noises and the variables 
$\(\varphi, \psi\)$ and $c$ are independent of each other. The parameter $c$ that allows introducing heteroscedasticity 
must be strictly positive, while the distribution of $(\phi, \psi)$ has to fulfill some hypotheses, which will be made 
explicit hereafter, in order to ensure the stationarity of the stochastic process $\{\hat X_t\}_{t \in {\mathbb Z}}$ and of the aggregated excess realized volatility $\{\bar X_{n,t}\}_{t \in {\mathbb Z}}$.

The appeal of the dynamic (\ref{eq:ARDyn}) rests on the simple interpretation that can be made of the structural 
coefficient $(\varphi,\psi)$ and of its distributional properties. It is obvious that the impact of the previous 
anticipation $\hat X_{i,t-1}$ on the current anticipation $\hat X_{i,t}$ depends on the value of the heterogeneity 
coefficient $\varphi_i$:
\begin{enumerate}
\item when $\varphi_i$ is close to (but less than) one, the impact of the previous anticipation on the current 
anticipation is very important and, unless a very large piece a information arrives -- {\it i.e.} unless one 
observes a large $\epsilon_t$ and/or a large $\eta_{i,t}$ -- the current value of the expected excess 
log-volatility $\hat X_{i,t}$ will be very close to the previous one. On the overall, the agent believes in the 
continuity of the previous market conditions;
\item when $\varphi_i$ is close to (but larger than) minus one, the impact of the previous anticipation still 
remains more important than the arrival of a new piece of information, but the agent exhibits a systematic 
tendency to believe in the reversal of the volatility since her anticipation appears to be the opposite 
of the one she made the day before;
\item eventually, when $|\varphi_i|$ is close to 0, the anticipations are only scarcely related to those made 
the day before and mainly rely on the flow of incoming news.
\end{enumerate}
To sum up, in the two first situations, we can notice that the market is highly self-referential: the impact 
of a new incoming piece of information remains weak, all the more so the closer to one the magnitude of 
$\varphi_i$. On the contrary, when $|\varphi|$ is close to zero, the degree of reactivity of the market participants 
to a new incoming piece of information is high. Thus, the magnitude of $\varphi$ can be seen as a way to quantify 
the degree of efficiency of the market. The same considerations obviously holds for $\psi_i$, but instead of referring to the past anticipation it refers to the publicly observed past realized volatility.

\subsection{Properties of the (log-) volatility process}

Equation~(\ref{eq:ARDyn}) can be written in a more compact form as follows
\be
\label{eq:Dyn2}
\hat X_t = A \hat X_{t-1} + \epsilon_t \cdot C + \eta_t,
\ee
where $A= D + \frac{1}{n} \Psi \cdot 1_n'~$ is an $n \times n$ matrix
with $D= {\rm diag}(\varphi_1, \dots, \varphi_n)$, $\Psi = \(\psi_1, \cdots, \psi_n \)'$ and 
$1_n =\underbrace{\(1, \dots, 1\)'}_{n~times}$, while $C = \(c_1, \dots, c_n\)'$ and 
$\eta_t = \( \eta_{1,t}, \dots, \eta_{n,t}\)$. Besides, the range of the distribution of 
$(\varphi,\psi)$ is assumed to be such that the spectral radius of $A$ is less the one. This condition is necessary and sufficient to ensure the stationarity of the vectorial process $\{\hat X_{t}\}_{t \ge 0}$.

Under the assumption $||A|| <1$, {\em that will be assumed in all the sequel of this article}, 
the stationary and causal solution of equation~(\ref{eq:Dyn2}) is given by
\be
\hat X_t = \sum_{k=0}^\infty A^k C \epsilon_{t-k} + \sum_{k=0}^\infty A^k  \eta_{t-k},
\ee
so that, as proved in appendix \ref{app:CV_beta}:
\begin{proposition}
\label{propMain}
In the limit of a large number of economic agents, $n \rightarrow \infty$, the excess realized log-volatility $\{\bar X_t\}_{t \in \mathbb Z}$ follows an infinite order auto-regressive moving average process
\be
\label{eq:mdsofuh}
 \(1 - \sum_{k=0}^\infty {\rm E}\[\psi \phi^k\] L^{k+1}\)\bar X_t = {\rm E}\[c\] \cdot \( \sum_{k=0}^\infty {\rm E}\[\phi^k\] L^k\) \epsilon_t,
\ee
where $L$ denotes the lag operator.
\end{proposition}
Not surprisingly, the flows of private information $\{\eta_{i,t}\}$ does not come into play since, under our assumptions, it does not convey any piece of information, on average across agents. 

As a byproduct of the proposition above, we see that the solution to equation~(\ref{eq:mdsofuh}) can conveniently be expressed in terms of an infinite order moving-average provided that $\sum_{k=0}^\infty \E\[\psi \phi^k\] z^{k+1} \neq 1$ for all $z$ inside the unit circle:
\be
\label{eq:slkdn2}
\bar X_t =  \E\[c\] \cdot \sum_{k=0}^\infty  \tilde \beta_k \epsilon_{t-k},
\ee
where the $\tilde \beta_k$'s are formally given by the coefficients of the power series
\bea
\label{eq:beta_tilde}
\sum_{k=0}^\infty \tilde \beta_{k} x^{k} 
&:=& \ds \E\[ \ds\frac{1}{1- x \cdot \varphi}\] \(\ds 1-x\E\[\frac{\psi}{1- x \cdot \varphi} \]\)^{-1}~ ,\\
&=& \ds \E\[ \ds\frac{1}{1-x\varphi}\] \(\ds 1-\E\[\frac{x g(\phi)}{1-x\varphi} \]\)^{-1}~ ,
\eea
where $g(\phi) = \E[\psi | \phi]$.

The process $Y_t := \( \sum_{k=0}^\infty \E\[\phi^k\] L^k\) \epsilon_t$, is stationary if and only if $\sum _{k=0}^\infty \E\[\phi^k\]^2 < \infty$, which requires that the law of $\phi$ be concentrated on $(-1,1)$. It is however not sufficient, as shown by \citeasnoun{goncalves88}. Furthermore, $Y_t$ exhibits long memory provided that $\sum_{k=0}^\infty \vert \E\[\phi^k\] \vert=\infty$. Restricting our attention to the case where $Y_t$ is said to exhibit long memory if its spectral density 
\be
f_Y(\lambda) = \frac{\sigma_\varepsilon^2}{2 \pi}\left| \sum_{k=0}^\infty \E\[\phi^k\] e^{-i k \lambda} \right|^2 \sim \lambda^{-2d_Y}, \quad d_Y \in \(-\frac{1}{2}, \frac{1}{2}\),
\ee
we conclude that the density of $\phi$ -- if it exists -- should behave as $(1-\phi)^{-d_Y}$ as $\phi$ goes to one.
Actually, $Y_t$ exhibits long-memory  if $d_Y \in (0,1/2)$ and anti-persistence if $d_Y \in (-1/2,0)$. The upper bound on $d_Y$ is necessary in order for the process to be second order stationary.

As for the filter $A(L):=1 - \sum_{k=0}^\infty \E\[\psi \phi^k\] L^{k+1}$ in the left-hand side of (\ref{eq:mdsofuh}), it is well defined if $\sum_{k=0}^\infty \E\[\psi \phi^k\]^2 < \infty$ which, by Cauchy-Schwartz inequality, is satisfied if $\E\[\psi^2\] < \infty$ and $\sum_{k=0}^\infty \E\[\phi^{2k}\] < \infty$. This requirement is, however, not necessary. As a further hypothesis, we will assume that $\sum_{k=0}^\infty \E\[\psi \phi^k\]^2 x^{k+1} \neq 1$ for all  $x \in[-1, 1]$ which ensures that $A(z)$ does not vanish for any $z$ inside the unit circle. The behavior of $A(z)$, as $z$ goes to one, depends on the conditional expectation of $\psi$ given $\phi$, namely $g(\phi) := \E\[\psi | \phi\]$. Either $A(z)$ remains finite as $z \to 1$, and $X_t$ exhibits long memory if and only if $Y_t$ itself exhibits long memory or $A(z)$ diverges hyperbolically, as $z \to 1$, and the memory (short or long) of the log-volatility process results from a mix between the properties of the distribution of $\phi$ and of the conditional expectation $g(\phi)$. More precisely, we can state that (see the proof in Appendix~\ref{App:Prop2})
\begin{proposition}
\label{prop:memory}
Under the assumptions above, if the density of $\phi$ is $f(\phi) \sim (1-\phi)^{-\alpha}$, $\alpha \in (0,1/2)$, and if the conditional expectation of $\psi$ given $\phi$ is $g(\phi)\sim (1-\phi)^{\beta}$, $\beta > \alpha-1/2$, as $\phi \to 1^-$, the spectral density of the log-volatility process behaves as $f_X(\lambda) \sim  \lambda^{-2 \min\{\alpha, \beta\}}$, as $\lambda \to 0$.
\end{proposition}

As an illustration of this result, let us assume that $\phi \stackrel{law}{=}$ Beta$(-\alpha, 1+\alpha)$. We deduce that 
$ \E\[ \frac{1}{1-x\phi}\] = (1-x)^\alpha.$ 
If, in addition, we assume that $\E[\psi|\phi] = (1-\phi)^\beta$, we get
$ \E\[ \frac{g(\phi)}{1-x\phi}\] = \frac{\Gamma(1+\alpha + \beta)}{\Gamma(1+\alpha)\Gamma(1 + \beta)}
F(1,-\alpha;1+\beta;x),$ where $F$ is the hypergeometric function (see \citeasnoun{abramowitz65} for the definition). Thus, equation (\ref{eq:beta_tilde}) writes
\be
\label{eq:beta_F}
\sum_{k=0}^\infty \tilde \beta_{k}x^k = \frac{(1-x)^\alpha}
{\ds 1-\frac{\Gamma(1+\alpha + \beta)}{\Gamma(1+\alpha)\Gamma(1 + \beta)}x F(1,-\alpha;1+\beta;x)}.
\ee

Under the assumption that $\beta>1$, which assures that the denominator in equation (\ref{eq:beta_F}) remains bounded for all $|x| \le 1$, the series $\sum_{k=0}^\infty \tilde \beta_{k}x^k $ behaves like $C(1-x)^\alpha$ in the neighborhood of 1
and thus diverges hyperbolically if $\alpha<0$. 
The expression of the spectral density is then deduced from equations (\ref{eq:slkdn2}) and (\ref{eq:beta_F}):
\be
\label{eq:sd_fX}
f_X(\lambda) = \frac{\E\[c\]^2 \sigma_\epsilon^2}{2\pi}
\frac{|1 - e^{-i\lambda}|^{2\alpha}}
{\left|1 - \frac{\Gamma(1+\alpha+\beta)}{\Gamma(1+\alpha)\Gamma(1+\beta)}e^{-i\lambda}F(1,-\alpha;1+\beta,e^{-i\lambda})\right|^2}
\ee
and, since $f_X(\lambda) \underset{\lambda \rightarrow 0}{\sim} C' |\lambda|^{2\alpha}$, the auto-correlation function 
reads $\rho(h) \sim K |h|^{-2\alpha-1}$ as $|h|\rightarrow \infty$, provided that the second order stationarity condition $\alpha > -1/2$ holds.

\begin{landscape}
\begin{figure}
\begin{center}
\includegraphics[width=10cm]{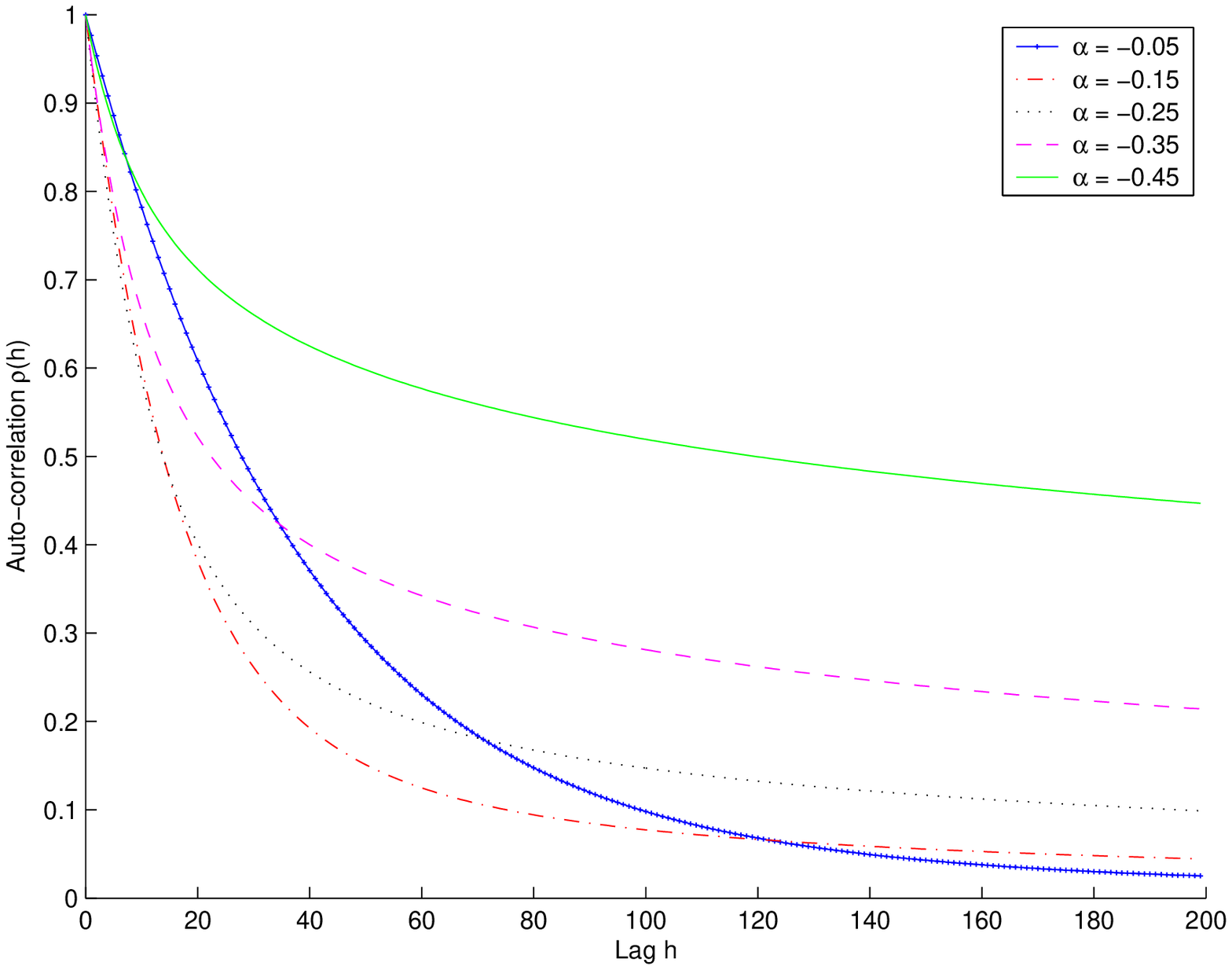}
\includegraphics[width=10cm]{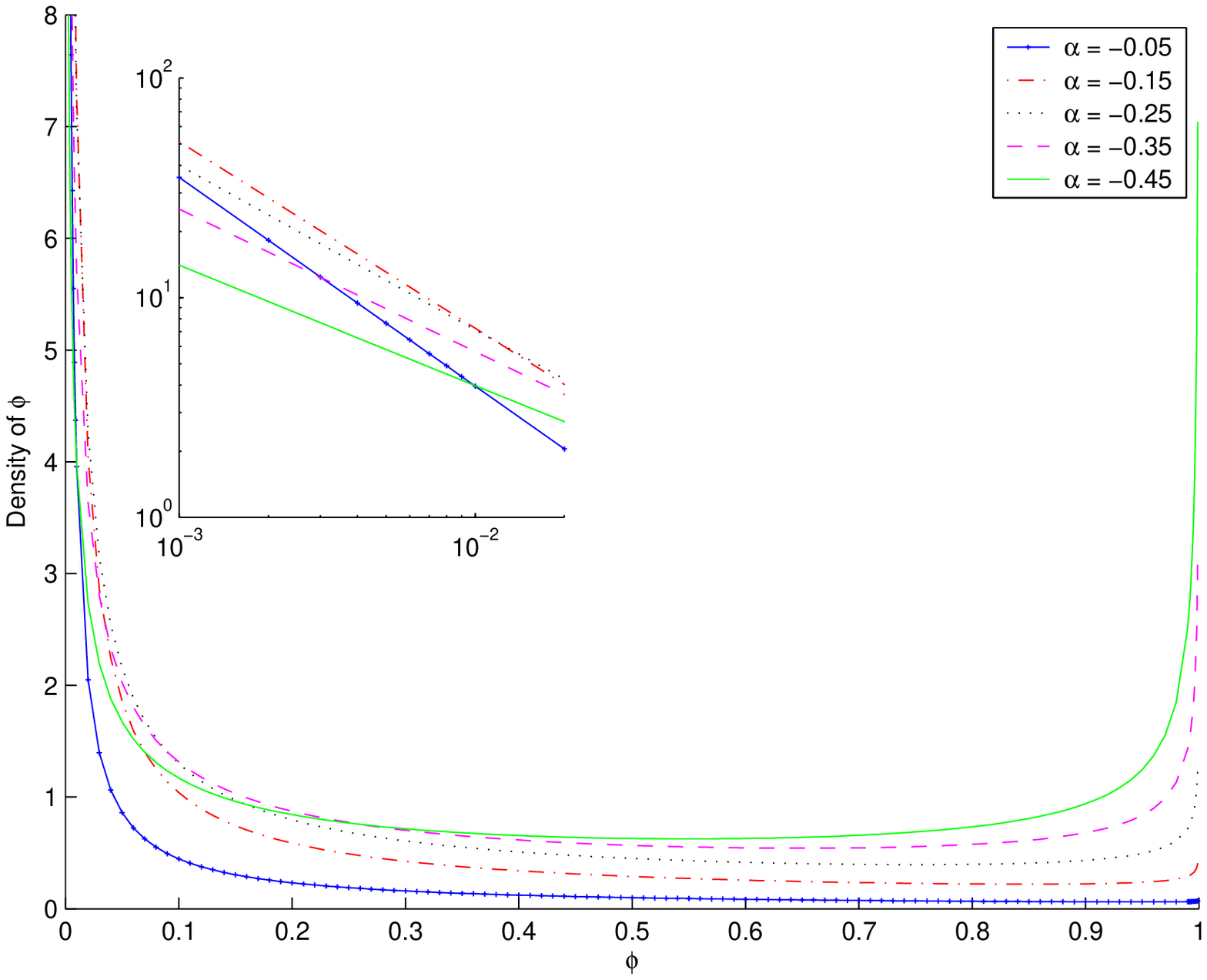}
\end{center}
\caption{\label{fig:ACF_fft_alpha} The left panel and the right panel represent respectively
the auto-correlation function over 200 days and the density of the heterogeneity coefficient $\alpha$ 
for different values of $\alpha$ ranging in $[-0.45,-0.05]$ and $\beta = 1.5$. The law of $\phi$
is a Beta($-\alpha$,$1+\alpha$) and its density equal to
$f(\phi) = \frac{1}{B(-\alpha,1+\alpha)}\phi^{-\alpha-1}(1-\phi)^\alpha$ explains the divergence near 0 and 1.} 
\end{figure}
\end{landscape}

For the sake of completeness, let us discuss the role played by the two parameters $\alpha$ and $\beta$ in the model.
As $\phi$ is assumed to follow a Beta($-\alpha,1+\alpha$) law, then 
$\E[\phi] = -\alpha \in [-0.5,0]$. It implies that the anticipation 
of an agent is not mainly based on her past anticipation. Recall that $|\phi|$ should be close to one in order to observe a strongly self-referential behavior. In other words, the agents attach more importance
to the realization of the past volatility and to the information flows, on average.
The closer $\alpha$ to 0, the less importance is given to the past anticipation. Consequently, as the
anticipations of the agents are more based on common facts (past realized volatility and informations),
then the heterogeneity of the agents' beliefs decreases. As a conclusion the long memory should decrease.
This phenomenon is shown on the left panel of figure \ref{fig:ACF_fft_alpha}, 
where auto-correlation functions for a fixed $\beta=1.5$
and different values of $\alpha$ ranging in $[-0.45,-0.05]$ are drawn. 

\vspace{1cm}
\centerline{\it [Insert figure~\ref{fig:ACF_fft_alpha} about here]}
\vspace{1cm}

The right panel of figure \ref{fig:ACF_fft_alpha} shows the densities of $\phi$ for the same set of values of $\alpha$. This figure confirms that 
the closer $\alpha$ to -0.5, the stronger is the divergence near 1. It means that the relative proportion
of agents attaching importance to their past anticipation increases and thus the long memory increases. 
Furthermore, we notice that the speed of the decrease of the auto-correlation function and the divergence of 
the density function are strongly related. It means that $\alpha$ impacts both on short and long memory, which is not a surprise since -- with our simple parameterization -- $\alpha$ controls both the behavior of the density of $\phi$ close the one and in the neighborhood of zero.

As for $\beta$, in order to illustrate its role, 
we focus on the other heterogeneity parameter $\psi$ whose mean is
$\E[\psi] = \frac{\Gamma(1+\alpha+\beta)}{\Gamma(1+\alpha)\Gamma(1+\beta)}$. 
The limits of $\E[\psi]$ when $\beta$ tends to 1 and to infinity are respectively
$\E[\psi] \underset{\beta \rightarrow 1}{\longrightarrow} 1+\alpha$ and 
$\E[\psi] \underset{\beta \rightarrow +\infty}{\longrightarrow} 0 $.
So, the larger $\beta$, the closer $\psi$ to 0 on average. 

\begin{figure}
\begin{center}
\includegraphics[width=10cm]{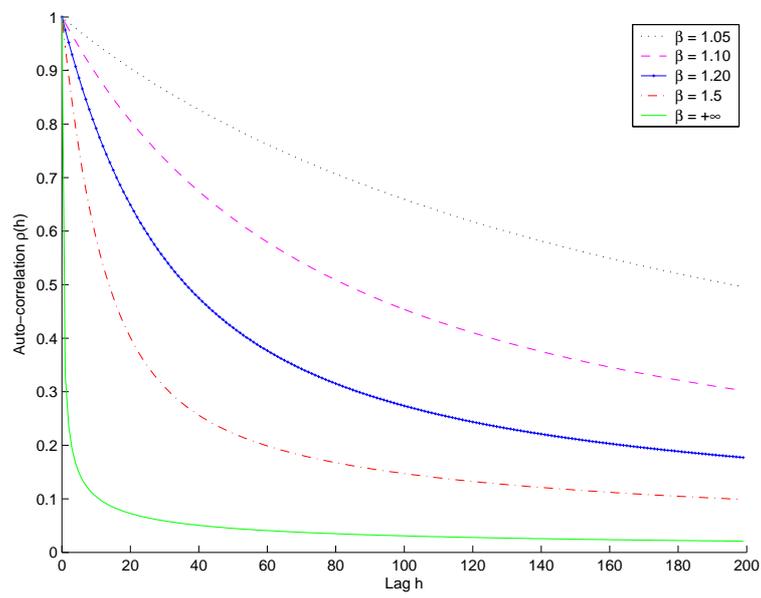}
\end{center}
\caption{\label{fig:ACF_fft_beta} Auto-correlation function over 200 days for $\beta = 1.5$ and different values of $\alpha$.} 
\end{figure}

Now, by aggregation of the system (\ref{eq:ARDyn}), we obtain that the realization of the volatility on day $t$
is equal to a quantity depending on the past anticipations plus the past realization of the volatility
times the mean of $\psi_i$ plus the information flows.
Thus the aggregation of equation (\ref{eq:ARDyn}) could be interpreted as an an AR(1) process
where $\E[\psi]$ would be the auto-regressive coefficient.
Even if the quantity depending on the past anticipations impacts on the short memory as we previously
pointed out, all things being equal furthermore, 
when $\E[\psi]$ decreases there are less short-term correlations and then the short memory falls.
It is the reason why when $\E[\psi]$ gets close to 0 (that is equivalent to $\beta$ tends to infinity), 
the short memory drops (see figure \ref{fig:ACF_fft_beta}). 

\vspace{1cm}
\centerline{\it [Insert figure~\ref{fig:ACF_fft_beta} about here]}
\vspace{1cm}

\subsection{Examples}

Let us now discuss in details several typical examples of dynamics encompassed by our model and relate them with the agents' behaviors.

\subsubsection{Case 1: Rational agents}

As recalled in the previous section, rational agents base their anticipations at time $t$ on the innovations 
$\hat X_{i,s} - \bar X_{n,s}$, $s \le t-1$. Thus, setting $\psi_i = - \phi_i$, for all $i=1, \ldots, n$ 
allows us to model the behavior of the realized volatility when the investors are fully rational agents 
since it allows us to retrieve the expression of the optimal predictor (\ref{eq:pred}), in the particular case 
where the agents focus on the last innovation $\hat X_{i,t-1} - \bar X_{n,t-1}$ only. In this case, 
the dynamics on the individual anticipation writes
\be
\hat X_{i,t} = \varphi_i \cdot \(\hat X_{i,t-1} -  \bar X_{n,t-1}\)+ c_i \cdot \varepsilon_t + \eta_{i,t} 
\qquad t\geq 0, \qquad i=1,2,\dots,n,
\ee
and by equation (\ref{eq:beta_tilde}), we immediately obtain $\ds \sum_{k=0}^\infty \tilde \beta_k x^k = 1$, so that
\be
\bar X_t = \E\[c\] \epsilon_t.
\ee
Notice, in passing, that we do not need each agent to be rational by setting $\psi_i = - \phi_i$, but only that 
$\E[\psi|\phi] = - \phi$ which means that the agents are not necessarily individually rational but only on average.
Thus, if the agents are rational (individually or in average), the excess of the realized log-volatility only reflects 
the aggregated information, publicly available at time $t$, namely $\E\[c\] \epsilon_t$. This result 
is not surprising insofar as rational investors perform optimal anticipations and, therefore, the 
realized volatility can only convey the piece of information not present in the past innovation 
$\hat X_{i,t-1} - \bar X_{n,t-1}$. Then, since the information flow is assumed to be a white noise, 
the realized volatility should also be a white noise in the case where all the investors were rational.

\subsubsection{Case 2 : Absence of heterogeneity}

As a second basic example, let us assume that the investors only focus on the past realized volatility and completely neglect their past anticipations, {\em i.e.}, $\phi=0$. Consequently, the dynamics of the individual anticipations writes
\be
\hat X_{i,t} = \psi_i \cdot \bar X_{n,t-1} + c_i\varepsilon_t + \eta_{i,t} 
\qquad t\geq 0, \qquad i=1,2,\dots,n.
\ee
By summation over all the agents $i$, and in the limit $n \rightarrow \infty$, one gets
\be
\bar X_t = \E \[\psi\] \cdot \bar X_{n,t-1} + \E \[c\] \varepsilon_t
\qquad t\geq 0,
\ee
which is a simple AR(1) process exhibiting only short memory since
\be
\rho(h) = \E \[\psi\]^{|h|}.
\ee
Thus, as noticed in the previous section, we remark again that the heterogeneity coefficient $\phi$ is responsible for the long memory while $\psi$ 
mainly impacts the short memory. It clearly shows that the long memory phenomenon is rooted in the self-referential 
behavior of the investors, and more precisely in the self-reference to theirs own past anticipations and not 
to the past realized volatility.

\subsubsection{Case 3: Absence of reference to the past realized volatility}

At the opposite of the previous case, let us consider that each agent only bases 
her present anticipation on her past anticipation and on the flow of incoming news, but that she neglects 
the past realized volatility so that the dynamics of the individual anticipations reads
\be
\hat X_{i,t} = \varphi_i \cdot \hat X_{i,t-1} + c_i\varepsilon_t + \eta_{i,t} 
\qquad t\geq 0, \qquad i=1,2,\dots,n,
\ee
and the expression of the $\tilde \beta_k$'s then simplifies to 
\be
\tilde \beta_{k}  = \E\[ \phi^k\],
\ee
which allows us to write the excess realized log-volatility and its auto-correlation function as
\be
\label{eq:rho2}
\bar X_t =  \E\[c\]\sum_{k=0}^\infty \E \[\varphi^k\] \epsilon_{t-k}, \qquad \text{and} \qquad
\rho(h) =\frac{\ds \sum_{k=0}^\infty \E[\varphi^k] \E[\varphi^{k+h}]}
{\ds \sum_{k=0}^\infty ( \E[\varphi^k])^2},
\ee
provided that the stationarity condition $\ds \sum_{k=0}^\infty ( \E[\varphi^k])^2 < \infty$ holds.

It turns out that the properties of this dynamic has been investigated by \can{goncalves88}. Based on their result, one can remark, in passing, that this the correlation function, namely the auto-correlation function of the aggregated time series, is very different from the average correlation function $\bar \rho$. Indeed, before the aggregation, the auto-correlation function 
is of each individual agent's expectation is
\begin{equation}
\rho_i(h) = \varphi_i^{|h|}, \qquad i=1,2,\dots, \qquad h=0,1,\dots
\end{equation}
so that
\begin{equation}
\bar \rho(h) = \E[\varphi^{|h|}]	\qquad h=0,1,\dots
\end{equation}

Besides, the dynamics exhibits long memory if (and only if) the density of $\phi$ diverges at one in order to get a hyperbolic decay of the correlation function~(\ref{eq:rho2}). In other words, the density should behave like $(1-x)^{-d}$, $d< 1/2$, in the neighborhood of one, that is to say like the
density of the Beta law for instance, in order to obtain $\rho(h) \sim \lambda \cdot |h|^{2d-1}$, as $|h|$ goes to infinity. 
In such a situation, one can conclude that a significant part of the agents base their anticipations on the one they 
have made the day before.

\subsubsection{Case 4: Deviation from rationality}

Let us now focus on a particular case of bounded rationality agents. For, we assume\footnote{Remark that most of the results in the section still hold if we only assume $\E\[\psi|\phi\] = -\alpha \cdot \varphi +\bar \varphi'$.}
$\psi = -\alpha \cdot \varphi +\bar \varphi'$, where $\alpha$ and $\bar \phi'$ are two constants while $\varphi$ follows a ``stretched'' Beta$(p,q)$ law, namely a Beta$(p,q)$ law extended over the entire range $[-1,1]$. As we expect that the fraction of investors characterized by a heterogeneity parameter $\phi_i$ close to one be larger and, in fact, diverges at 1, the parameter $p$ must be larger than one: $p>1$. On the contrary, we expect that the fraction of agents characterized by a heterogeneity parameter $\phi_i$ close to {\em minus} one remains finite, so that $q$ must be smaller than one: $0<q<1$. The case of rational agents is obviously encompassed by this representation, since it corresponds to 
the situation where $\alpha=-1$ and $\bar \varphi'=0$.

In this setting, the dynamics of the anticipation of the agent $i$ becomes
\be
\hat X_{i,t} = \varphi_i \cdot \hat X_{i,t-1} + (\bar\varphi' -\alpha\varphi_i) \cdot \bar X_{n,t-1}+ c_i \cdot \varepsilon_t 
+ \eta_{i,t} \qquad t\geq 0, \qquad i=1,2,\dots,n.
\ee
and, from equation (\ref{eq:beta_tilde}), the excess log-volatility $\bar X_t$ follows the infinite order moving-average 
(\ref{eq:slkdn2}) with coefficients $\tilde \beta_k$'s given by the relation

\be
\label{eq:Sbkxk_dfr}
\sum_{k=0}^\infty \tilde \beta_{k} x^{k} = 
\frac{(1-x)^{q-1} G(p,q,x)}{1 -\alpha + (\alpha-x\bar \phi')(1-x)^{q-1}G(p,q,x)},
\ee
where
\be
G(p,q,x) =
\begin{cases}
(1+x)^{-q} \cdot F(p+q-1,q;p+q,\frac{2}{1+x^{-1}}) &  \text{if}~ x\ge 0,\\ 
(1-x)^{-q} \cdot F(1,q;p+q,\frac{2}{1-x^{-1}}) & \text{if}~ x<0,
\end{cases}
\ee
with $F(.,.;.,.)$ is the 
hypergeometric function. Two conditions are required in order for $G(p,q,x)$ to remain finite for all $|x| \le 1$: $p>1$ and $0<q<1$. Since $0<q<1$, the numerator diverges. Let us study the behavior of the denominator,
and figure out the conditions for its convergence.
\be
\sum_{k=0}^\infty \tilde \beta_{k} x^{k} = 
\frac{G(p,q,x)}{(1 -\alpha)(1-x)^{1-q} + (\alpha-x\bar \phi')G(p,q,x)}
\underset{x\rightarrow 1}{\longrightarrow} \frac{1}{\alpha-\bar\phi'},
\ee
which yields the following condition for long memory 
\be
\bar\phi'=\alpha.
\ee
The dynamics of the anticipation of the agent $i$ then becomes
\be
\hat X_{i,t} = \varphi_i \cdot \hat X_{i,t-1} + \alpha(1 -\varphi_i) \cdot \bar X_{n,t-1}+ c_i \cdot \varepsilon_t 
+ \eta_{i,t} \qquad t\geq 0, \qquad i=1,2,\dots,n.
\ee
When $\phi_i$ is close to 1 then $\alpha(1 -\varphi_i)$ is close to 0
and inversely when $\phi_i$ is close to 0 then $\alpha(1 -\varphi_i)$ is close to $\alpha$. Finally, when
$\phi$ is close to -1, then $\alpha(1 -\varphi_i)$ is close to $2\alpha$.
There is a kind of balance for an agent to the weight she gives to her past anticipation and to the past realized volatility.
The more importance she gives to her past anticipation, the less the past realization matters. And inversely.
Besides, when an agent decides to base her anticipation contradictory to her past anticipation (corresponding to $\phi=-1$), 
for example she thinks she was wrong, then she also gives and important weight to the past 
realization via the coefficient $2\alpha$. This is a reasonable behavior.

Replacing $\bar\phi'$ by $\alpha$ in equation (\ref{eq:Sbkxk_dfr}) leads to the simplified expression
\be
\label{eq:fXsbpq}
\sum_{k=0}^\infty \tilde \beta_{k} x^{k} = 
\frac{(1-x)^{q-1} G(p,q,x)}{1 -\alpha + \alpha(1-x)^{q}G(p,q,x)}.
\ee
So, under the assumption that the denominator in equation (\ref{eq:fXsbpq}) remain bounded for all $|x| \le 1$, which requires that
\be
\frac{1-p}{q}<\alpha<1,
\ee
the series $\sum_{k=0}^\infty \tilde \beta_{k}x^k $ behaves like $C(1-x)^{q-1}$ in the neighborhood of 1
and thus diverges hyperbolically if $q<1$. In such a case, the auto-correlation function behaves like $K |h|^{1-2q}$ as $|h|\rightarrow \infty$, provided that the second order stationarity condition $q > 1/2$ holds.

Now, we will study the influence of the parameters on the shape of the auto-correlation function. First
we focus on the short memory case where $\bar\phi' \neq \alpha$, then on the long memory case
where $\bar\phi' = \alpha$. 

\begin{figure}
\begin{center}
\includegraphics[width=10cm]{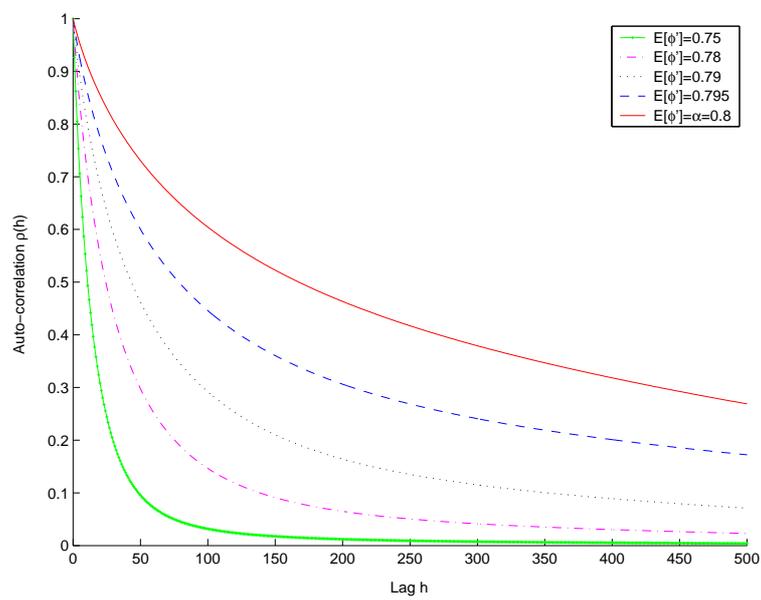}
\end{center}
\caption{\label{fig:ACF_fft_p5_q075_alpha08} Auto-correlation function over 500 days for $p = 5$, $q = 0.75$, $\alpha = 0.8$ 
and different values of $\bar\phi'$ ranging in $[0.60,0.79]$.} 
\end{figure}

\paragraph{Short memory}
In the case $\bar\phi' \neq \alpha$, the condition for the denominator not to vanish becomes
$\alpha>\max\(\bar\phi', \frac{\(1+\frac{q}{p-1}\)\bar\phi' + 2}{1 - \frac{q}{p-1}}\)$ if $\frac{q}{p-1}>1$
and $\bar\phi' < \alpha < \frac{\(1+\frac{q}{p-1}\)\bar\phi' + 2}{1 - \frac{q}{p-1}}$ if $\frac{q}{p-1}<1$.
We set $p=5$, $q=0.75$ and $\alpha=0.8$. Figure \ref{fig:ACF_fft_p5_q075_alpha08} shows how the auto-correlation function
behaves when $\bar\phi'$ tends to $\alpha$. We observe that the nearest $\bar\phi'$ is to $\alpha$
the slower the decrease of the auto-correlation function.

\vspace{1cm}
\centerline{\it [Insert figure~\ref{fig:ACF_fft_p5_q075_alpha08} about here]}
\vspace{1cm}

As a special case, let us set $\alpha = 1$, as in the case of rational agent, but $\bar \phi' \neq 0$. It is, in some sense, the simplest way to account for the departure from rationality. The anticipation of the agent $i$ then writes
\be
\hat X_{i,t} = \varphi_i \cdot \( \hat X_{i,t-1} - \bar X_{n,t-1}\) + \bar \varphi' \cdot \bar X_{n,t-1}+ c_i \cdot \varepsilon_t + \eta_{i,t} 
\qquad t\geq 0, \qquad i=1,2,\dots,n,
\ee
which means that the agent performs her anticipation on the basis of the past innovation $\hat X_{i,t-1} - \bar X_{n,t-1}$ but, in addition, also pays attention to the past realized volatility in itself, as pointed out by the presence of the term $\bar \varphi' \cdot \bar X_{n,t-1}$.

Equation (\ref{eq:beta_tilde}) yields
\be
\sum_{k=0}^\infty \tilde \beta_{k} x^{k} = \frac{1}{1 - x \cdot \bar \phi'},
\ee
which shows that the dynamics of the (excess) realized log-volatility is nothing but a simple short memory AR(1) process
\be
\bar X_t - \bar \phi' \cdot \bar X_{t-1}=\E[c] \epsilon_t,
\ee
whose auto-correlation function is given by
\be
\rho(h) = \bar {\varphi'}^{|h|}
\ee
Thus, if $\bar \phi'=0$, the volatility exhibits exactly the same behavior as in the case of fully rational agents (case 1), which shows again that the rationality at the individual level is not a necessary condition for the volatility, at the aggregated level, to behave as if the agent were individually rational.

\begin{landscape}
\begin{figure}
\begin{center}
\includegraphics[width=10cm]{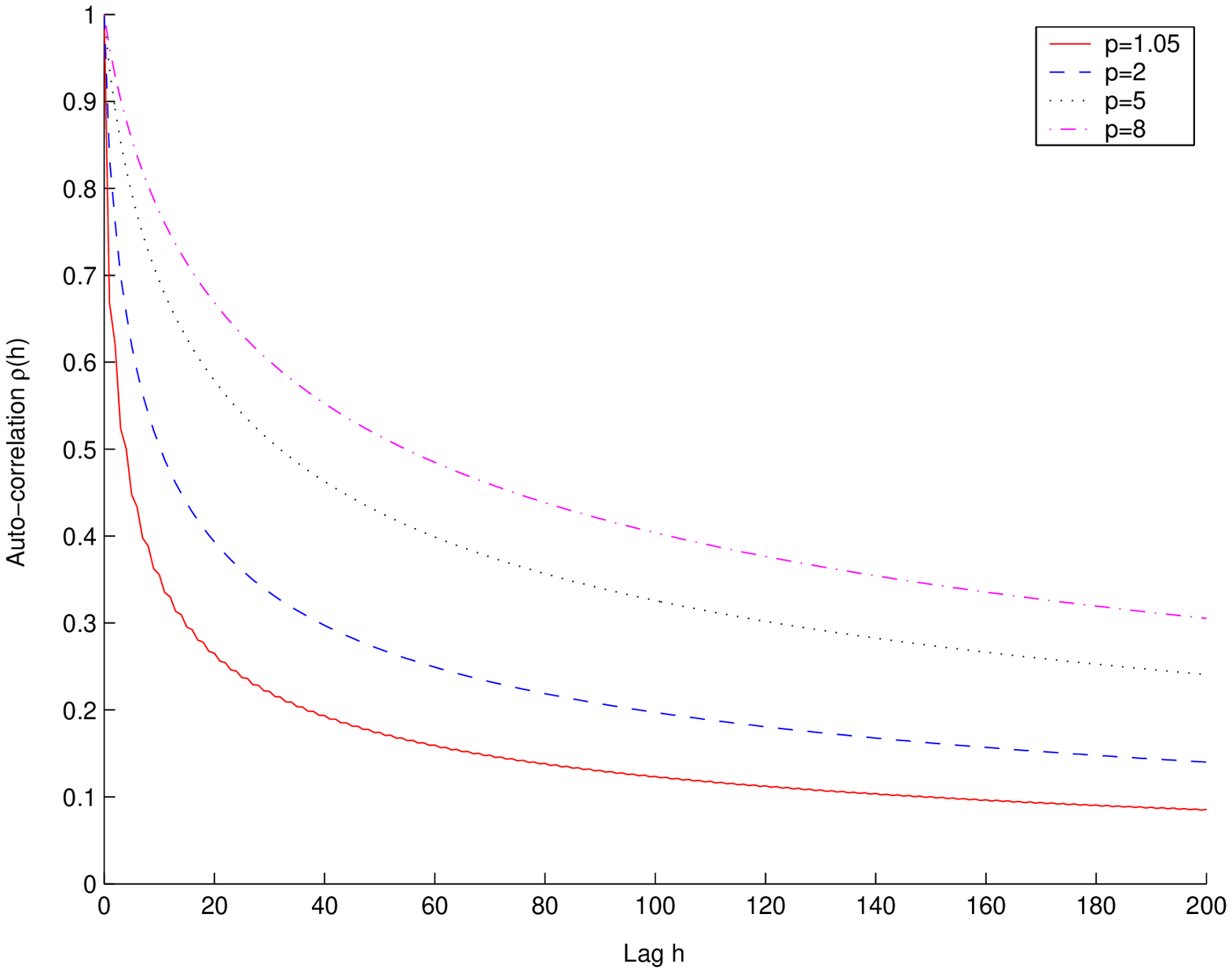}
\includegraphics[width=10cm]{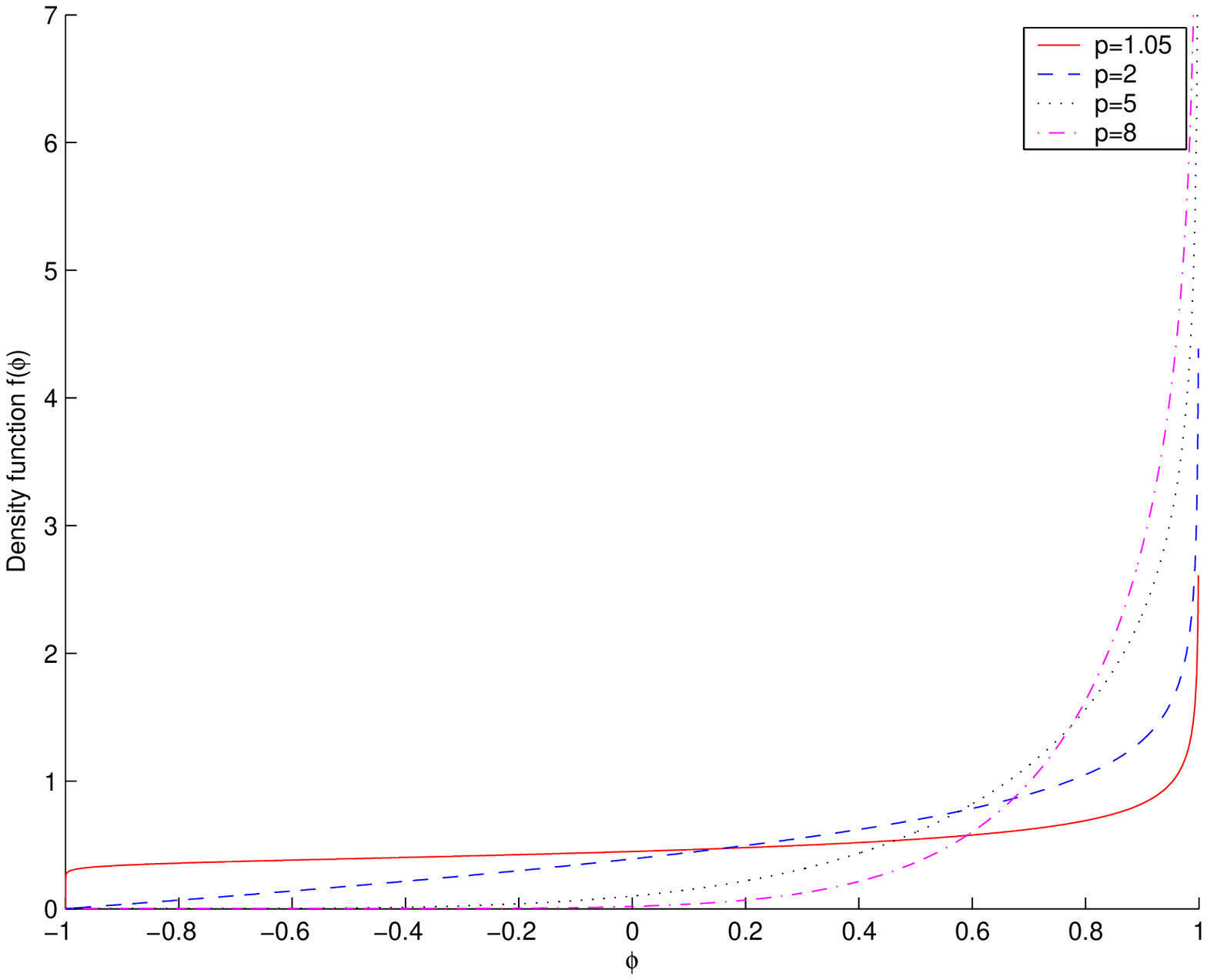}
\end{center}
\caption{\label{fig:ACF_fft_q075_alpha03} The left panel and the right panel represent respectively
the auto-correlation function over 200 days and the density of the heterogeneous coefficient 
for $q = 0.75$, $\alpha = 0.3$ and different values of $p$ ranging in $[1.05 , 8]$. The law of $\phi$
is a Beta($p$,$q$) extended over $[-1,1]$ and its density equal to
$f(\phi) = \frac{1}{2^{p+q-1}B(p,q)}(1+\phi)^{p-1}(1-\phi)^{q-1}$ 
explains the divergence near 0 and 1.}
\end{figure}
\end{landscape}

\paragraph{Long memory} Let us now focus on the case where the model exhibits long memory, namely when $\bar\phi' = \alpha$. We are interested in the role of the parameters $p$, $q$ and $\alpha$
in the short and long memories. 
On figure \ref{fig:ACF_fft_q075_alpha03}, the role of $p$ is depicted. The two other parameters $q$ and $\alpha$
are fixed respectively to 0.75 and 0.3 and $p$ ranges between $1.05$ and $8$. On the left panel the auto-correlation
functions are drawn over 200 lags, and on the right one the corresponding density functions. It shows that the larger $p$, 
the slower the decrease of the auto-correlation function at short lags. Thus, $p$ only influences the short term behavior of the auto-correlation function.

\vspace{1cm}
\centerline{\it [Insert figure~\ref{fig:ACF_fft_q075_alpha03} about here]}
\vspace{1cm}

\begin{landscape}
\begin{figure}
\begin{center}
\includegraphics[width=10cm]{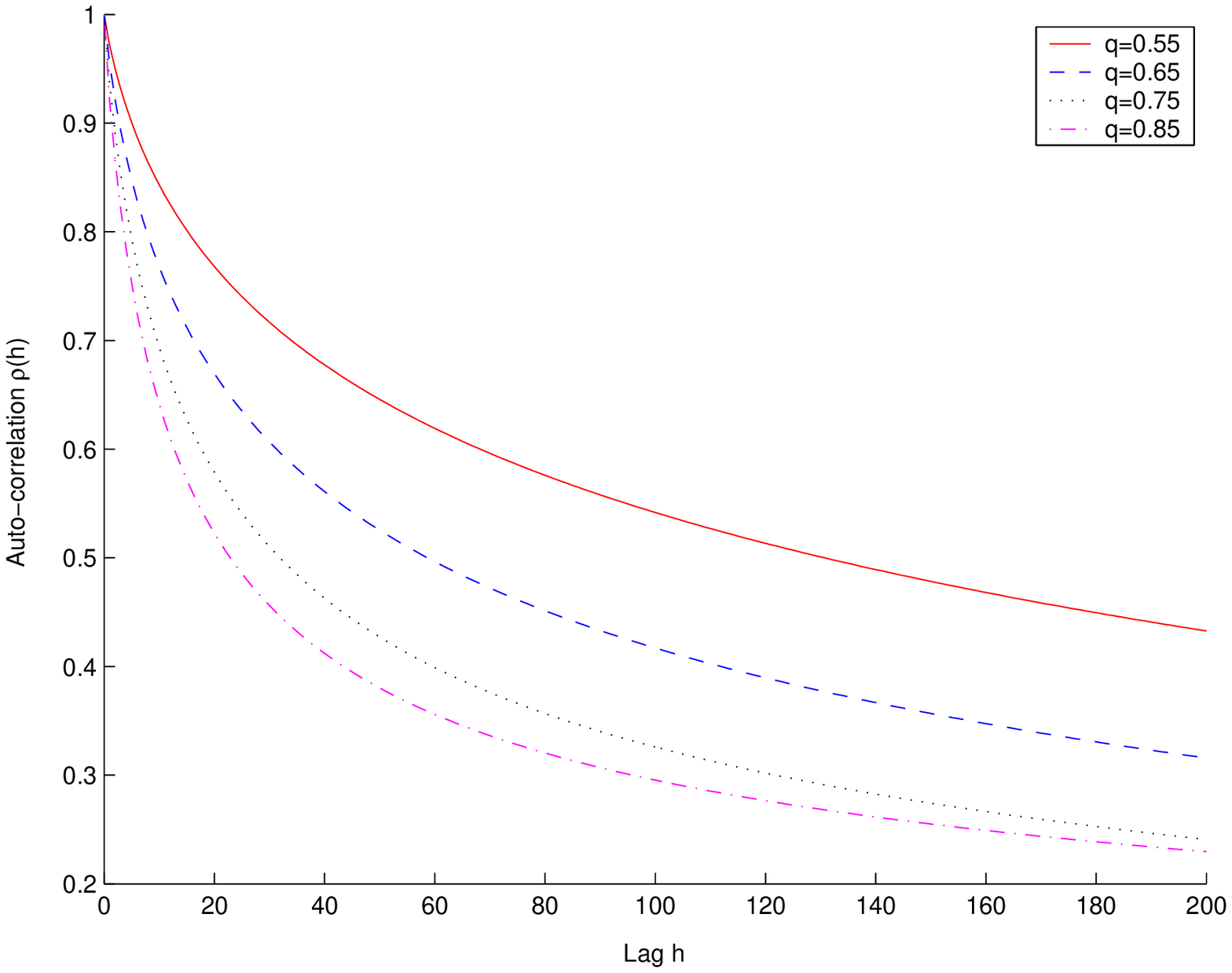}
\includegraphics[width=10cm]{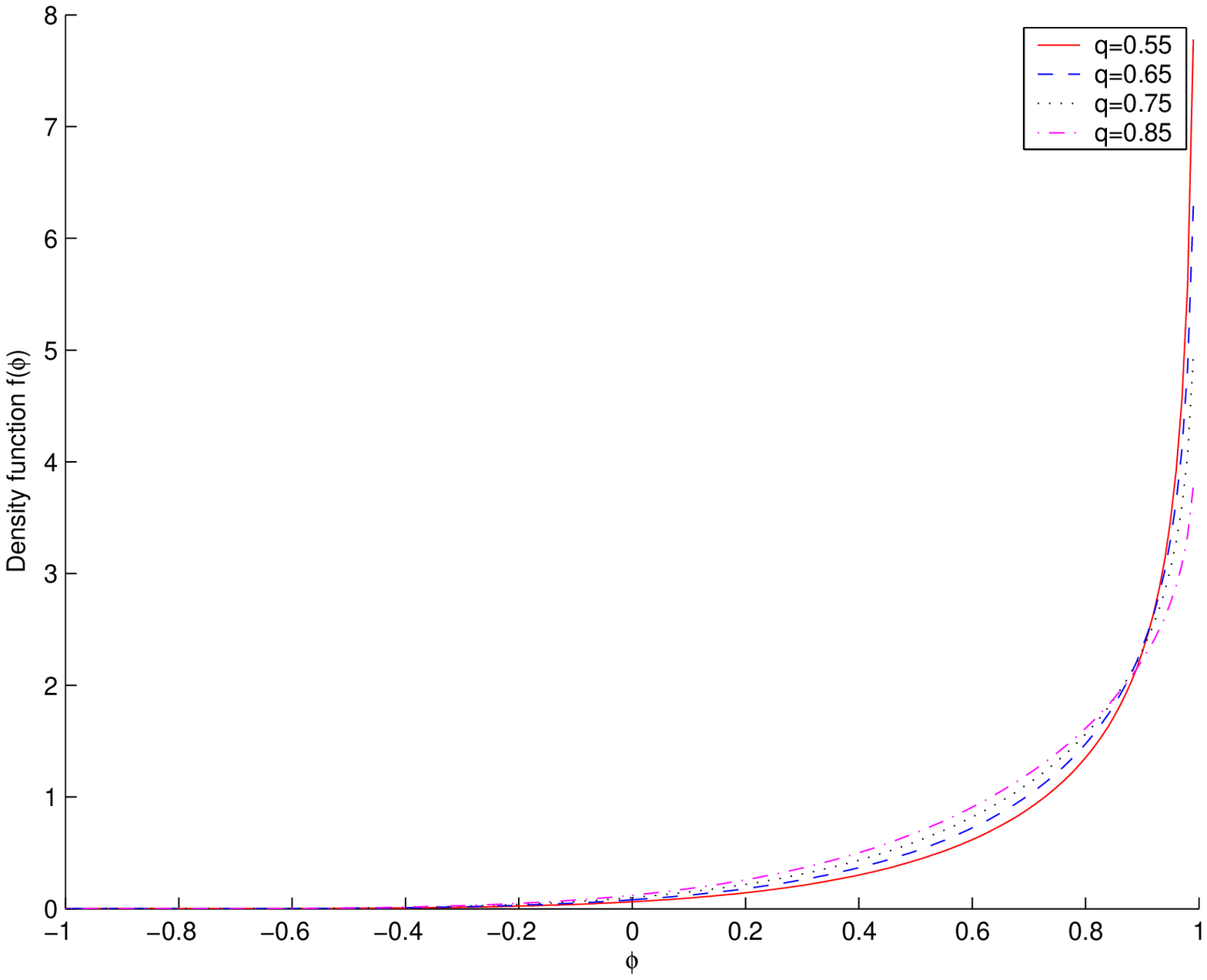}
\end{center}
\caption{\label{fig:ACF_fft_p5_alpha03} The left panel and the right panel represent respectively
the auto-correlation function over 200 days and the density of the heterogeneous coefficient 
for $p = 5$, $\alpha = 0.3$ and different values of $q$ ranging in $[0.55 , 0.85]$. The law of $\phi$
is a Beta($p$,$q$) extended over $[-1,1]$ and its density equal to
$f(\phi) = \frac{1}{2^{p+q-1}B(p,q)}(1+\phi)^{p-1}(1-\phi)^{q-1}$ 
explains the divergence near 0 and 1.} 
\end{figure}
\end{landscape}

On figure \ref{fig:ACF_fft_p5_alpha03}, the role of $q$ is studied. The two other parameters $p$ and $\alpha$
are fixed respectively to 5 and 0.3 while $q$ ranges between  $0.55$ and $0.85$.
The closer $q$ is to 0.5, i.e. to the lower bound for second order stationarity,  the faster is the divergence of the density function in the neighborhood of 1 and
the slower is the decay of the auto-correlation function. Consequently, $q$ impacts on the long memory.

\vspace{1cm}
\centerline{\it [Insert figure~\ref{fig:ACF_fft_p5_alpha03} about here]}
\vspace{1cm}

\begin{landscape}
\begin{figure}
\begin{center}
\includegraphics[width=10cm]{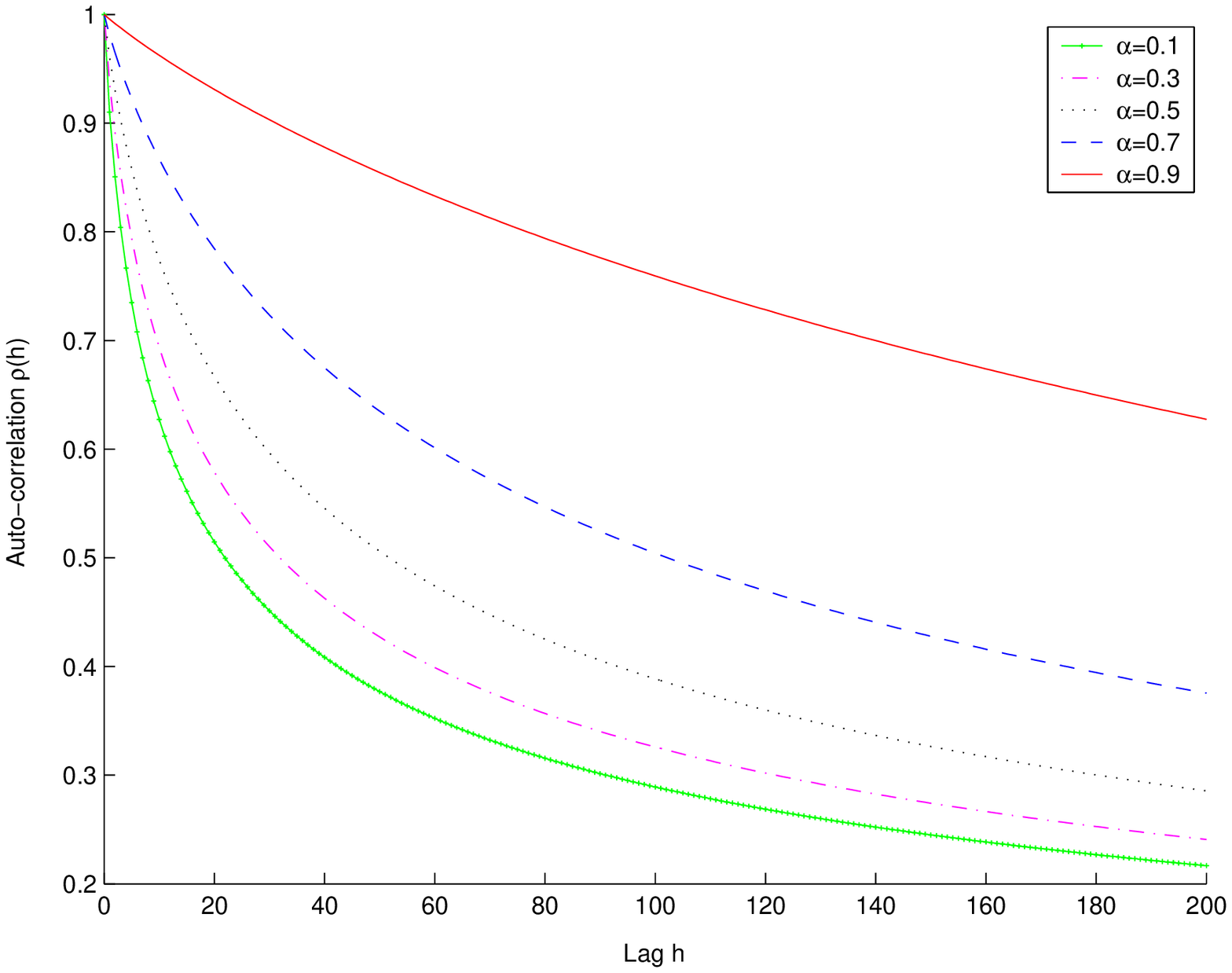}
\includegraphics[width=10cm]{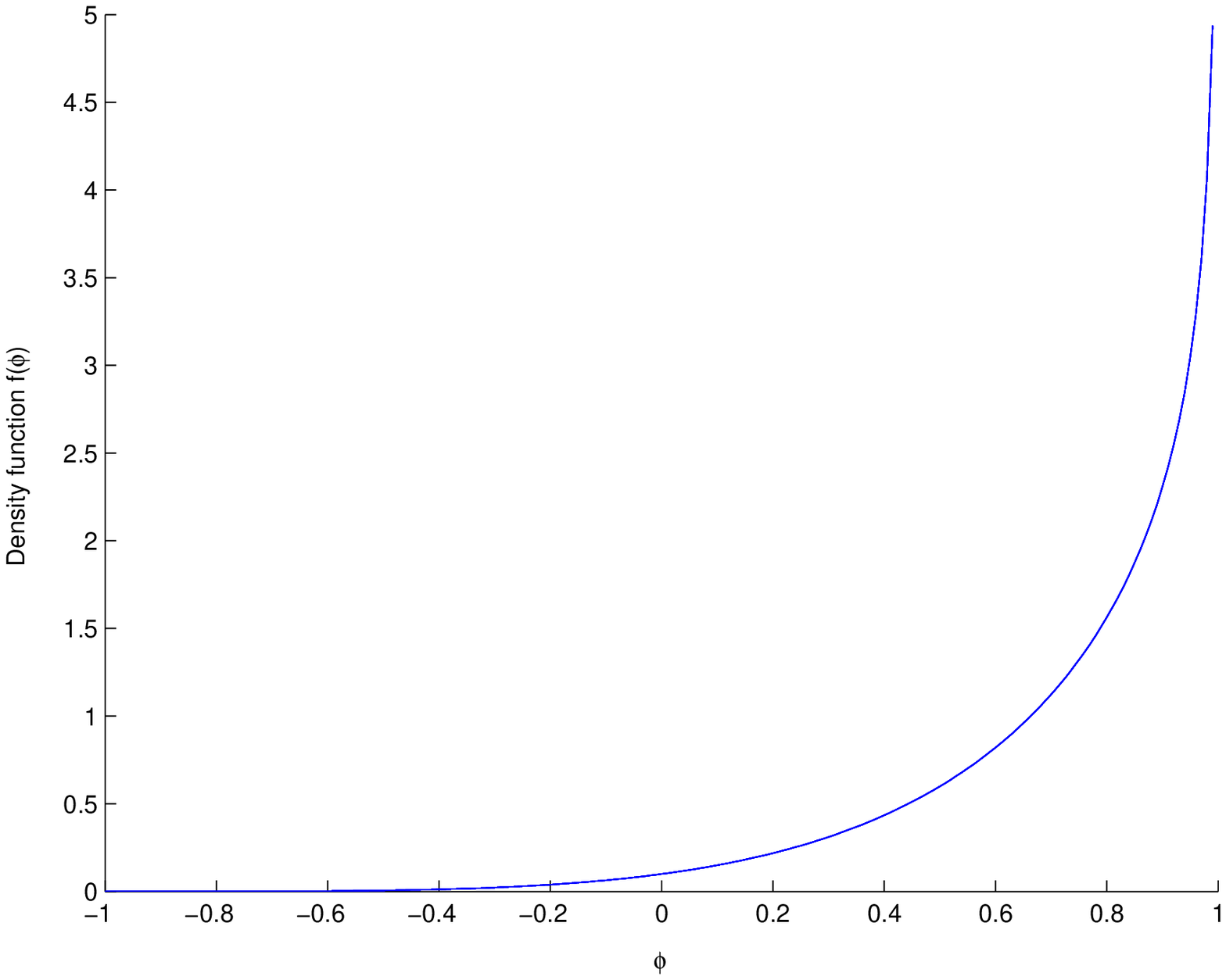}
\end{center}
\caption{\label{fig:ACF_fft_p5_q075} The left panel and the right panel represent respectively
the auto-correlation function over 200 days and the density of the heterogeneous coefficient 
for $p = 5$, $q = 0.75$ and different values of $\alpha$ ranging in $[0.1 , 0.9]$. The law of $\phi$
is a Beta($p$,$q$) extended over $[-1,1]$ and its density equal to
$f(\phi) = \frac{1}{2^{p+q-1}B(p,q)}(1+\phi)^{p-1}(1-\phi)^{q-1}$ 
explains the divergence near 0 and 1.} 
\end{figure}
\end{landscape}

Figure \ref{fig:ACF_fft_p5_q075} shows the role of $\alpha$. The two other parameters $p$ and $q$
are fixed respectively to 5 and 0.75 and $\alpha$ ranges betwwen $0.1$ and $0.9$. We observe that $\alpha$
has an important impact on the short memory. Indeed, the smaller $\alpha$, the slower the short-lag decrease.

\vspace{1cm}
\centerline{\it [Insert figure~\ref{fig:ACF_fft_p5_q075} about here]}
\vspace{1cm}

Let us take now into account a larger class of agents' behavior in the case of $\bar\phi' = \alpha$.
To this aim, we modify the density function: we still keep
the beta law but we add another law which have a bell-like density. The bell will be able to move all over [-1,1].
For example, if the bell is set around zero, it means that the agents do not give importance 
to their past anticipation. 
\be
f(\phi) = w \frac{1}{2^{p+q-1}B(p,q)}(1+\phi)^{p-1}(1-\phi)^{q-1} 
+ (1-w) \frac{1}{K(m,\sigma)}(1+\phi)(1-\phi)\exp\(-\frac{1}{2}\frac{(\phi-m)^2}{\sigma^2}\),
\ee
where $K$ is the normalizing constant (that can be expressed in closed-form) law and $w$ in $[0,1]$ is the weight given to the stretched Beta law.The conditions of long memory remain the same, only the conditions of stationarity change. They can be easily expressed in closed-form but their expression is rather cumbersome; That is why we will not give them here.

\begin{landscape}
\begin{figure}
\begin{center}
\includegraphics[width=10cm]{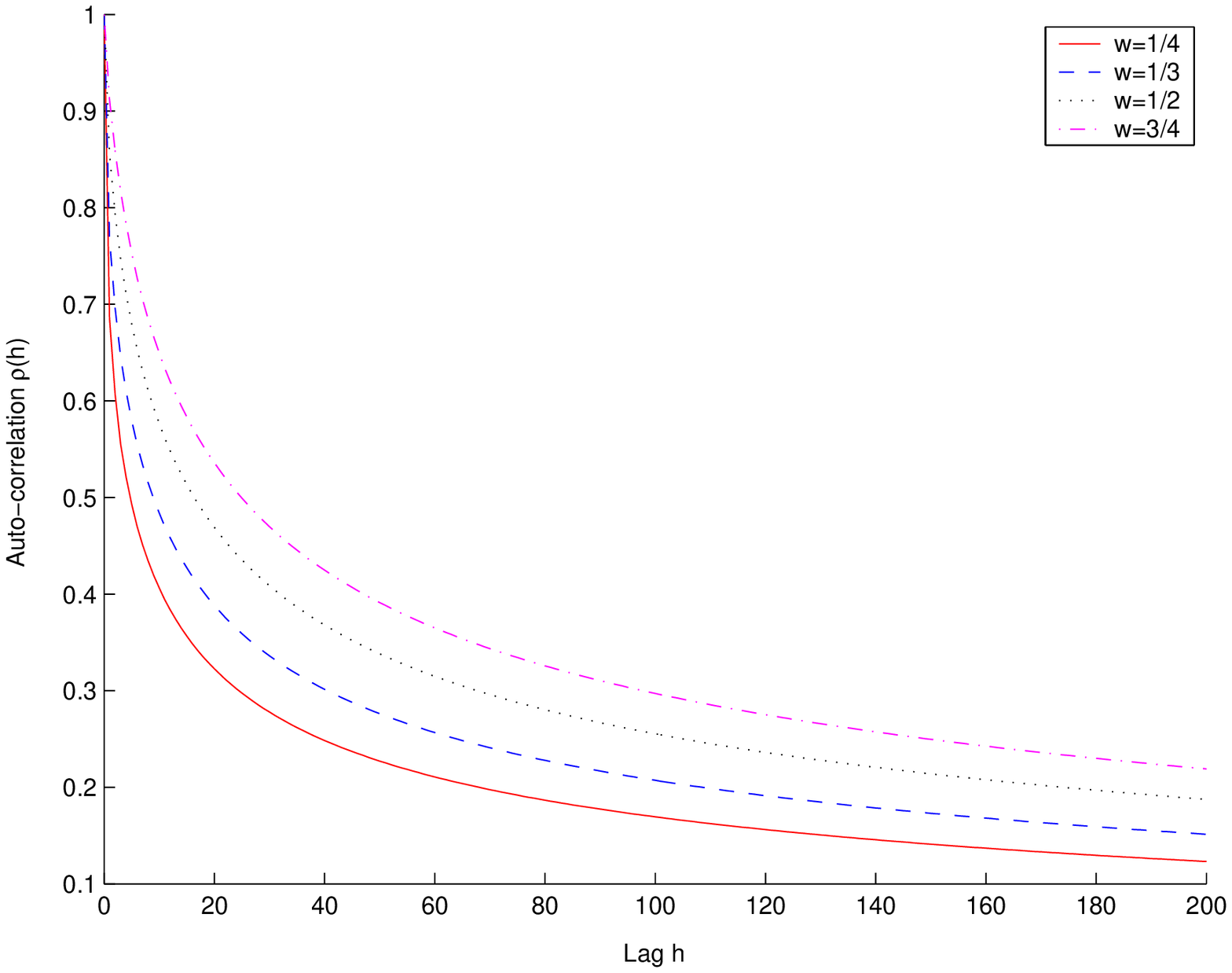}
\includegraphics[width=10cm]{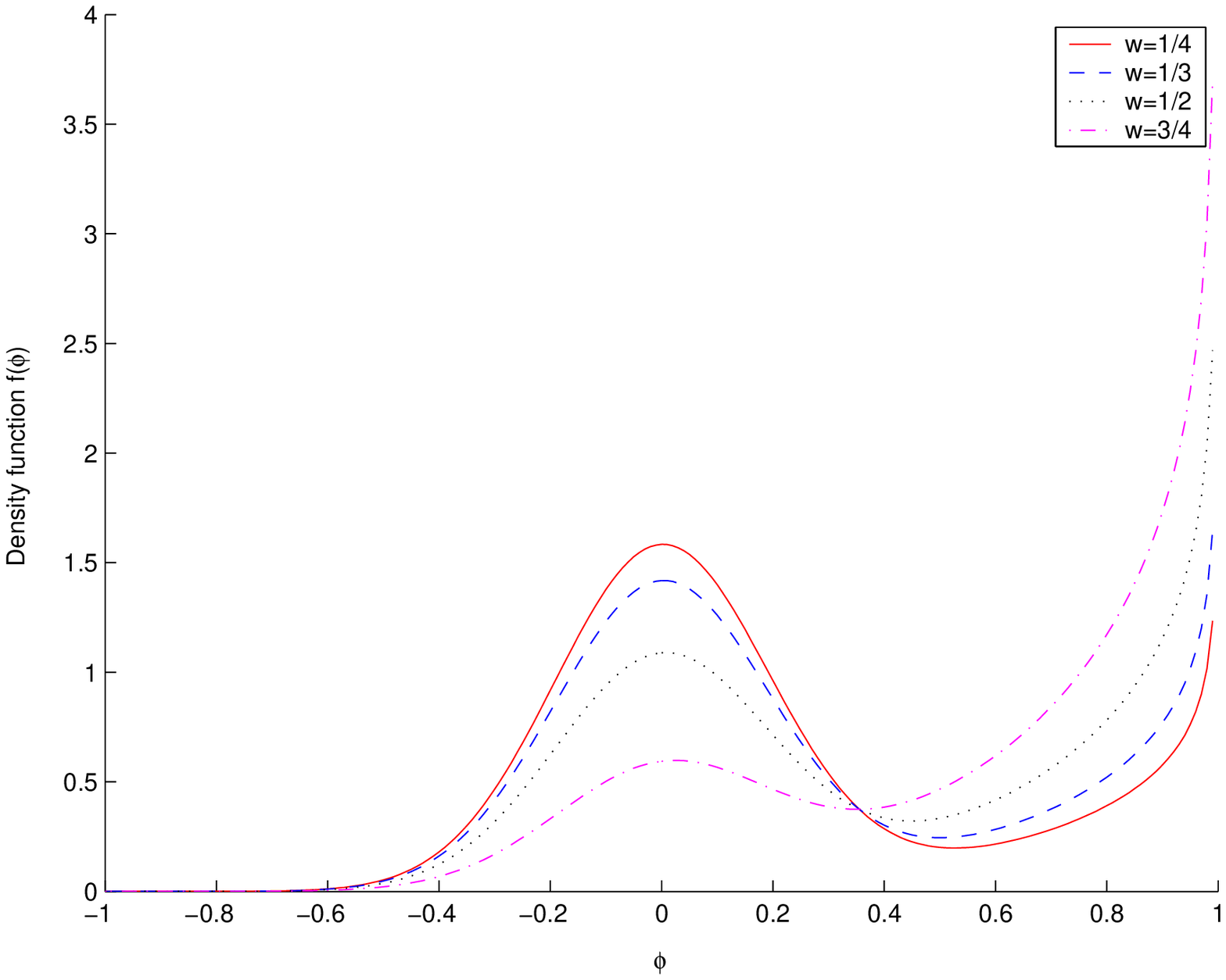}
\end{center}
\caption{\label{fig:ACF_fft_p5_q075_alpha03_m0_v02} The left panel and the right panel represent respectively
the auto-correlation function over 200 days and the density of the heterogeneous coefficient 
for $p = 5$, $q = 0.75$, $\alpha=0.3$, $m=0$, $\sigma=0.2$ and different values of $w$ ranging in $[1/4 , 3/4]$. 
The density of the law of $\phi$ is equal to
$f(\phi) = w \frac{1}{2^{p+q-1}B(p,q)}(1+\phi)^{p-1}(1-\phi)^{q-1} + (1-w) \frac{1}{K}(1+\phi)(1-\phi)\exp\(-\frac{1}{2}\frac{(\phi-m)^2}{\sigma^2}\)$ 
explains the divergence near 0 and 1.} 
\end{figure}
\end{landscape}

On figure \ref{fig:ACF_fft_p5_q075_alpha03_m0_v02}, the role of the relative weight of the singular part the density, $w$, is studied. The parameters $p$ and $q$
are fixed respectively to 5 and 0.75, $\alpha=0.3$, $m=0$, $\sigma=0.2$ and $w$ is ranged over $[1/4,3/4]$.
The weight given to each density impacts on the short but not the long memory. 
The larger the relative part of the bell-like density compared to the stretched Beta one, the faster the decay of the short memory. In terms of agents' behavior, it means that the more the agents base their anticipations on the incoming information flows and on the past realization of the volatility, the smallest the short-term correlations are.

\vspace{1cm}
\centerline{\it [Insert figure~\ref{fig:ACF_fft_p5_q075_alpha03_m0_v02} about here]}
\vspace{1cm}

\begin{landscape}
\begin{figure}
\begin{center}
\includegraphics[width=10cm]{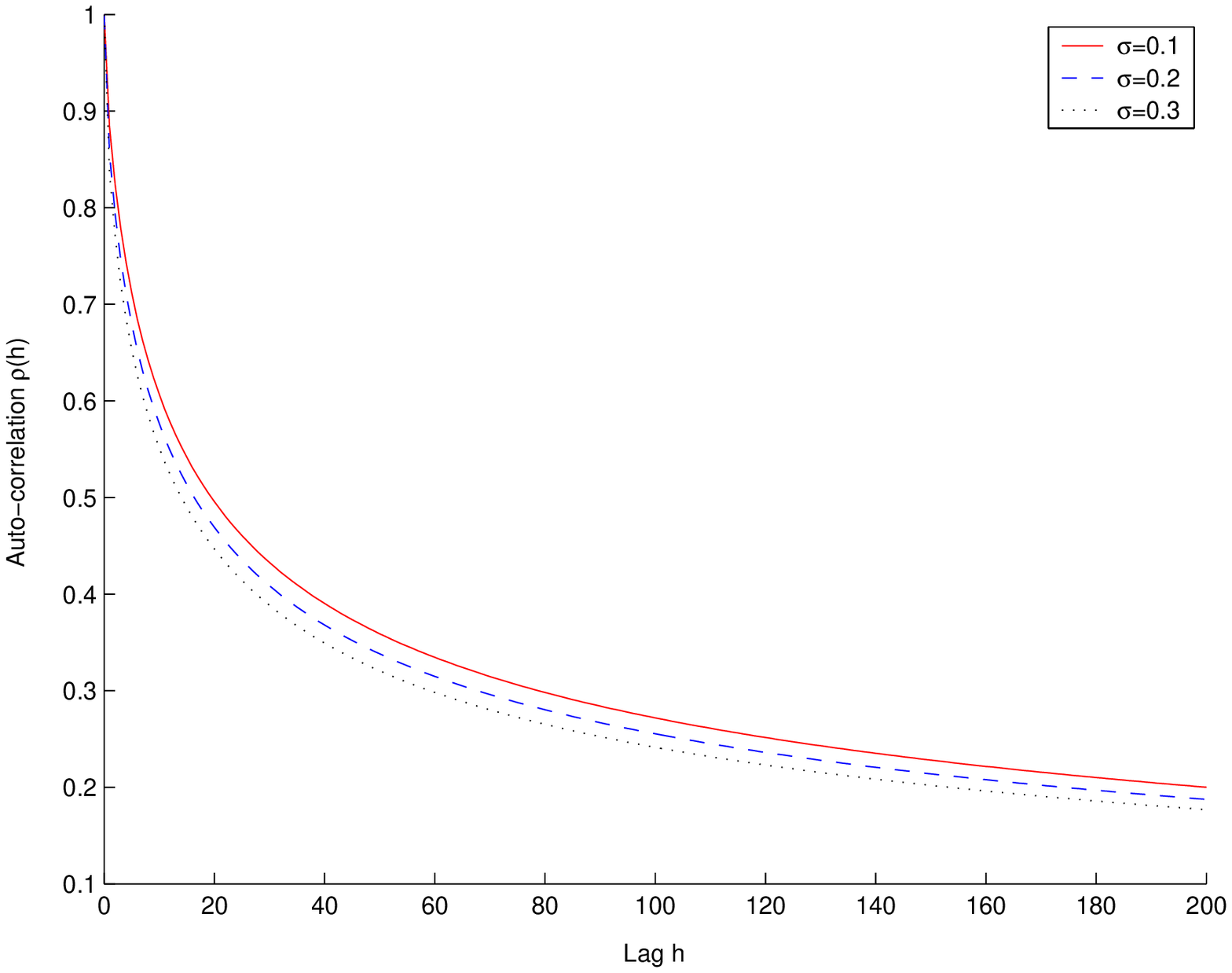}
\includegraphics[width=10cm]{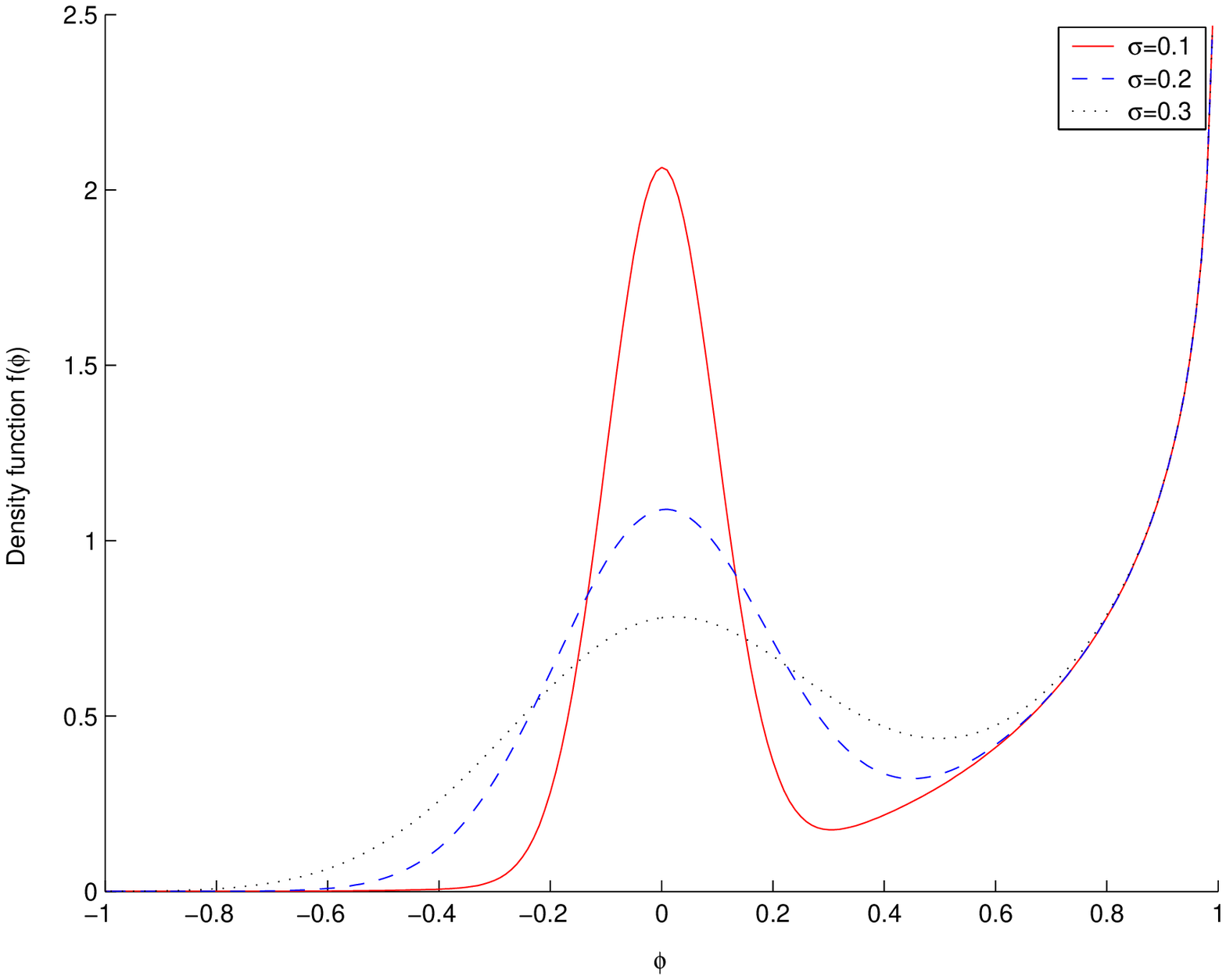}
\end{center}
\caption{\label{fig:ACF_fft_p5_q075_alpha03_m0_w05} The left panel and the right panel represent respectively
the auto-correlation function over 200 days and the density of the heterogeneous coefficient 
for $p = 5$, $q = 0.75$, $\alpha = 0.3$, $m = 0$, $w = 1/2$ and different values of $\sigma$ ranging in $[0.1 , 0.3]$. 
The density of the law of $\phi$ is equal to
$f(\phi) = w \frac{1}{2^{p+q-1}B(p,q)}(1+\phi)^{p-1}(1-\phi)^{q-1} + (1-w) \frac{1}{K}(1+\phi)(1-\phi)\exp\(-\frac{1}{2}\frac{(\phi-m)^2}{\sigma^2}\)$ 
explains the divergence near 0 and 1.} 
\end{figure}
\end{landscape}

Figure \ref{fig:ACF_fft_p5_q075_alpha03_m0_w05}, shows the impact of the width of the peak of the bell controlled by $\sigma$. The parameters $p$ and $q$
are fixed respectively to 5 and 0.75, $\alpha=0.3$, $m=0$, $w=1/2$ and $\sigma$ ranges between $0.1$ and $0.3$.
The intensity of the peak of the bell-like density has only a little impact on the short memory. The sharper
the bell is, the more important is the short memory.

\vspace{1cm}
\centerline{\it [Insert figure~\ref{fig:ACF_fft_p5_q075_alpha03_m0_w05} about here]}
\vspace{1cm}

\begin{landscape}
\begin{figure}
\begin{center}
\includegraphics[width=10cm]{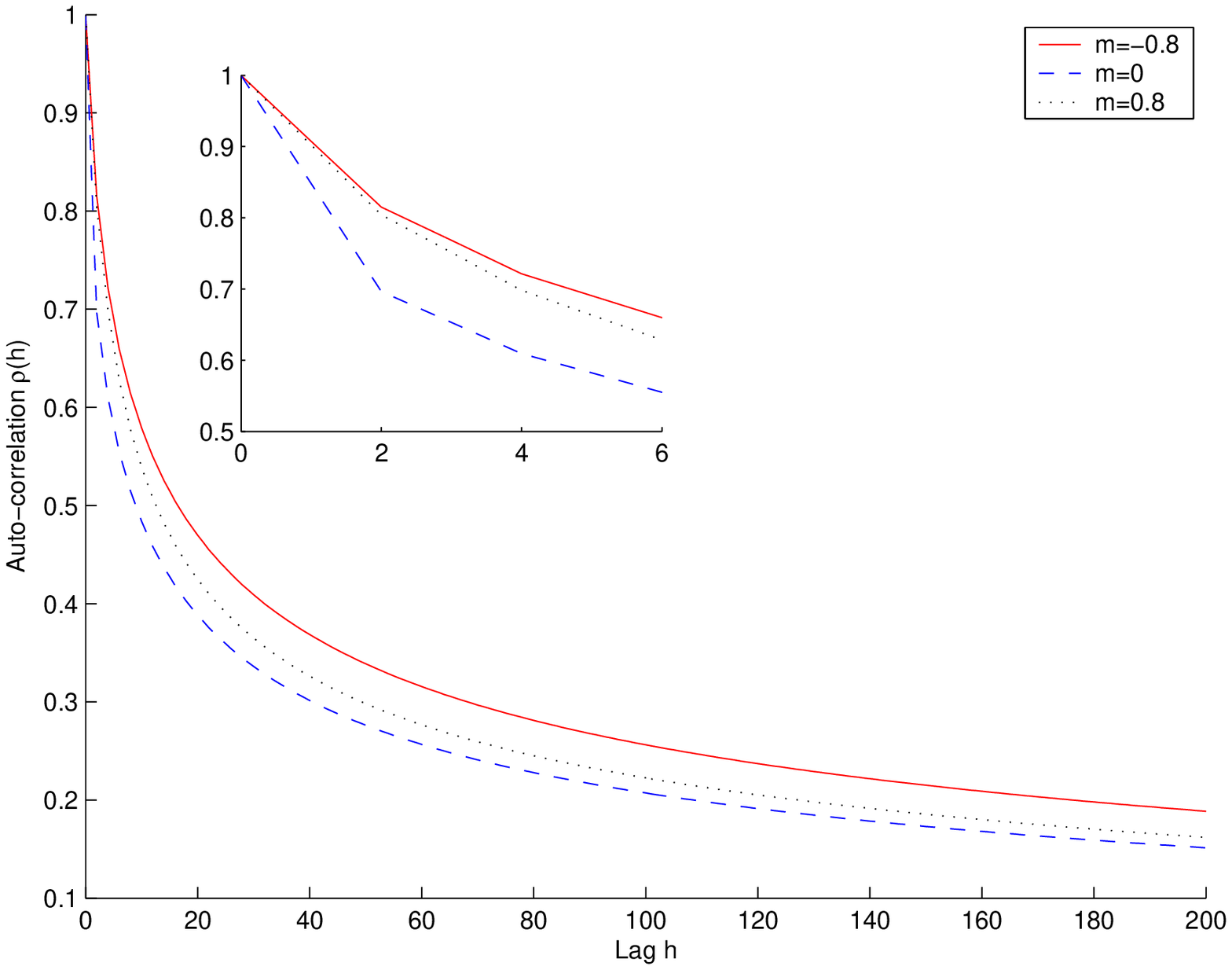}
\includegraphics[width=10cm]{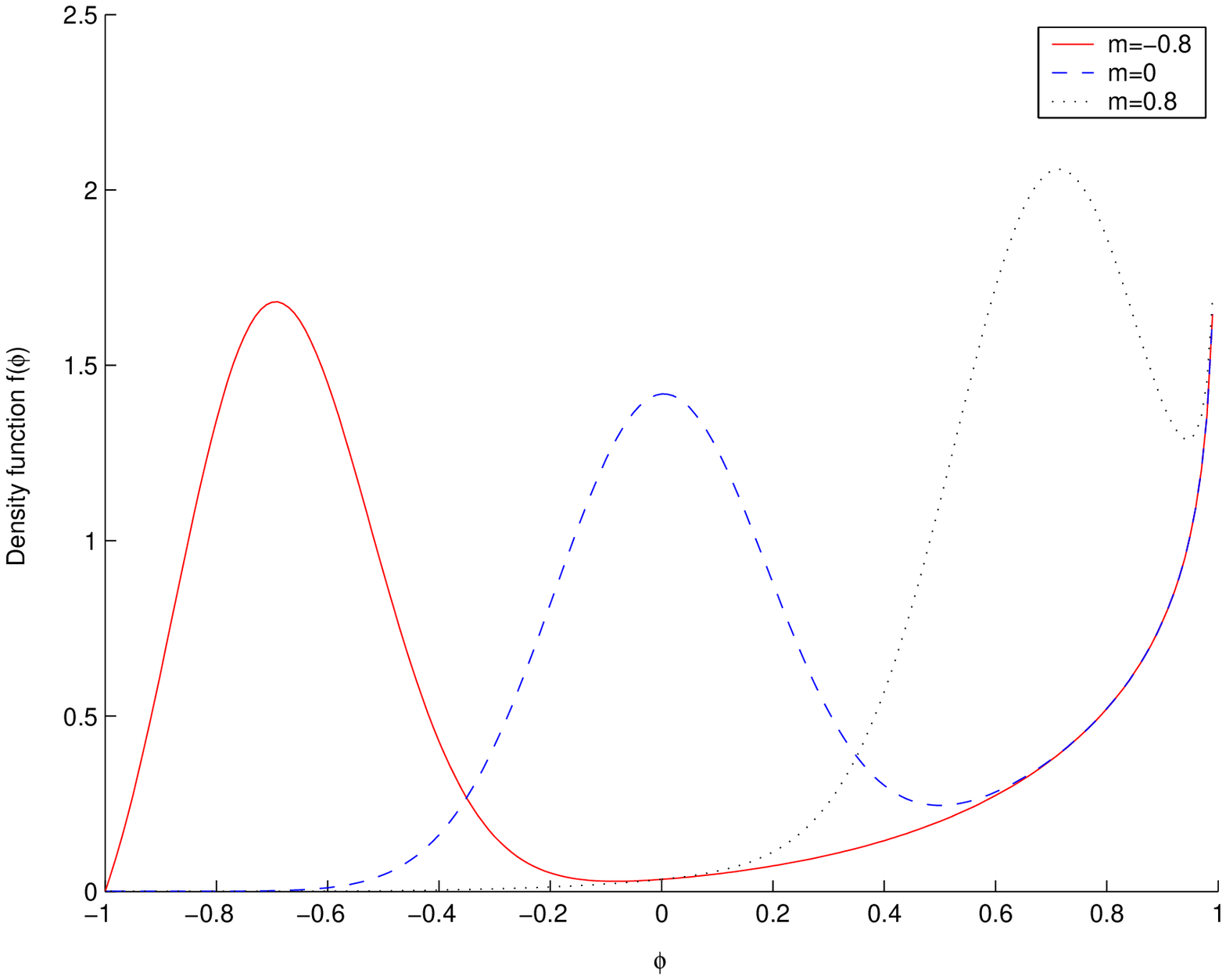}
\end{center}
\caption{\label{fig:ACF_fft_p5_q075_alpha03_v02_w033} The left panel and the right panel represent respectively
the auto-correlation function over 200 days and the density of the heterogeneous coefficient 
for $p = 5$, $q = 0.75$, $\alpha = 0.3$, $\sigma = 0.2$, $w = 1/3$ and different values of $m$ ranging in $[-0.8, 0.8]$. 
The density of the law of $\phi$ is equal to
$f(\phi) = w \frac{1}{2^{p+q-1}B(p,q)}(1+\phi)^{p-1}(1-\phi)^{q-1} + (1-w) \frac{1}{K}(1+\phi)(1-\phi)\exp\(-\frac{1}{2}\frac{(\phi-m)^2}{\sigma^2}\)$ 
explains the divergence near 0 and 1.} 
\end{figure}
\end{landscape}

On figure \ref{fig:ACF_fft_p5_q075_alpha03_v02_w033}, the role of $m$ which determines the position of the bell is studied.
The parameters $p$ and $q$
are fixed respectively to 5 and 0.75, $\alpha=0.3$, $\sigma=0.2$, $w=1/3$ and $m$ is ranged over $[-0.8,0.8]$.
The translation of the bell plays a part on short memory. Over the very first lags, we notice that the
closer the bell is to -1, the faster the decrease of the auto-correlation function is. But after a few lags,
this situation does not hold anymore. 

\vspace{1cm}
\centerline{\it [Insert figure~\ref{fig:ACF_fft_p5_q075_alpha03_v02_w033} about here]}
\vspace{1cm}


\section{Statistical Inference}

Let us now turn to the question of the estimation of the parameters of the model. As exemplified in the previous section, the long memory phenomenon can be ascribed to the hyperbolic divergence of the density of the heterogeneity parameter $\phi$ in the neighborhood of $1^-$. It is thus convenient to split the density of the heterogeneity variable $\varphi$ in two terms: a regular one and a singular one. For this reason, we will model the law of the realized log-volatility of financial assets as the mixture of a regular law, with density $f_1$, and of a Beta$(2,1-d)$ law ($d \in ]0;1[$) with density $f_2$, that will allow capturing the hyperbolic divergence of the density of $\phi$ at one. As a consequence, the density of $\varphi$ can be written as
\begin{equation}
\label{eq:f}
f(x) = w f_1(x) + (1-w) f_2(x), \quad w \in [0,1],
\end{equation}
where $f_1$ is any regular, {\it i.e.} continuous and bounded, density function defined on $[-1,1]$ and
\be
\label{eq:f2}
f_2(x) = (1-d)(2-d)\frac{x}{(1-x)^{d}} 1_{[0;1]}(x).
\ee

Given a density $f_1(x)$ which admits an expansion in terms of a Fourier series, an assumption that will be made 
in all the sequel of this article,
\be
\label{eq:Fourier_f1}
f_1(x) = \frac{1}{2} + \sum_{n=1}^\infty a_n \cos (n \pi x)  + \sum_{n=1}^\infty b_n \sin (n \pi x), \quad x \in [-1,1],
\ee
the expression of the auto-covariance and auto-correlation functions of the realized log-volatility can be numerically 
calculated by use of the expression of the moment of order $k$ of the heterogeneity variable $\varphi$
\be
\E \[ \varphi^k\] = w \E_1 \[\varphi^k\] + (1-w) \E_2\[\varphi^k\],
\ee
where $\E_1[\cdot]$ and $\E_2[\cdot]$ denotes the expectations with respect to $f_1$ and $f_2$ respectively, 
which yields
\be
\E \[ \varphi^k\] = w \(\frac{1}{2} A_k +  \sum_{n=1}^\infty a_n B_{n,k} + \sum_{n=1}^\infty b_n C_{n,k} \) 
+ (1-w) \cdot \frac{\Gamma(k+2)}{\Gamma(k + 3 -d)}\Gamma(3-d),
\ee
where the expressions of $A_k$, $B_k$ and $C_k$ are given in appendix~\ref{app:acf}.

\subsection{Asymptotic Normality of the estimator}

For simplicity, and for ease of the exposition, we will assume that the coefficients $(a_n, b_n)$ of the Fourier expansion (\ref{eq:Fourier_f1}) vanishes beyond the rank $q$, so that the density $f_1$ reads
\be
\label{eq:f1q}
f_1(x) = \frac{1}{2} + \sum_{n=1}^q a_n \cos (n \pi x)  + \sum_{n=1}^q b_n \sin (n \pi x), \quad x \in [-1,1].
\ee
In addition, we focus on the case where $\E\[\psi|\phi\]=-\alpha(\phi-1)$, which ensures that the long range memory of the volatility is controlled by the parameter $d$ of the Beta law, as shown in the previous section.

Let us denote by $\theta = (a_1, \dots, a_q, b_1, \ldots, b_q,\alpha, w, \sigma_\varepsilon, d)'$ the $2q + 4$ dimensional vector of 
the parameters involved in our problem and by $\gamma (h; \theta)$ the value of the auto-covariance function  
at lag $h$ for the distribution of heterogeneity defined by equations (\ref{eq:f}), (\ref{eq:f2}) and 
(\ref{eq:f1q}) and for the parameter value $\theta$.

Given the $T$-sample of the logarithm of the realized volatilities $\{\omega_1, \dots, \omega_T\}$, treated as 
observed data, it seems natural to estimate $\theta$ by minimization of the weighted difference between the 
sample auto-covariance function
\be
\hat \gamma_T(h) = \frac{1}{T} \sum_{i=1}^{T-h} \(\omega_{i} - \bar \omega_T\) 
\(\omega_{i+h} - \bar \omega_T \), \qquad h=0,1, \dots
\ee
and $\gamma(h; \theta)$, the auto-covariance function at the parameter value $\theta$.  We could 
thus consider the {\em classical minimum distance} estimator $\hat \theta_{T,L}$ 
\cite{newey94} solution to
\begin{equation}
\min_{\theta \in \Theta} \[\hat \gamma_{T,L} - \gamma_L(\theta)\]' W_L^{-1} \[\hat \gamma_{T,L} - \gamma_L(\theta)\],
\end{equation}
where $\gamma_L(\theta) = \( \gamma(1;\theta), \dots, \gamma(L;\theta)\)'$, $\hat \gamma_{T,L} = \(\hat \gamma_T(1), 
\dots, \hat\gamma_T(L)\)'$
and $W_L$ is any symmetric positive definite $L \times L$ matrix, while $\Theta$ 
is the parameter set
\be
\Theta = \left\{ \theta = (a_1, \dots, a_q, b_1, \ldots, b_q, \alpha, w, \sigma, d)';~ f_1 \ge 0,~ 0 < 
d < 1~ \text{and}~ 0 \leq w\leq 1 \right\}.
\ee

However, when dealing with long-memory time series, {\it i.e} such that $\sum_{h=0}^\infty |\gamma(h)| =\infty$, 
the asymptotic properties of the sample estimates $\hat \gamma_{T,L}$ are not really suitable. Indeed, 
as recalled by \can{hosking96}, the limiting distribution of $\gamma_T(h)$ when $d>0$ is such that
\be
T^{1-2d} \( \hat \gamma_T(h) - \gamma(h)  \) \stackrel{{\cal L}}{\longrightarrow} {\cal R},
\ee
where ${\cal R}$ denotes the modified Rosenblatt distribution. In particular, the mean of the Rosenblatt 
distribution is not equal to zero and can even be much larger than its standard deviation for $d$ larger 
than or of the order of one fourth. 

As a consequence, it is desirable to rely on the minimization of another criteria with more suitable 
asymptotic properties.  In fact, irrespective of the value of $d \in (-\frac{1}{2}, \frac{1}{2})$, the 
limit distribution of any subset of the variables $D_h=\sqrt{T} \[ \( \hat \gamma_T(h) - \hat \gamma_T(0) \) 
- \(\gamma(h) -\gamma(0) \)\]$, $h  \ge 1$ is a multivariate normal  with zero mean and asymptotic 
covariance matrix (see \can[th. 5]{hosking96})
\bea
\[\Sigma^2\]_{kl} &=& \lim_{T\rightarrow \infty} \Cov \(D_k, D_l \),\\
&=& \frac{1}{2} \sum_{s=-\infty}^{\infty} \[\gamma(s) - \gamma(s-k) - \gamma(s-l) + \gamma(s-k+l) \]^2.
\eea
In view of this asymptotic result, a convenient estimator of the parameter $\theta$ is given by the solution to
\be
\min_{\theta \in \Theta} \[\hat \eta_{T,L} - \eta_L(\theta)\]' W_L^{-1} \[\hat \eta_{T,L} - \eta_L(\theta)\],
\end{equation}
where 
\be
\eta_L(\theta) = \( \gamma(1;\theta) - \gamma(0;\theta), \dots, \gamma(L;\theta)- \gamma(0;\theta)\)',
\ee
and
\be
\quad \hat \gamma_{T,L} = \(\hat \gamma_T(1) - \hat \gamma_T(0),  \dots, \hat\gamma_T(L) - \hat \gamma_T(0)\)'.
\ee

The consistency and the asymptotic normality of the estimator $\hat \theta_{T,L}$ follows from the 
general asymptotic properties of the classical minimum distance estimators. Concerning the asymptotic normality, 
we can state the following result
\begin{proposition}
\label{prop1}
Assuming that $\theta_{T,L} \stackrel{P}{\longrightarrow} \theta_0$, for all $L \ge 2q + 4$, as $T \longrightarrow \infty$, 
\be
\sqrt{T} \(\hat \theta_{T,L} - \theta_0 \) \stackrel{{\cal L}}{\longrightarrow} {\cal N} \(0, A_L(\theta_0) \),
\ee
with
\be
A_L(\theta_0) = \[G_L(\theta_0)' W_L G_L(\theta_0)\] ^{-1}  G_L(\theta_0)' W_L \Sigma_L^2(\theta_0) W_L G_L(\theta_0)
  \[G_L(\theta_0)' W_L G_L(\theta_0)\] ^{-1},
\ee
where
\be
G_L(\theta_0) = \left. {\rm grad}_\theta \eta_L(\theta) \right|_{\theta=\theta_0},
\ee
and 
\be 
\[\Sigma_L^2(\theta_0)\]_{lk} = \frac{1}{2} \sum_{s=-\infty}^{\infty} \[\gamma\(s;\theta_0\) 
- \gamma\(s-k;\theta_0\) - \gamma\(s-l;\theta_0\) + \gamma\(s-k+l;\theta_0\) \]^2, \label{eq:bartlett2}
\ee
$l,k = 1, \dots, L$.
\end{proposition} 

\begin{proof} By theorem 5 in \can{hosking96}, $\ds \sqrt{T} \(\hat \eta_{T,L} 
- \eta_L(\theta_0)\) \stackrel{{\cal L}}{\longrightarrow} {\cal N} \(0, \Sigma_L^2(\theta_0) \)$, 
thus, the result follows straightforwardly from theorem 3.2 in \can{newey94}.
\end{proof}

For fixed $L$, the minimum of the asymptotic variance $A_L(\theta_0)$ is reached when 
$W_L = \[\Sigma_L^2(\theta_0) \]^{-1}$, so that
\be
\sqrt{T} \(\hat \theta_{T,L} - \theta_0 \) \stackrel{{\cal L}}{\longrightarrow} {\cal N} 
\(0, G_L(\theta_0)' \[\Sigma_L^2(\theta_0) \]^{-1} G_L(\theta_0)\).
\ee
On the other hand, given $W_L$, one can get an optimal value $L^*$ of $L$ as the solution to 
\be
L^* = \arg\min_{L \in \{2q+4, \dots, T-1 \}} ||A_L(\theta_0) ||^2.
\ee

\begin{proposition}
\label{prop2} Under the assumptions in proposition~\ref{prop1}, denoting by $D(\theta)$ the vector
\be
D(\theta) = \[
\begin{array}{c}
w \cos (\pi x)\\
\vdots\\
w \cos (q \pi x)\\
w \sin (\pi x)\\
\vdots\\
w \sin (q \pi x)\\
0\\
f_1(x) - f_2(x)\\
0\\
(w-1) f_2(x) \[\ln(1-x) + \frac{3-2d}{(1-d)(2-d)} \]
\end{array}
\],
\ee
the estimator $\hat f(x)$ of the density $f(x)$ is asymptotically Gaussian
\be
\sqrt{T} \(\hat f(x) - f(x) \) \stackrel{{\cal L}}{\longrightarrow} {\cal N} \( 0, D(\theta_0)' A(\theta_0) D(\theta_0) \).
\ee
\end{proposition}

\begin{proof}
$D(\theta)$ is nothing but the gradient of $f(x)$ with respect to $\theta$. Thus, by use of 
the Delta method \cite{vaart00}, the result follows straightforwardly from proposition~\ref{prop1}.
\end{proof}


\section{\label{sec:computations}Empirical results}

In this section we present the conclusions drawn from the calibration of our model whose implementation is discussed in appendix~\ref{App:Conv}. To this aim, we use 
the intraday prices of ten middle and fourteen large capitalization stocks traded on the NYSE or the Nasdaq 
from 01/01/1994 to 12/31/2003 (which represents 2518 trading days). The description of the data, provided by TickData, 
is given in table~\ref{tab:data}.

\vspace{1cm}
\centerline{\it [Insert table~\ref{tab:data} about here]}
\vspace{1cm}

We first estimate the daily realized-volatility by use of the estimator~(\ref{eq:SSR}), as already discussed 
in section~\ref{sec1}. We stress that this variable will be considered as an observed variable in all the sequel. 
Before going further, it is worth to notice that we should get 390 one-minute prices for all assets since 
the quotations begin at 9:30 am and end at 16:00 pm. However, we have much less observations for the middle 
capitalization stocks than expected: typically 133 per day, on average. In addition, the number of intraday 
quotations  is not constant over time. Indeed, the number of available data is much smaller from 1994 to 1997 
than from 1998 to 2003 (respectively 45 and 192 for the middle capitalizations and 282 and 374 for the large 
capitalizations). Nevertheless, we has chosen to estimate the realized volatility over the whole time period 
ranging from 01/01/1994 to 12/31/2003.

\subsection{Calibration of the model over the whole time interval}

Let us underline that the estimation of the density of the heterogeneity variable $\phi$ provides us 
access to the parameter $d$ that characterizes the long-memory behavior of the time series of the realized 
log-volatility. It is then interesting to compare the values of $d$ obtained with our model and those obtained 
by other methods like the rescaled rang method set by \can{hurst51} and the regression method introduced \can{geweke83}. Notice that the former method has been refined by \citename{mandelbrot72} \citeyear{mandelbrot72,mandelbrot75}, \can{mandelbrot79} and later by \can{lo91}. However, his generalization involves an additional parameter whose value has a great influence on the results \cite{teverovsky99}. As a consequence, we have only resorted to the classical rescaled range statistic.

\begin{landscape}
\begin{figure}
\begin{center}
\includegraphics[width=11cm]{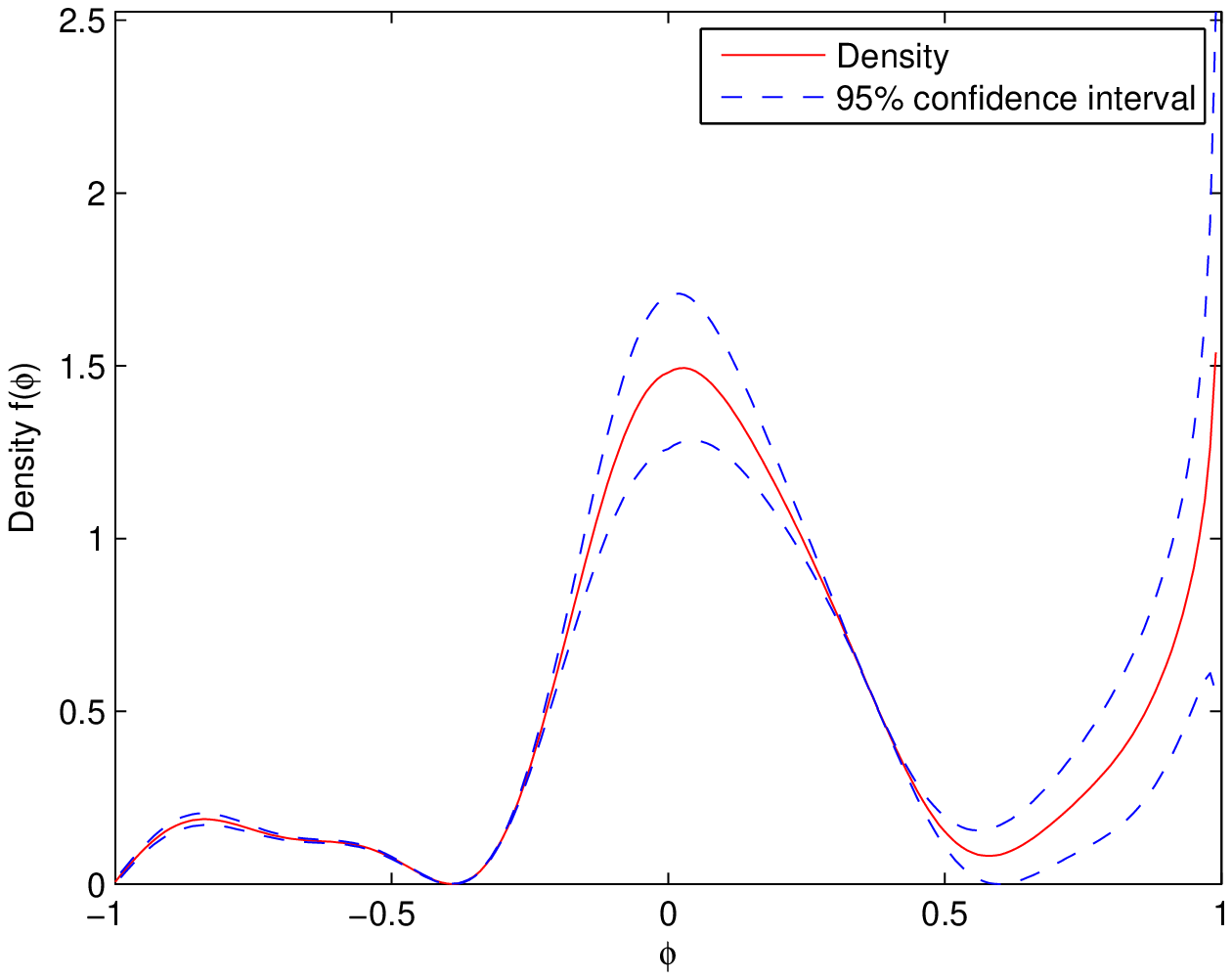} 
\includegraphics[width=11cm]{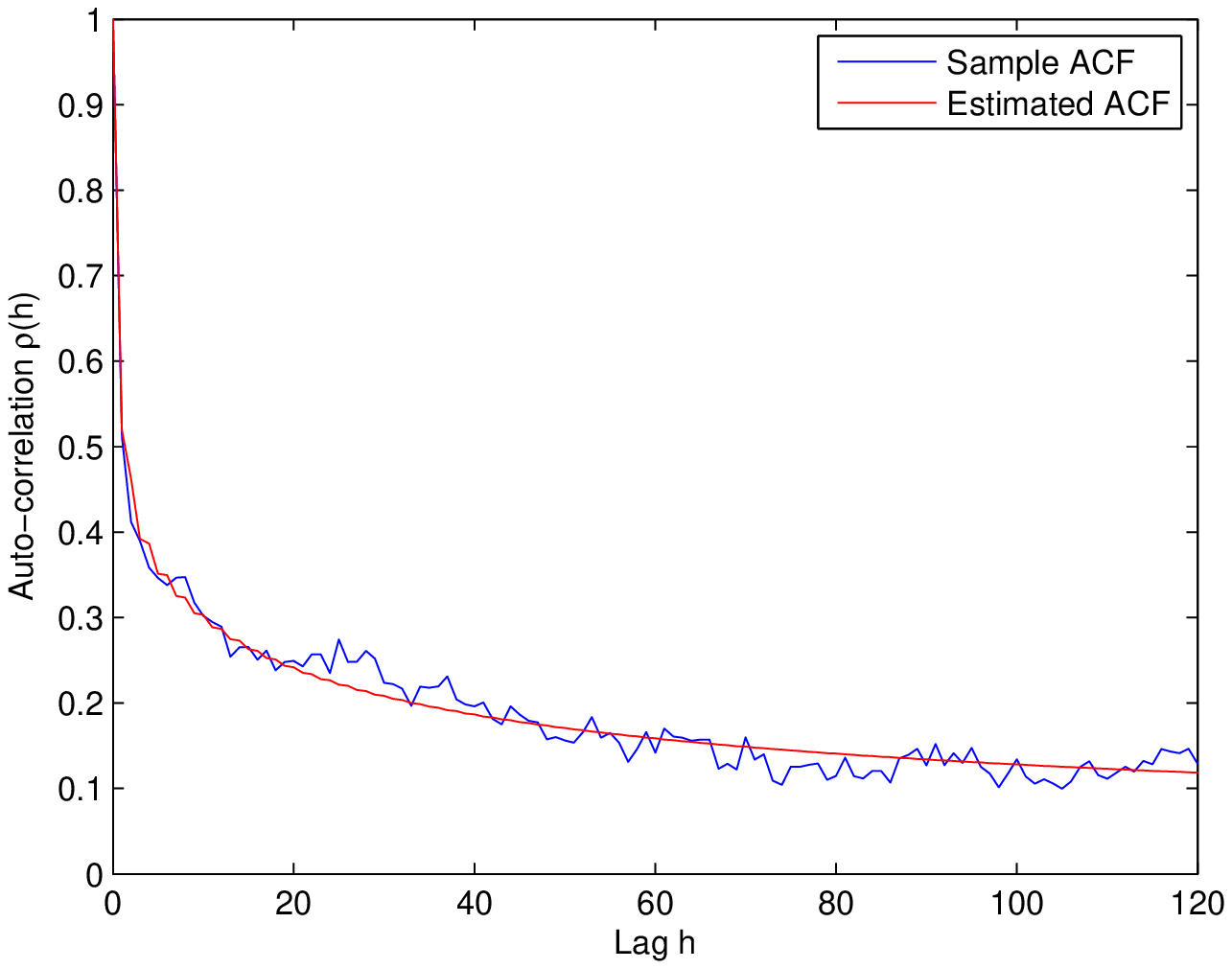}
\end{center}
\caption{\label{fig:FNF_optim_LVR}Estimation of the density with its 95 \% confidence interval (on the left panel) and the auto-correlation function from 
lag 0 to lag 120 (on the right panel) for FNF.} 
\end{figure}
\end{landscape}

First we show the graphical results obtained for a middle capitalization: Fidelity National Financial inc. (FNF, 
see figure~\ref{fig:FNF_optim_LVR}).
On the right panel the estimated auto-correlation function fits very well the sample one. The density of the heterogeneity variable $\phi$ is depicted on the left panel (plain curve) with the pointwise 95\% confidence interval (dashed curves). This shape with three distinct masses (one close to -1, an other close to 1 and the last one around 0)
is representative of half of the assets (The other half is depicted afterwards). 
In terms of agents' behavior, it means that there is mainly three kinds of agents. First, looking at 
the central part of the density function, we can conclude that most of the agents base their anticipations 
on the incoming flow of information and on the past realization (depending on the value of $\alpha$). 
Some others believe in the anticipation they performed the day before, 
which is related to the diverging part of the curve, in the neighborhood of one, while only a few, shown
by the part of the density near -1, systematically take the opposite of their previous anticipations.

\vspace{1cm}
\centerline{\it [Insert figure~\ref{fig:FNF_optim_LVR} about here]}
\vspace{1cm}

\begin{landscape}
\begin{figure}
\begin{center}
\includegraphics[width=11cm]{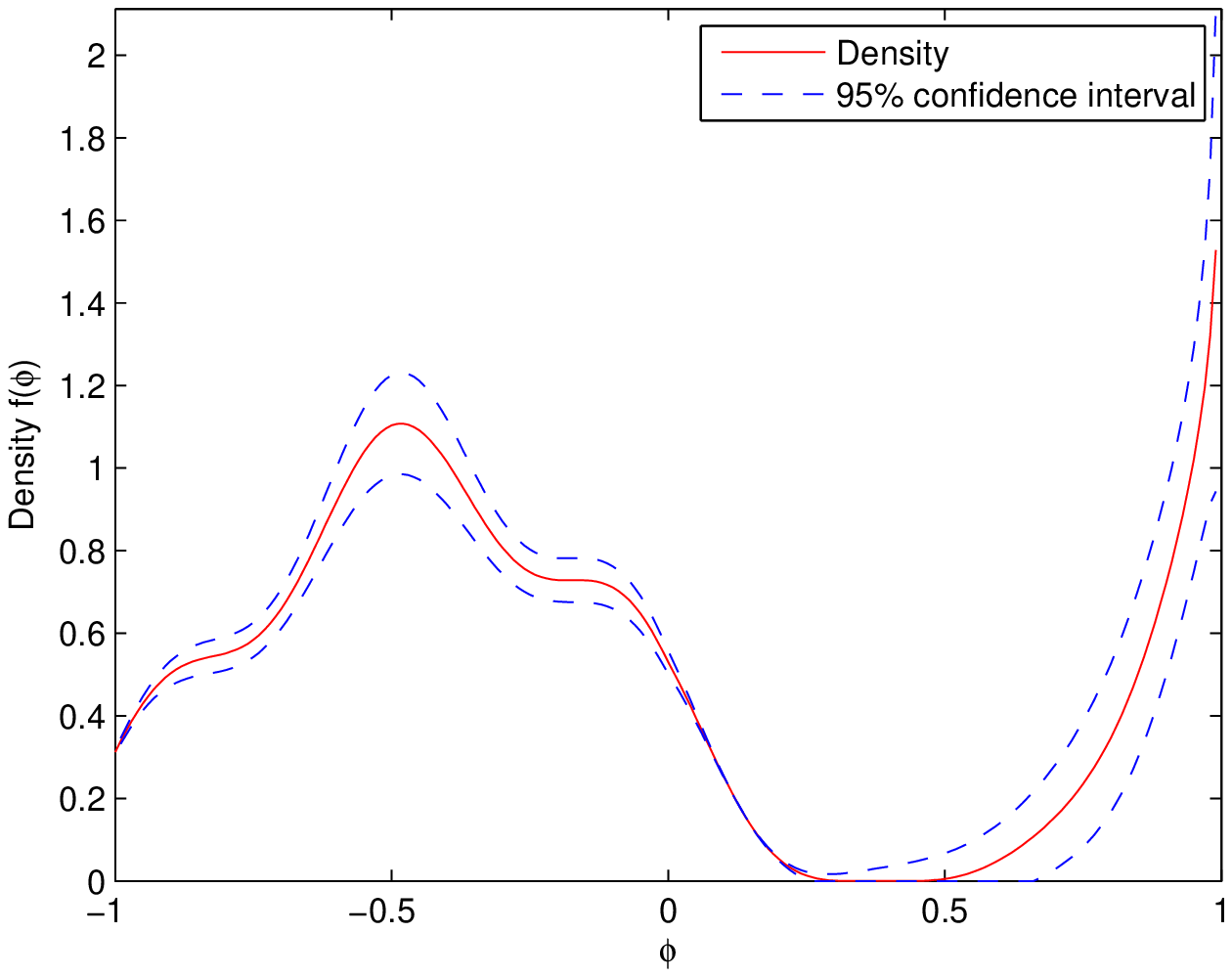} 
\includegraphics[width=11cm]{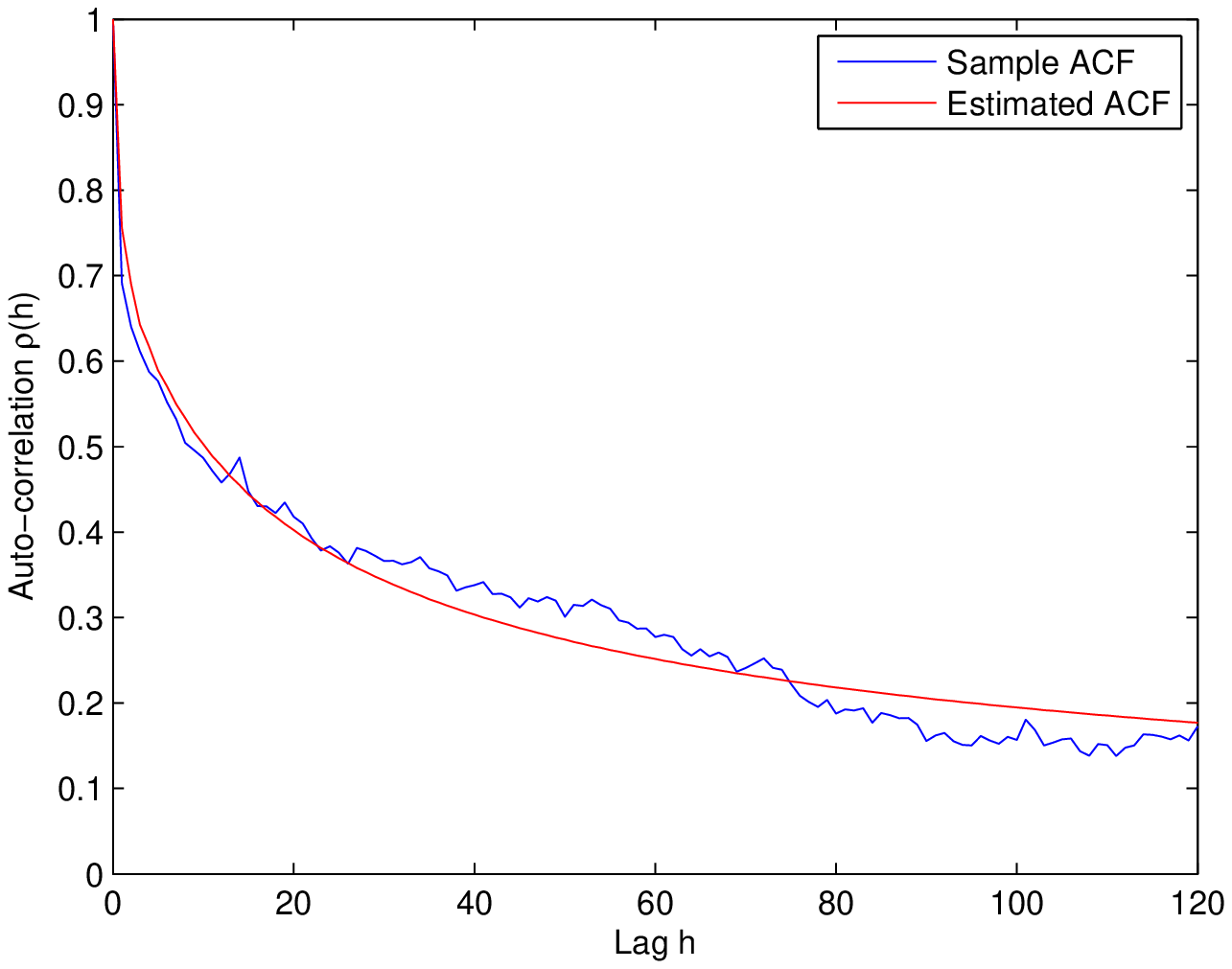}
\end{center}
\caption{\label{fig:MSFT_optim_LVR}Estimation of the density with its 95 \% confidence interval (on the left panel) 
and the auto-correlation function from lag 0 to lag 120 (on the right panel) for Microsoft.} 
\end{figure}
\end{landscape}

Secondly we show what we graphically obtain for the other half of the assets with the example of a large
capitalization: Microsoft (MSFT, see figure~\ref{fig:MSFT_optim_LVR}). Instead of noticing  
three distinct masses, we only see two. The agents confident in their past anticipation still remains, but it
is now more vague with the second category. In fact the curve is symmetric with a peak centered around -0.5. 
We still find agents who do not trust anymore in the anticipation they performed the previous day (the 
part of the density function near -1), and others who only care about incoming news (the part of the density
function near 0), and between these two situations there are many agents who both take into account the news 
and the opposite of their past anticipation. This is quite logical if an agent realizes that her past 
anticipation was far from the realization of the volatility, then she takes the opposite of her previous
anticipation and also becomes more careful in the incoming news.

\vspace{1cm}
\centerline{\it [Insert figure~\ref{fig:MSFT_optim_LVR} about here]}
\vspace{1cm}

These two typical shapes of the distribution of heterogeneity does not appear to be related to the size of the firms nor to any specific industry. However, the small number of assets per industry in our database (maximum five) prevents us from drawing definitive conclusions.

The results of the estimations of the long memory parameter $d$ by our model and by the two semi-parametric
methods are shown in table \ref{tab:d_all_assets}. It is quite obvious that, on the one hand,
each estimators agrees to say that long memory is present in almost all series while, on the other hand,
each estimator gives, for a same asset, different values of $d$. 
Most of the values obtained by our model range between $0.2$ and $0.4$ and a few are negative, instead of
between $0.3$ and $0.45$ with Geweke and Porter-Hudak's estimator. At least the long memory parameter obtained by the 
classical rescaled range analysis method is mainly contained between $0.40$ and $0.55$.

It is common knowledge that beyond $d=0.5$ the series is no longer stationary. As the estimations by the Hurst
method are greater than, but close to, 0.5, we may say that it is due to the uncertainty and, maybe, the inaccuracy 
of the method.
Moreover, the differences noticed between these results may be explained by the difficulty to use the estimators. Indeed, it is well-known that Geweke and Porter-Hudak's estimator is quite sensitive to the presence of short memory. On the contrary, our estimates should be considered as more robust {\it vis-a-vis} the presence of short-memory insofar as our model takes it into account explicitly.

\subsection{Study of the bubble burst effect} 

The study of the bubble burst effect is motivated by the fact that we will be able to
get informations about the agents' behavior during a prosperity period (before the bubble burst)
and during a recession period (after the bubble burst). In particular we will be able to answer 
the question:``Do the agents behave differently before and after a bubble burst ?''

To perform the computations, we randomly selected 5 large capitalizations because the bubble phenomenon is far more
pronounced in their price evolution than it is for middle capitalizations. 
We simply split the series when the maximum price is reached into two subseries.
The first one is defined as the pre-bubble burst period while the other one is the post-bubble burst period. 
Hereafter, the illustrations are drawn for a ``new technology'' company, Cisco Systems (CSCO), and a 
distribution company, Coca Cola (KO). The point is that the bubble phenomenon may be more pronounced
in the price evolution of a new technology company, which was more affected by the internet bubble, 
than a non technology firm.

\subsubsection{Study of a new technology asset: Cisco Systems}

\begin{landscape}
\begin{figure}
\begin{center}
\includegraphics[width=11cm]{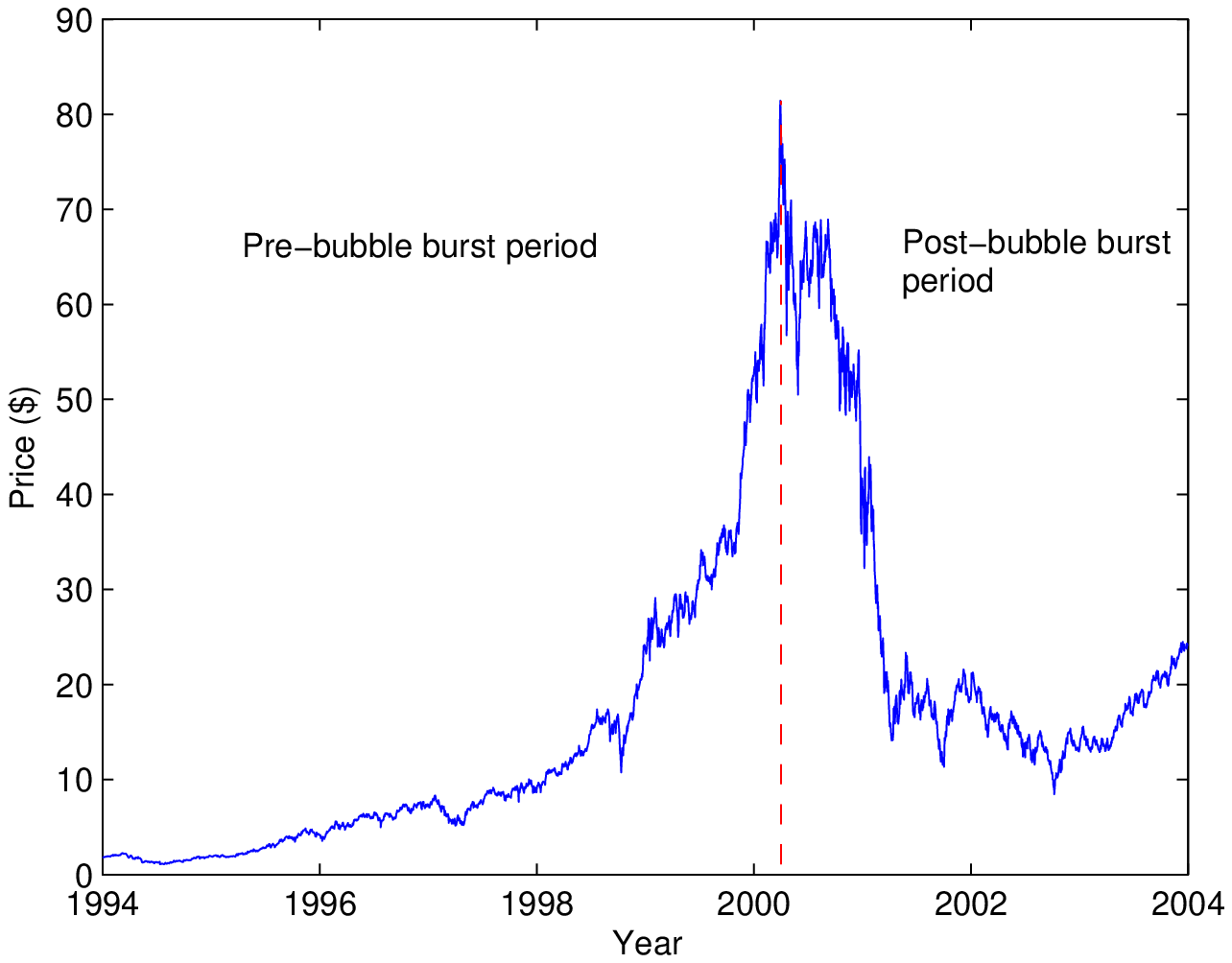} 
\includegraphics[width=11cm]{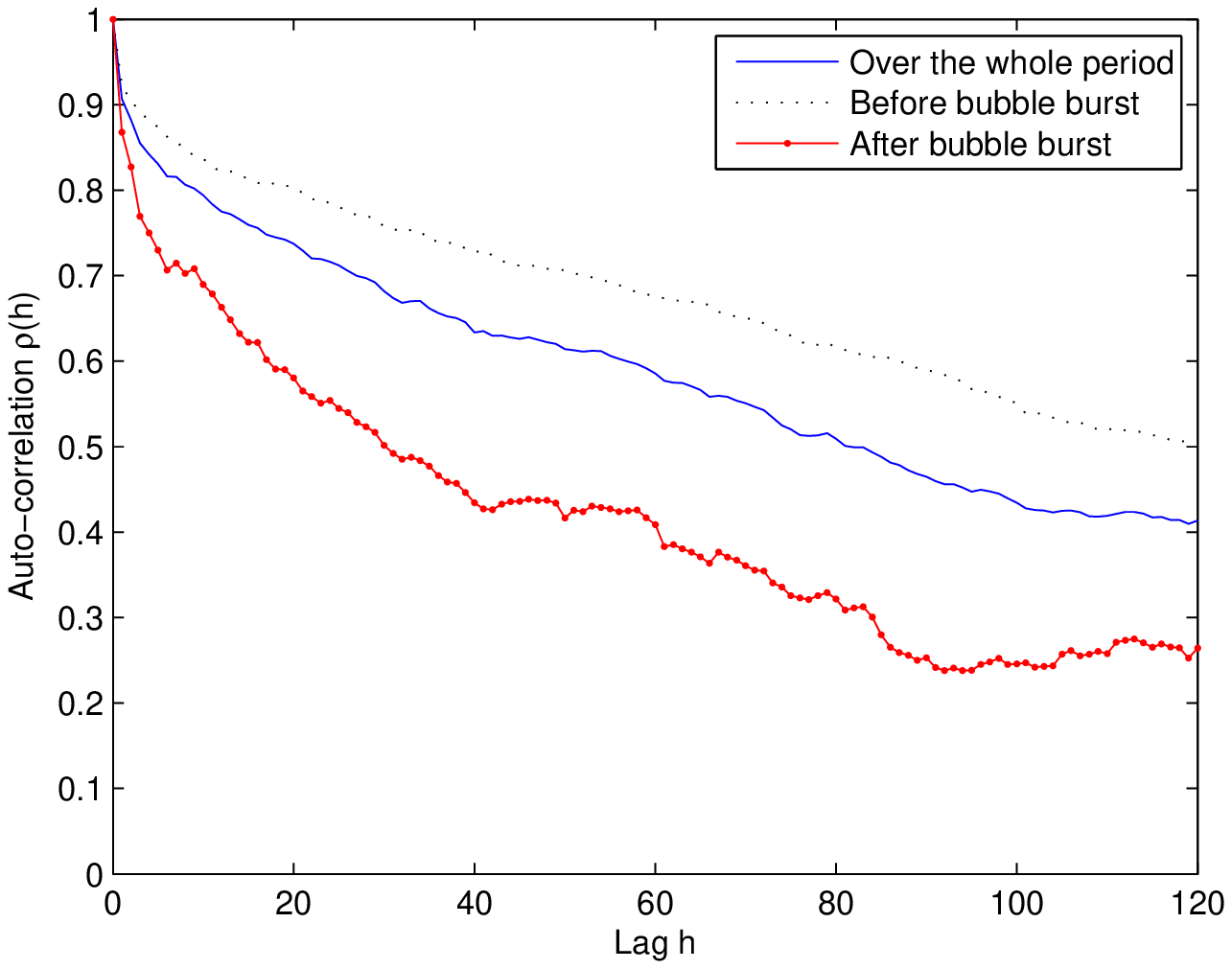}
\end{center}
\caption{\label{fig:bubble_CSCO}On the left panel is presented the evolution of the price of Cisco Systems
over the period considered in order to study the bubble burst effect. On the right panel, the auto-correlations
over different periods are drawn.} 
\end{figure}
\end{landscape}

On the left panel of figure \ref{fig:bubble_CSCO} we show the price evolution of CSCO between
01/01/1994 and 12/31/2003. 
On the right panel of figure \ref{fig:bubble_CSCO} we have drawn the sample auto-correlation functions 
of CSCO over the three periods. From it we easily deduce that on the one hand before the bubble burst the 
long memory parameter may be much larger than the one over the whole period, and on the other hand after 
the bubble burst, it must be a lot smaller.

\vspace{1cm}
\centerline{\it [Insert figure~\ref{fig:bubble_CSCO} about here]}
\vspace{1cm}

\begin{landscape}
\begin{figure}
\begin{center}
\includegraphics[width=11cm]{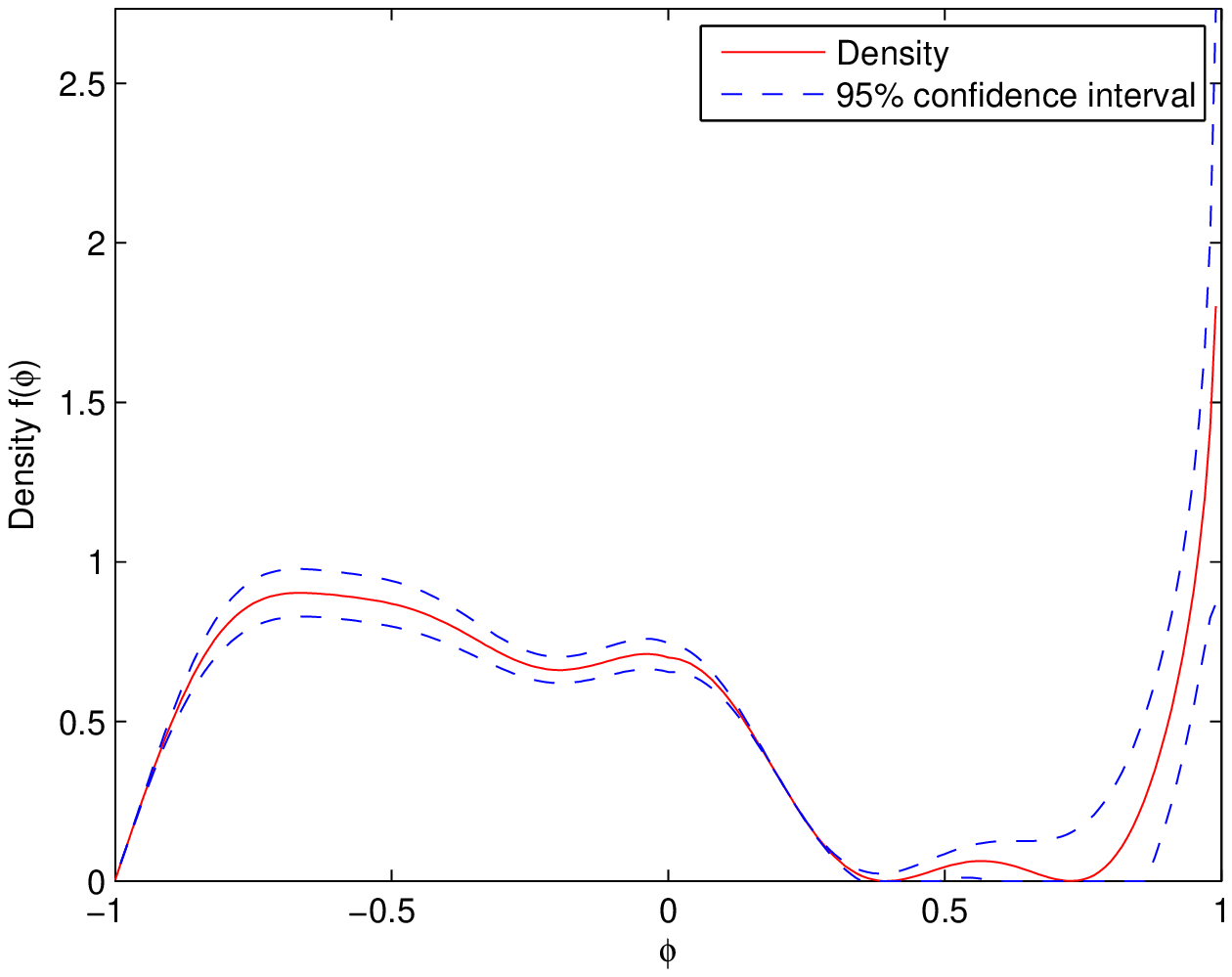} 
\includegraphics[width=11cm]{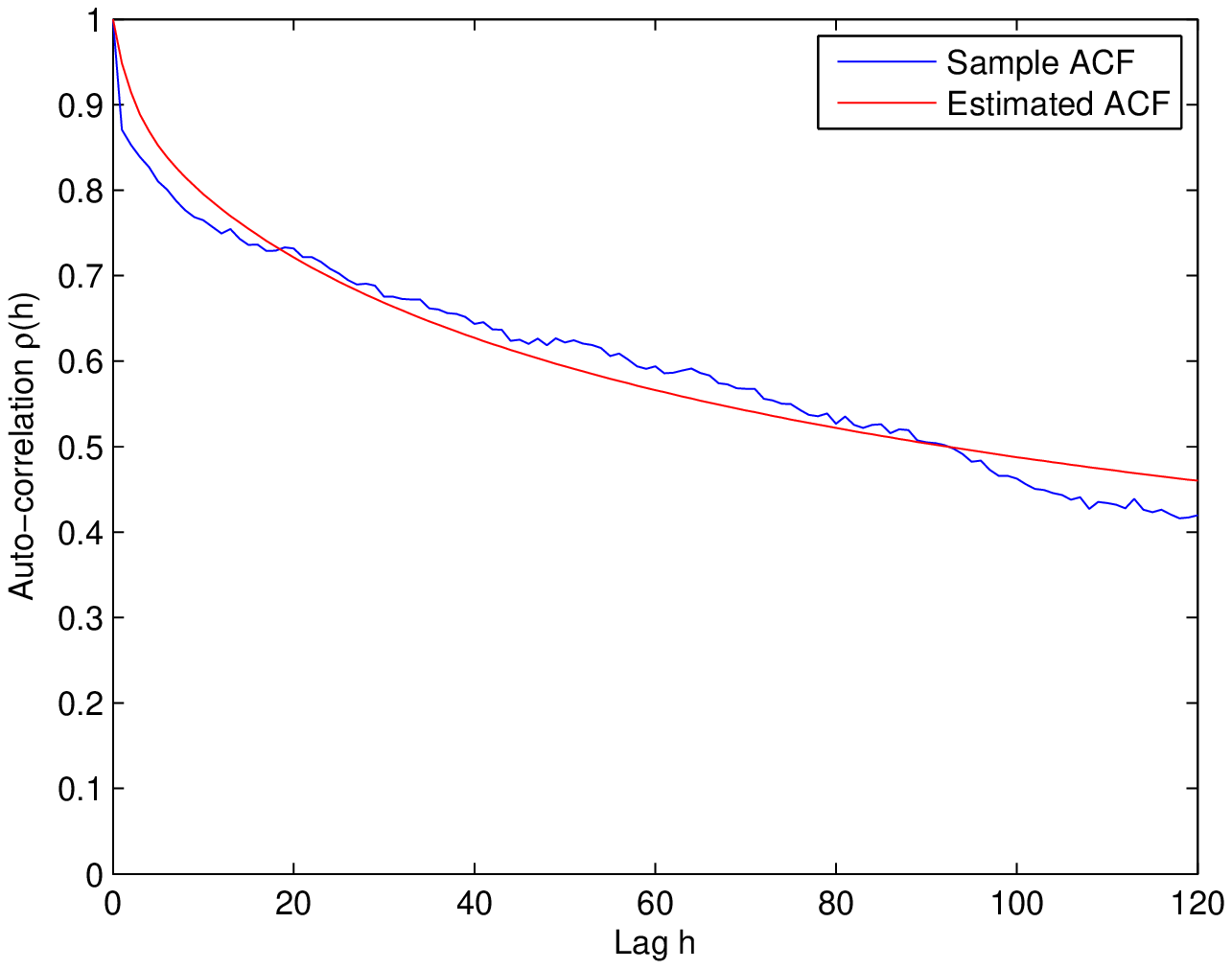}
\end{center}
\caption{\label{fig:bubble_CSCO_before}Density of the heterogeneity coefficient (on the left) and auto-correlation
of the realized log-volatility of Cisco Systems (on the right) over the pre-bubble burst period.} 
\end{figure}
\end{landscape}

\begin{landscape}
\begin{figure}
\begin{center}
\includegraphics[width=11cm]{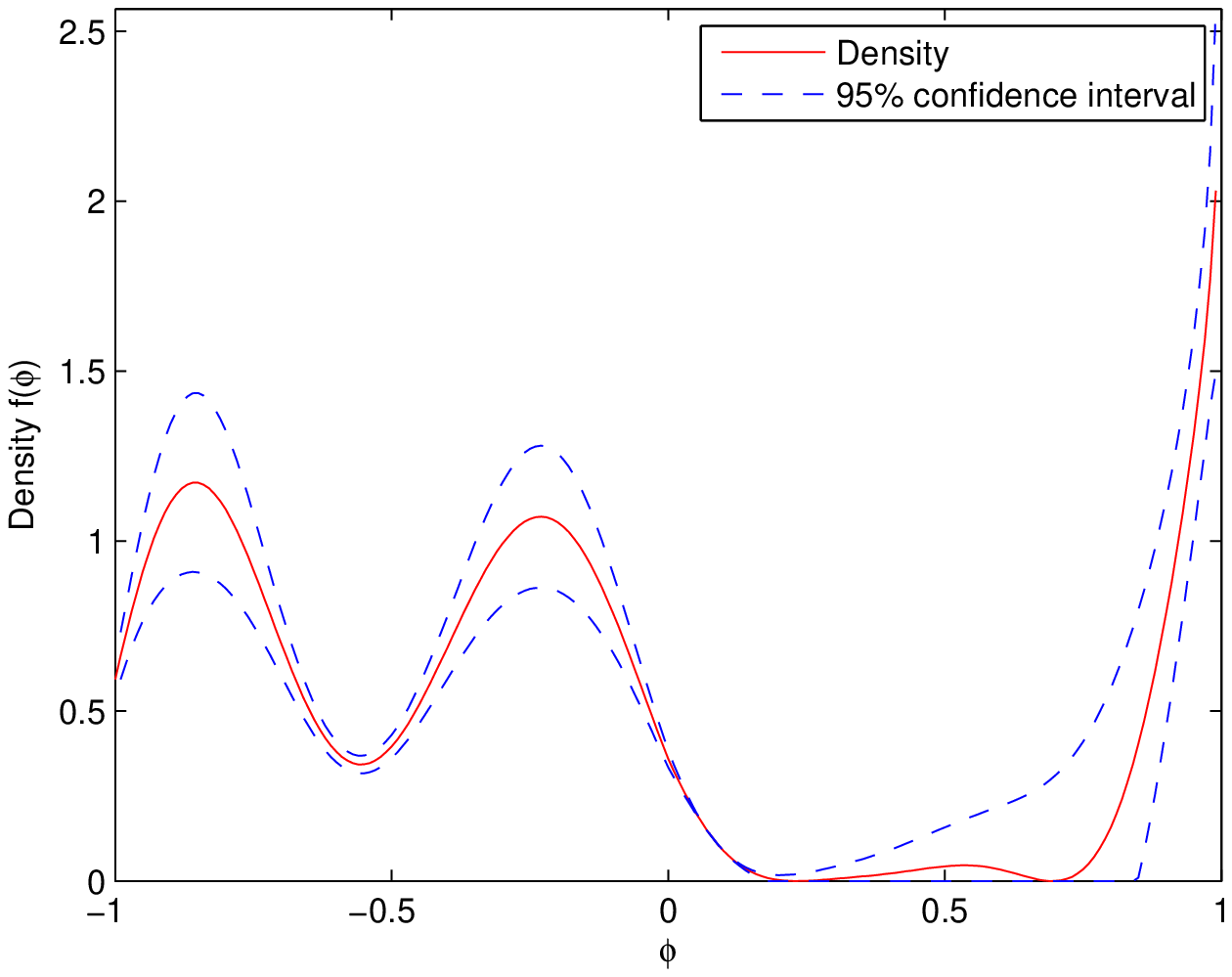} 
\includegraphics[width=11cm]{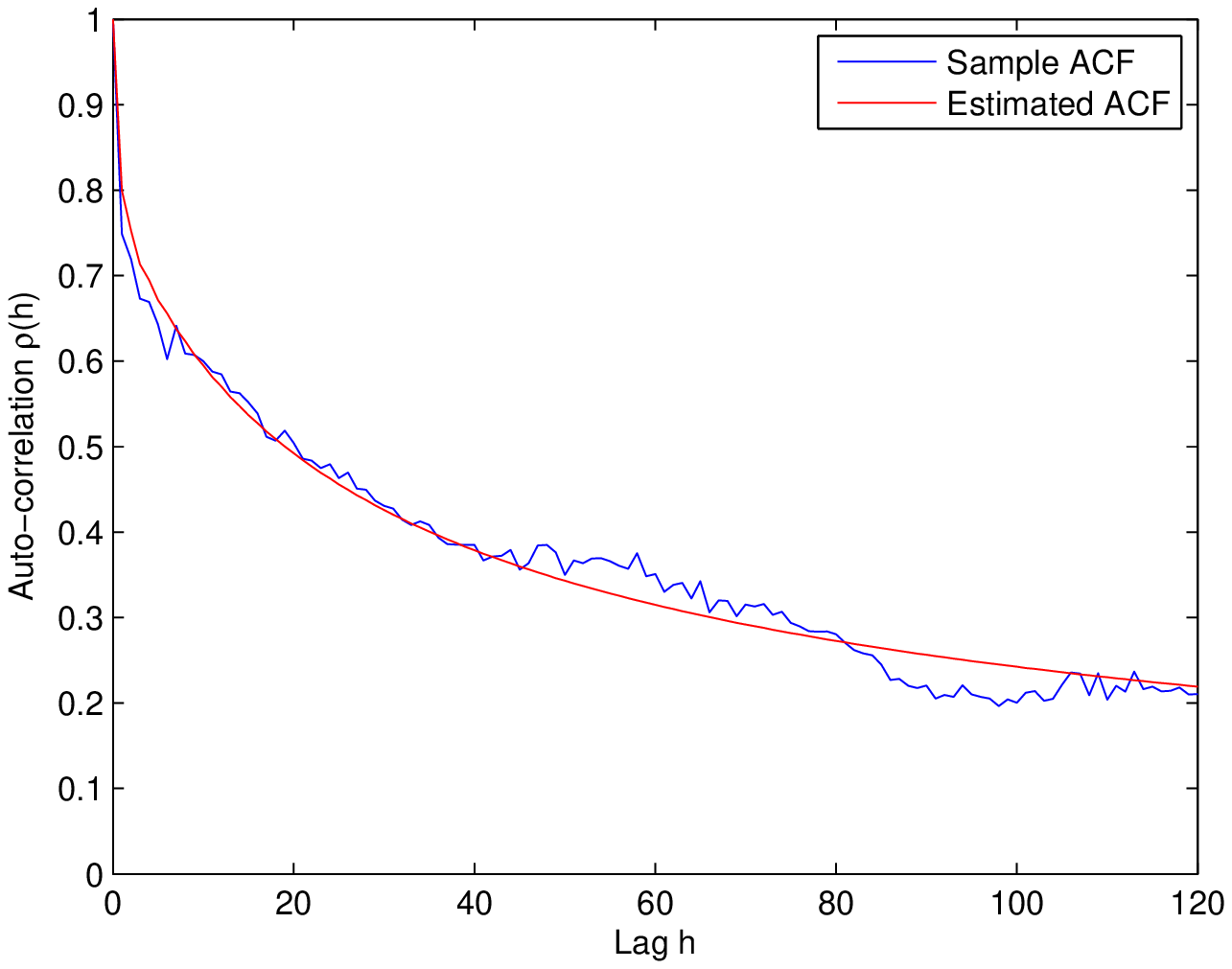}
\end{center}
\caption{\label{fig:bubble_CSCO_after}Density of the heterogeneity coefficient (on the left) and auto-correlation
of the realized log-volatility of Cisco Systems (on the right) over the post-bubble burst period.} 
\end{figure}
\end{landscape}

Let us complete these impressions with the densities and auto-correlation functions estimated
over the two subperiods (see figures \ref{fig:bubble_CSCO_before} and \ref{fig:bubble_CSCO_after}). 
We show on the left panel the estimated density with its 95\% confidence interval and on the right one 
the sample and estimated auto-correlation functions.

Concerning the pre-bubble burst period (see figure \ref{fig:bubble_CSCO_before}), on the right side,
we see that the estimated auto-correlation function fits pretty well to the sample one.
Moreover on the left side of this figure, the 95\% confident interval is very thin. We deduce that 
the optimization performed well. 

The shape of the density is quite similar as the one we obtained with Microsoft (see figure \ref{fig:MSFT_optim_LVR})
and implies a strong long memory.
To sum up, we get two masses: the one close to 1 represents agents who believe in their past anticipation;
the other one, from -1 to 0.2 includes several different behaviors. Nevertheless they have something in common: they 
do not replicate their past anticipation. Some are rational and use the incoming information flows and others not really.
They are inspired by the opposite of their past anticipation or a mix between the news and the contrary of the previous
anticipation.

\vspace{1cm}
\centerline{\it [Insert figure~\ref{fig:bubble_CSCO_before} about here]}
\vspace{1cm}

As for the post-bubble burst period, we first notice on figure \ref{fig:bubble_CSCO_after} that the estimated
auto-correlation function fits very well to the sample one (on the right panel) and that the 95\% confident
interval of the density function is good too. 

\begin{figure}
\begin{center}
\includegraphics[width=11cm]{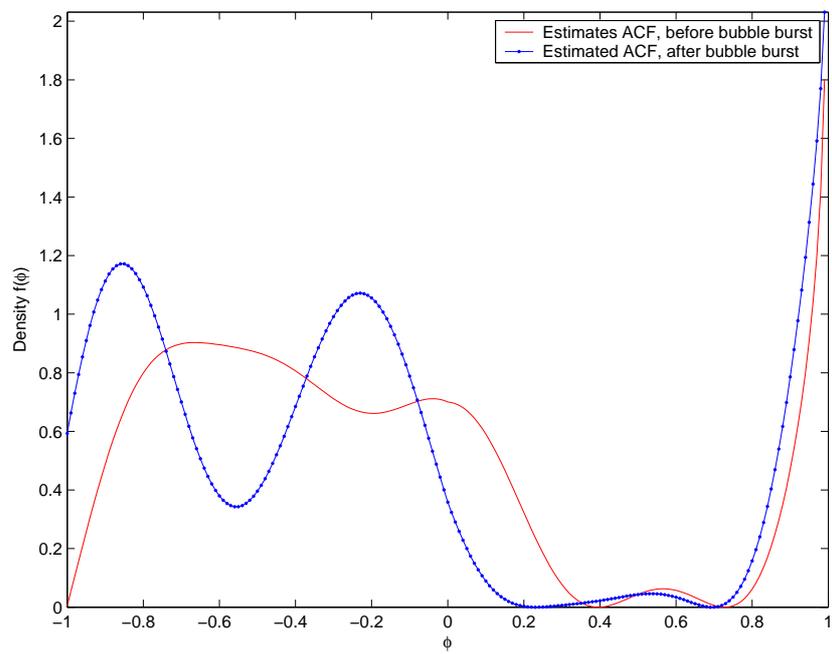} 
\end{center}
\caption{\label{fig:bubble_CSCO_comp}Pre bubble burst density compared to post bubble burst density of Cisco Systems.} 
\end{figure}

Here, the situation is a little bit different: the value of the long memory parameter $d$ is smaller (see table~\ref{tab:bubble_effects} hereafter).
Basically, it means that the proportion of agents who believe in the continuity of the previous market 
conditions is smaller. Figure \ref{fig:bubble_CSCO_comp} helps us compare the agents behaviors 
during the two subperiods.
We observe that after the bubble burst the mass from -1 and 0.2 changes. It was first compressed between -1 and 0,
then the mass has got divided into two masses. As a consequence, on the one hand the proportion of agents who
perform their anticipation basing their judgment on the contrary to their previous anticipation is greater. Those agents may recognize that they were wrong or it can reflect the confusion resulting from the increase in the level of uncertainty about the future evolution of the economic environment after the bubble burst.
On the other hand, the proportion of agents who mainly use information is greater too, in accordance with the increase in the level of uncertainty and therefore with the increasing need for information.

\vspace{1cm}
\centerline{\it [Insert figure~\ref{fig:bubble_CSCO_after} about here]}
\vspace{1cm}

\vspace{1cm}
\centerline{\it [Insert figure~\ref{fig:bubble_CSCO_comp} about here]}
\vspace{1cm}

\subsubsection{Study of a non technology asset : Coca Cola}

\begin{landscape}
\begin{figure}
\begin{center}
\includegraphics[width=11cm]{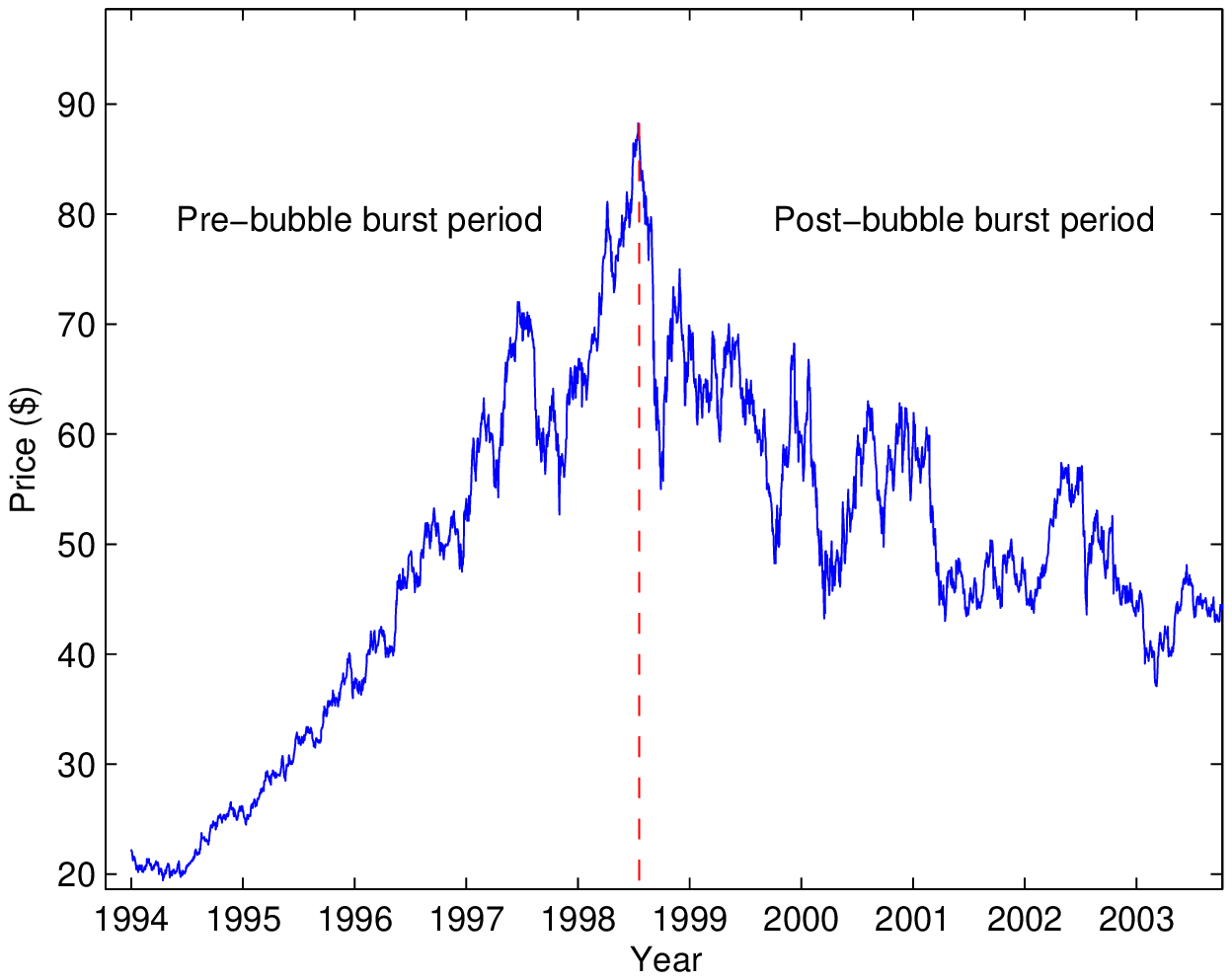} 
\includegraphics[width=11cm]{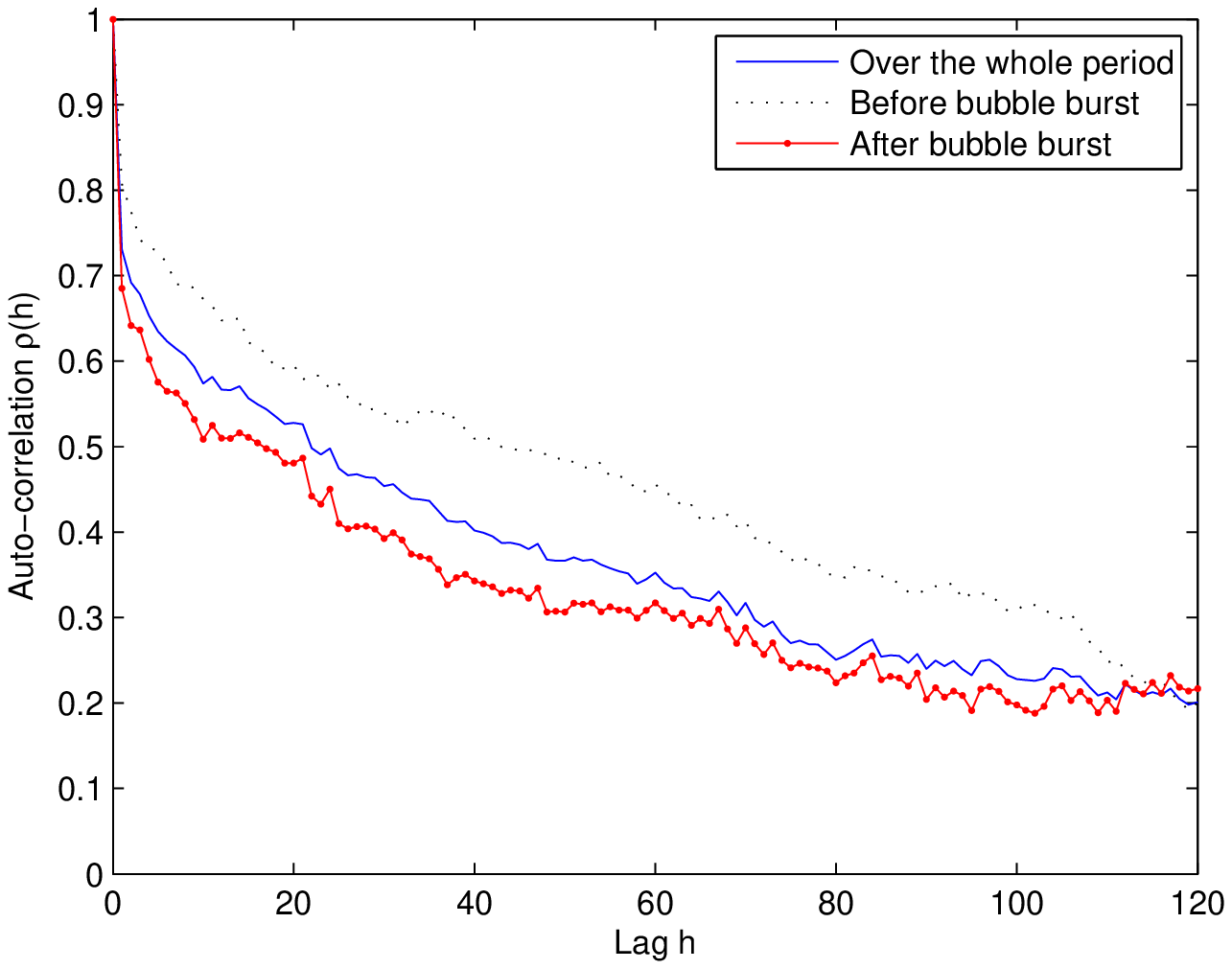}
\end{center}
\caption{\label{fig:bubble_KO}On the left panel is presented the evolution of the price of Coca Cola
over the period considered in order to study the bubble burst effect. On the right panel, the auto-correlations
over different periods are drawn.} 
\end{figure}
\end{landscape}

We reproduce for Coca Cola the same figures as for Cisco Systems.
We notice on the left side of figure \ref{fig:bubble_KO} that the bubble phenomenon is less obvious
as observed previously. The immediate consequence is shown on the left side of this figure where the sample
auto-correlation functions over the three periods are not as different as they are for CSCO. Then we 
expect to get closer long memory parameters, especially over the whole period and after the bubble burst.

\vspace{1cm}
\centerline{\it [Insert figure~\ref{fig:bubble_KO} about here]}
\vspace{1cm}

\begin{landscape}
\begin{figure}
\begin{center}
\includegraphics[width=11cm]{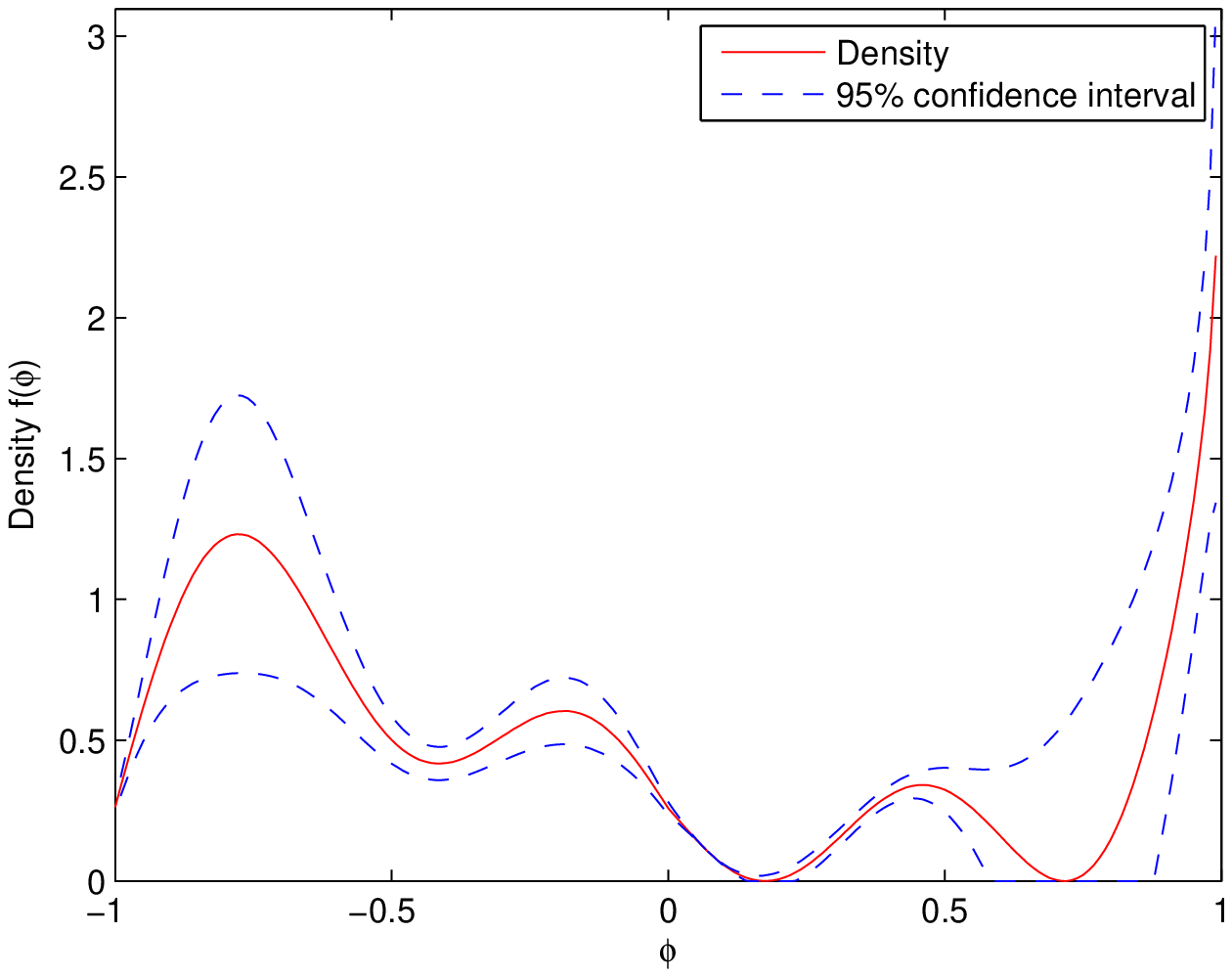} 
\includegraphics[width=11cm]{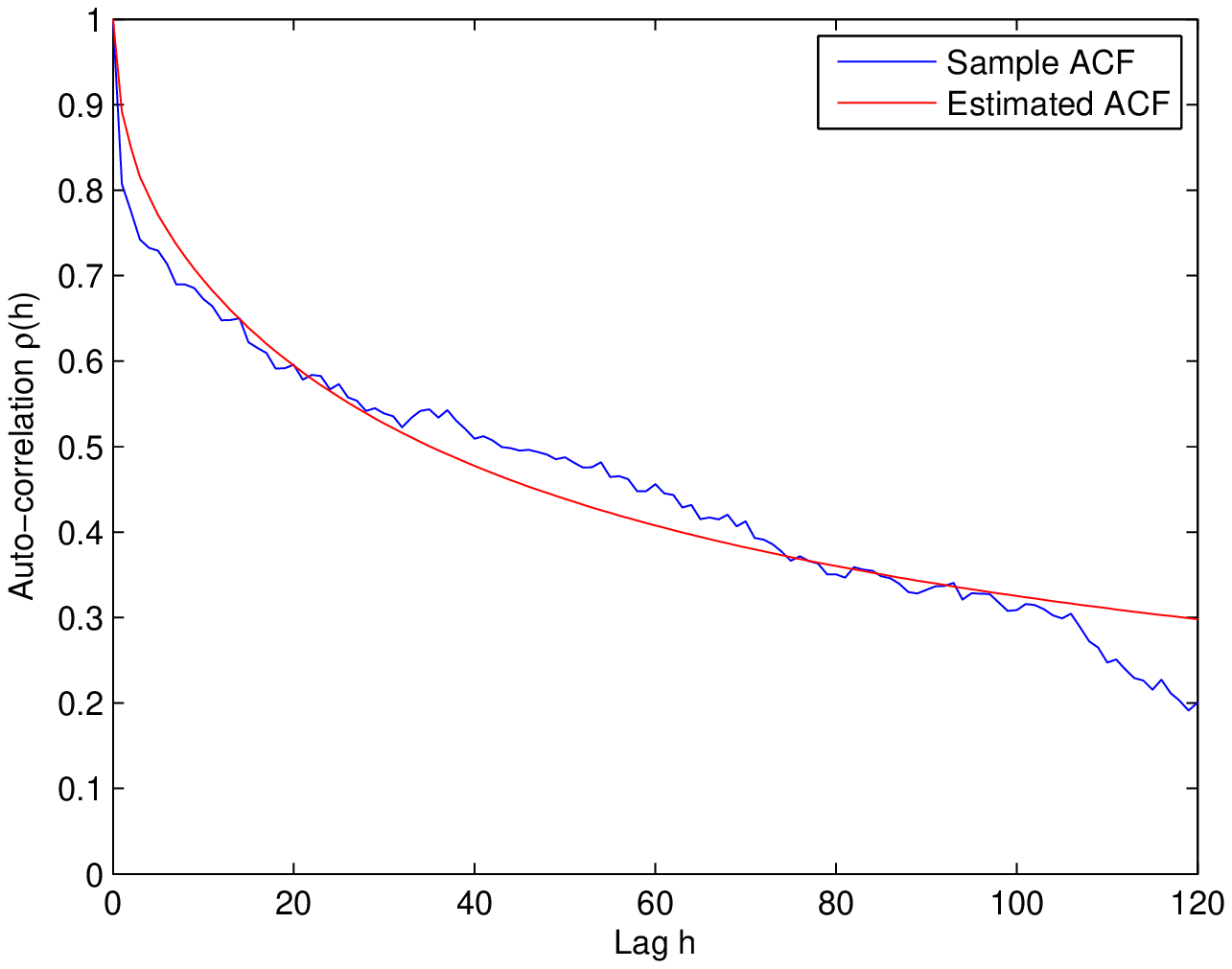}
\end{center}
\caption{\label{fig:bubble_KO_before}Density of the heterogeneity coefficient (on the left) and auto-correlation
of the realized log-volatility of Coca Cola (on the right) over the pre-bubble burst period.} 
\end{figure}
\end{landscape}

\begin{landscape}
\begin{figure}
\begin{center}
\includegraphics[width=11cm]{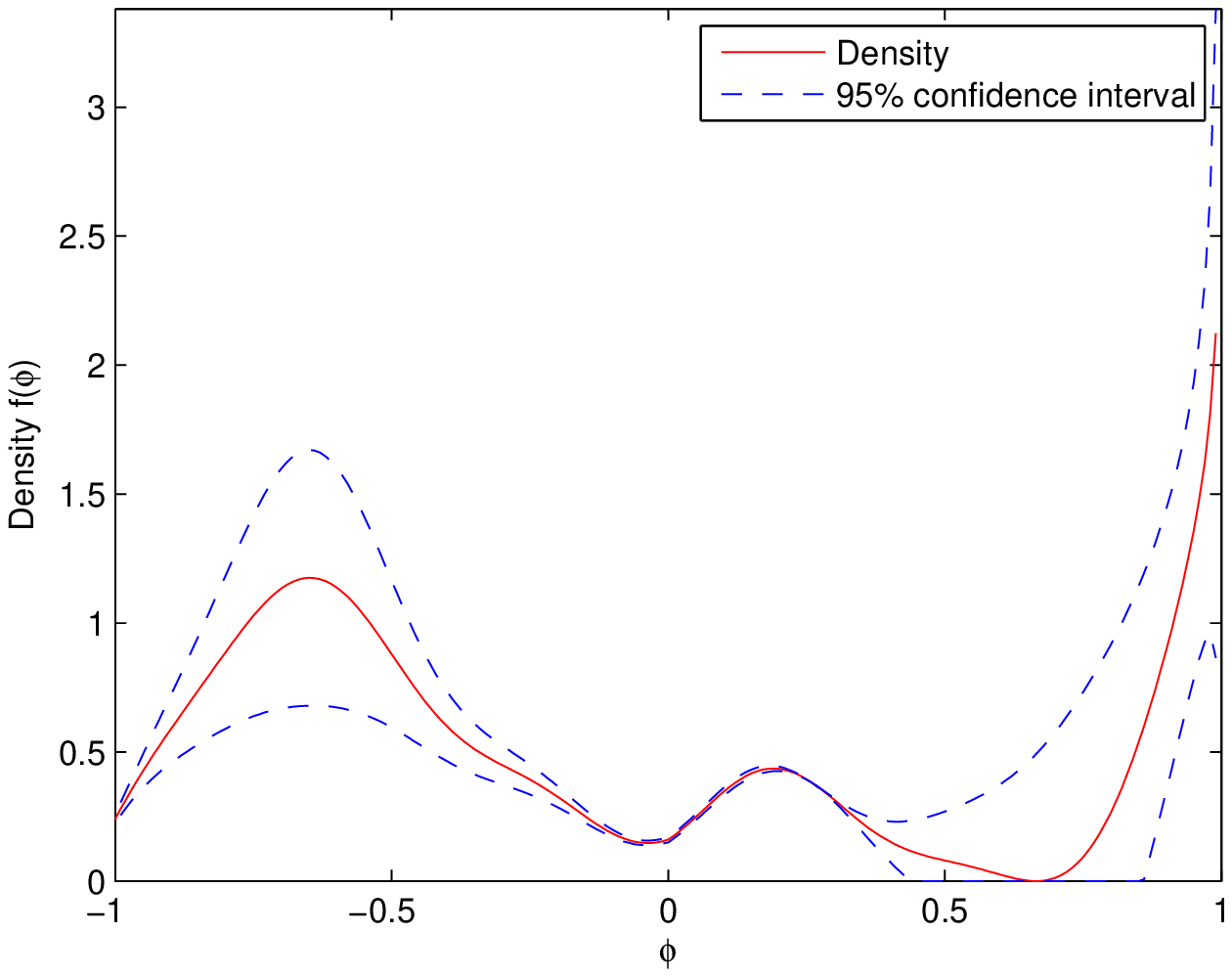} 
\includegraphics[width=11cm]{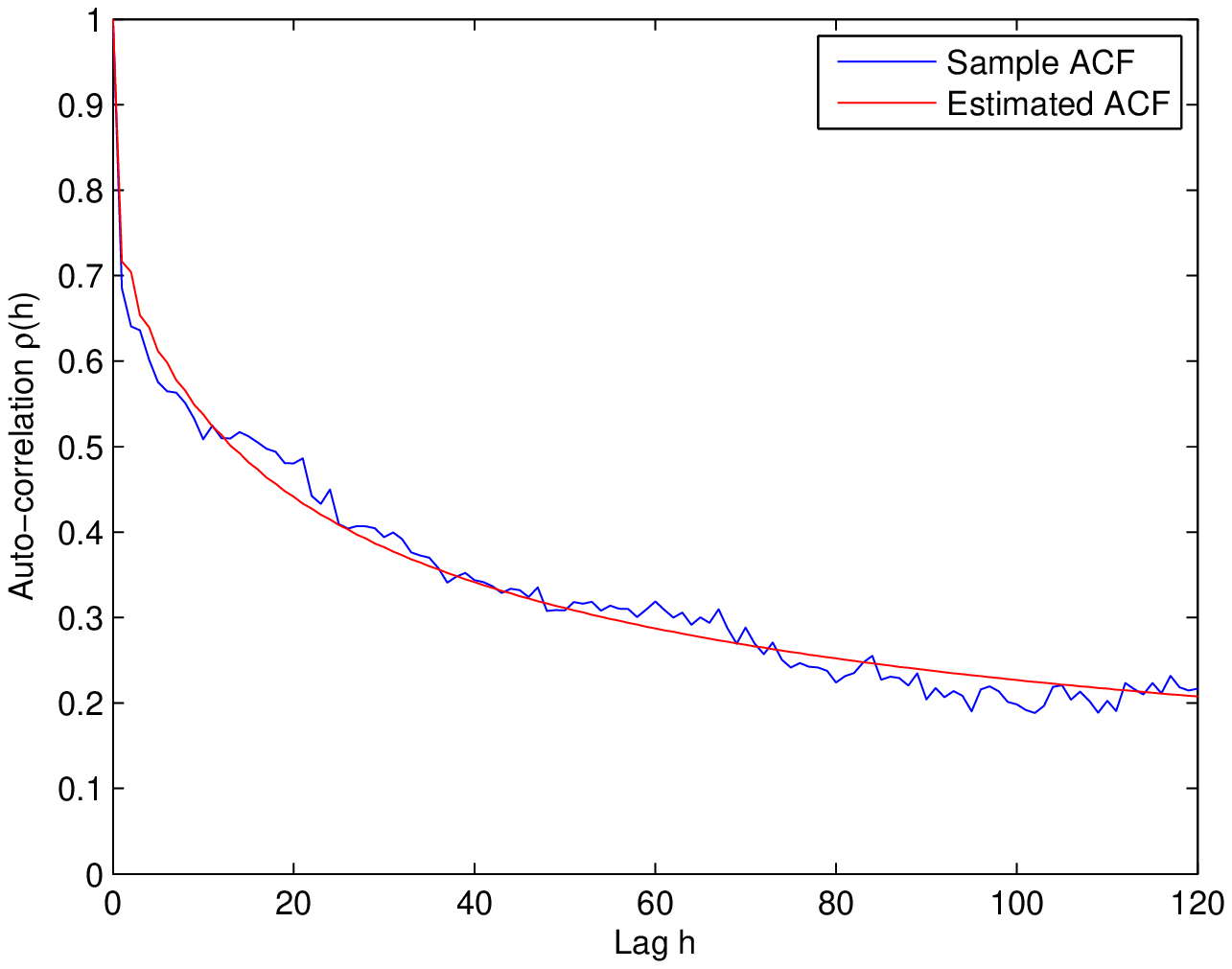}
\end{center}
\caption{\label{fig:bubble_KO_after}Density of the heterogeneity coefficient (on the left) and auto-correlation
of the realized log-volatility of Coca Cola (on the right) over the post-bubble burst period.} 
\end{figure}
\end{landscape}

Let us be more accurate with the densities and auto-correlation functions estimated
over the two subperiods (see figures \ref{fig:bubble_KO_before} and \ref{fig:bubble_KO_after}). 
For both of the figures, notice that the estimated auto-correlation functions on the right panels fit very well 
to the sample ones (in particular after the burst), and the 95\% confident intervals are good too.

\vspace{1cm}
\centerline{\it [Insert figure~\ref{fig:bubble_KO_before} about here]}
\vspace{1cm}

\vspace{1cm}
\centerline{\it [Insert figure~\ref{fig:bubble_KO_after} about here]}
\vspace{1cm}

\begin{figure}
\begin{center}
\includegraphics[width=11cm]{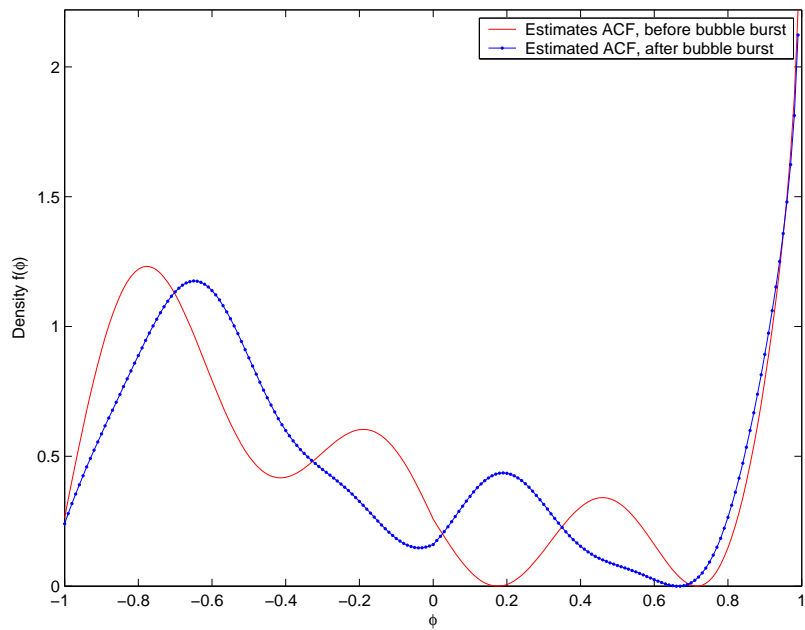} 
\end{center}
\caption{\label{fig:bubble_KO_comp}Pre bubble burst density compared to post bubble burst density of Coca Cola.} 
\end{figure}

As the shapes of these two densities look very similar let us put them together on the same graphic in order to analyze them (see figure \ref{fig:bubble_KO_comp}).
Looking at the divergence at one, the long memory parameter may be more or less the same. Contrary to the
evolution of CSCO density where the bulk of the post bubble burst density in the negative range was compressed, 
here it does not change that much. It only slightly expands along the positive axis after the bubble burst. The mass from 0.2 to 0.7 for the pre-bubble burst period which represent the agents who both believe in the continuity of the market
and also take into account the information flows has disappeared. It means that a smaller fraction of agents
still use their past anticipation.  

\vspace{1cm}
\centerline{\it [Insert figure~\ref{fig:bubble_KO_comp} about here]}
\vspace{1cm}

\subsubsection{General observations}

Let us look at the values of $d$. The values of the long memory parameter obtained by our model (see table \ref{tab:bubble_effects}) 
confirm our visual impressions. Indeed, before the burst, the long memory parameter is often greater 
than the one over the whole period for 4 out of 5 cases (in the 5th they are nearly similar). 
On the opposite, one observes that after the burst, $d$ falls for all the assets except for Coca Cola.

We have to mention that the very large standard deviation of the $d$-estimate for Microsoft during the second period should raise doubts about the convergence of our algorithm in this case. A reason might be the small number of available data. Indeed, for the post bubble burst period we only have 
about 900 days for 4 out of 5 assets whereas we have about 1600 data available for the pre bubble burst period. 
In the case of a smaller data number, the global optimum is more difficult to reach. Consequently the result may be wrong.

Nonetheless, to sum up, the results we obtained lead us to conclude that during a growth cycle, the number of
agents who are confident in the market and whose strategy remains almost the same day after day, is greater 
than the one during a decline cycle where the agents adopt more various behaviors. Indeed, some still base 
their anticipation on the previous one. Some do not believe in their past anticipation anymore
and take into account the information flow. Some admit they were wrong the day before and make a 
contrary anticipation. Others make a mix between a contrary anticipation and taking into account the news.

\section{Conclusion}

Based upon the recent literature on the aggregation theory, we have provided a model of realized 
log-volatility that aims at relating the behavior of the economic agents to the long memory of 
the volatility of asset returns. In spite of its simplicity, this model allows taking into account 
many agents' behavior and performs good estimates in general. The estimated coefficients often lead 
to auto-covariance and auto-correlation functions well fitted with their sample counterparts. 
In addition, the results derived from the study of the bubble burst effect -- namely a higher tendency 
to replicate the anticipations of the day before and to neglect the incoming information flow before 
the bubble burst than after -- are quite reasonable.
 


\clearpage
\appendix

\section{\label{app:HM} Proof of proposition~\ref{propMain}}
\label{app:CV_beta}

Under the assumption $||A|| <1$, the stationary solution of equation~(\ref{eq:Dyn2}) is given by
\be
\hat X_t = \sum_{k=0}^\infty A^k C \epsilon_{t-k} + \sum_{k=0}^\infty A^k  \eta_{t-k},
\ee
and the excess realized log-volatility writes
\be
\bar X_{n,t} = \sum_{k=0}^\infty \( \frac{1}{n} 1_n'A^k C\) \epsilon_{t-k} 
+ \sum_{k=0}^\infty \frac{1}{n} 1_n' A^k  \eta_{t-k}.
\ee
Accounting for the fact $A = D + \frac{1}{n} \Psi \cdot 1_n'$, $A^k$ is solution of 
\be
A^k = A^{k-1}  \( D + \frac{1}{n} \Psi \cdot 1_n'\),
\ee
so that one has to solve the recurrence equation
\be
1_n' A^k = \( 1_n' A^{k-1} \) D  + \beta_{k-1} 1_n',
\ee
where $\beta_k = \frac{1}{n} 1_n A^{k} \psi$. It is then a matter of simple algebraic 
manipulations to show that
\bea
\frac{1}{n}1_n' A^k C &=& \frac{1}{n}1_n' D^k C+ \frac{1}{n} \sum_{i=1}^k \beta_{i-1} \cdot 1_n' D^{k-i} C,\\
&=& \E\[c\] \cdot \E \[\varphi^k\]  + \E\[c\] \sum_{i=1}^k \beta_{i-1} \E \[\varphi^{k-i}\]\quad {\rm (a.s)}, 
\qquad {\rm 	as}~ n \rightarrow \infty, \label{eq:slkdn}
\eea
where
\bea
\beta_k &=& \frac{1}{n} 1_n' D^k \Psi + \sum_{i=1}^k \beta_{i-1} \cdot \( \frac{1}{n} 1_n' D^{k-i} \Psi\),\\
&=& \E \[ \psi \varphi^k \] + \sum_{i=1}^k \beta_{i-1} \cdot \E \[ \psi \varphi^{k-i} \]\quad {\rm (a.s)}, 
\qquad {\rm 	as}~ n \rightarrow \infty,\label{eq:rkth2}
\eea
while
\bea
\frac{1}{n}1_n' A^k \eta_{t_k} &=& \frac{1}{n}1_n' D^k \eta_{t_k} + \frac{1}{n} \sum_{i=1}^k \beta_{i-1} 
\cdot 1_n' D^{k-i} \eta_{t_k},\\
&=& \E \[\varphi^k\] \E\[\eta_t\]  + \sum_{i=1}^k \beta_{i-1} \E \[\varphi^{k-i}\] \E\[\eta_t\] 
\quad {\rm (a.s)}, \qquad {\rm 	as}~ n \rightarrow \infty,\\
&=& 0,
\eea
provided that $\E\[ \eta_t\]=0$.

So, $\bar X_t = \lim_{n \rightarrow \infty} \bar X_{n,t}$ is equal to
\be
\bar X_t =  \E\[c\] \cdot \sum_{k=0}^\infty  \( \E \[\varphi^k\]  +  \sum_{i=1}^k \beta_{i-1} 
\E \[\varphi^{k-i}\] \) \epsilon_{t-k}.
\ee

Let us simplify this expression by setting
\be
\label{eq:dskjk}
\tilde \beta_k = \E \[\varphi^k\] + \sum_{i=1}^k \beta_{i-1} \E \[\varphi^{k-i}\],
\ee
so that 
\be 
\bar X_t =  \E\[c\] \cdot \sum_{k=0}^\infty \tilde \beta_k \epsilon_{t-k}.
\ee

Now, multiplying equation (\ref{eq:dskjk}) by $x^k$ and summing over $k$ from zero to infinity, we get
\be
\sum_{k=0}^\infty \tilde \beta_{k+1} x^{k+1} = \sum_{k=0}^\infty \E\[\varphi^{k+1} \] x^{k+1} + 
\sum_{k=0}^\infty \( \sum_{i=0}^{k} \beta_i \E\[\varphi^{k-i} \]\) x^{k+1}.
\ee
As it is well-known that
\be
\sum_{k=0}^\infty \( \sum_{i=0}^{k} \beta_i \E\[\varphi^{k-i} \]\) x^{k+1} = 
x\(\sum_{k=0}^\infty \beta_{k+1} x^{k} \) \( \sum_{k=0}^\infty \E\[\varphi^{k} \] x^{k}\),
\ee
we obtain
\be
\label{eq:beta_tilde_k}
\sum_{k=0}^\infty \tilde \beta_{k} x^{k} = \sum_{k=0}^\infty \E\[\varphi^{k} \] x^{k}  
\left\{ 1 + x \sum_{k=0}^\infty \beta_{k} x^{k} \right\}.
\ee

Then, focusing on $\ds \sum_{k=0}^\infty \beta_{k} x^{k}$ and by use of equation (\ref{eq:rkth2}), the same argument yields
\be
\sum_{k=0}^\infty \beta_{k+1} x^{k+1} = \sum_{k=0}^\infty \E\[\psi\varphi^{k+1} \] x^{k+1} + 
\sum_{k=0}^\infty \( \sum_{i=0}^{k} \beta_i \E\[\psi\varphi^{k-i} \]\) x^{k+1},
\ee
\be
\sum_{k=0}^\infty \beta_{k} x^{k} = \sum_{k=0}^\infty \E\[\psi\varphi^{k} \] x^{k} + 
x \( \sum_{i=0}^{\infty} \beta_i x^{i}\) \( \sum_{i=0}^{\infty}\E\[\psi\varphi^{i} \] x^{i}\).
\ee
In other words
\be
\label{eq:beta_k}
\sum_{k=0}^\infty \beta_{k} x^{k} = \frac{\ds\sum_{k=0}^\infty \E\[\psi\varphi^{k} \] x^{k}}
{\ds 1 - x \sum_{k=0}^\infty \E\[\psi\varphi^{k} \] x^{k}},
\ee
which, by replacement in equation (\ref{eq:beta_tilde_k}), leads to
\be
\sum_{k=0}^\infty \tilde \beta_{k} x^{k} = 
\frac{\ds\sum_{k=0}^\infty \E\[\varphi^{k} \] x^{k}}{\ds 1 - x \sum_{k=0}^\infty \E\[\psi\varphi^k\] x^{k}},
\ee
and, provided that the permutation of the expectation and summation signs is allowed
\be
\sum_{k=0}^\infty \tilde \beta_{k} x^k = \frac{\E\[ \ds\frac{1}{1-x\cdot \varphi}\]}{\ds 1 -x \E\[ \frac{\psi}{1-x \cdot \varphi} \]},
\ee
which concludes the proof.\qed

\section{Proof of proposition~\ref{prop:memory} \label{App:Prop2}}

The expression of the spectral density of $X_t$ follows from proposition~\ref{propMain} and reads
\be
f_X(\lambda)=\frac{\E\[c\]^2 \cdot \sigma_\varepsilon^2}{2 \pi} \left| \frac{\displaystyle \sum_{k=0}^\infty {\rm E}\[\phi^k\] e^{-ik \lambda}}{\displaystyle 1 - e^{-i \lambda} \sum_{k=0}^\infty {\rm E}\[\psi \phi^k\] e^{-i k \lambda}} \right|^2 ~ .
\ee
Provided that the permutation of the expectation and summation signs is allowed, we can rewrite this relation as
\be
f_X(\lambda)=\frac{\E\[c\]^2 \cdot \sigma_\varepsilon^2}{2 \pi} \left|
\frac{\E\[ \ds\frac{1}{1-e^{-i \lambda} \cdot \varphi}\]}{\ds 1 - e^{-i \lambda} \cdot \E\[\frac{g(\varphi)}{1- e^{-i \lambda} \cdot \varphi} \]}
\right|^2.
\ee
Now, focusing on the term 
\be
N(x) = \E\[ \ds\frac{1}{1-x\cdot \varphi}\],
\ee
it follows from Karamata's theorem \cite{BGT89} that $N(x) \sim (1-x)^{-\alpha}$, as $x \to 1$, provided that the density of $\varphi$ satisfies $f(\varphi) \sim (1- \varphi)^{-\alpha}$, $\alpha \ge 0$. 

Similarly, provided that the density of $\varphi$ satisfies the previous assumption and that the conditional expectation $g(\varphi) \sim (1- \varphi)^{\beta}$, $\beta \ge \alpha - 1/2$, the term 
\be
D(x)=\E\[\frac{g(\varphi)}{1- x \cdot \varphi} \] \sim \left\{
\begin{array}{ll}
(1-x)^{-\alpha + \beta},& \quad \alpha > \beta \\
\E \[\frac{g(\varphi)}{1- \varphi}\],& \quad \alpha < \beta
\end{array}
\right.
\ee 
as $x \to 1$. 

So, when $\alpha$ is positive and larger than $\beta$, the ratio $N(x)/D(x) \sim (1-x)^{-\beta}$ so that the spectral density $f_X(\lambda) \sim \lambda^{-2 \beta}$. On the contrary, when $\beta$ is greater than of equal to $\alpha$, the ratio $N(x)/D(x) \sim (1-x)^{-\alpha}$ and the spectral density $f_X(\lambda) \sim \lambda^{-2 \alpha}$. \qed

\section{\label{app:acf_fft}Calculation of the auto-correlation function by the use of the Fast Fourier Transform}

The Auto-covariance function $\gamma_X(h)$ of a time series $\{X_t\}$ can be obtained by the spectral density
$f_X(\lambda)$ of $\{X_t\}$ according to
\be
\label{eq:acov_sd}
\gamma_X(h) = \int_{-\pi}^{\pi} e^{i \lambda h} f_X(\lambda) d\lambda = \int_0^{2\pi} e^{i \lambda h} f_X(\lambda) d\lambda.
\ee
Equation (\ref{eq:acov_sd}) in terms of a sum of integrals
\be
\label{eq:acov_sd_sum}
\gamma_X(h) = \sum_{k=0}^{N-1} \int_{\frac{2\pi}{N}k}^{\frac{2\pi}{N}(k+1)} e^{i \lambda h} f_X(\lambda) d\lambda.
\ee
By the trapezes method one can express an integral over a short domain such as
\be
\sum_{k=0}^{N-1} \int_{\frac{2\pi}{N}k}^{\frac{2\pi}{N}(k+1)} e^{i \lambda h} f_X(\lambda) d\lambda = 
\frac{\pi}{N} \( e^{i\frac{2\pi}{N}k}f_X\(\frac{2\pi}{N}k\) + e^{i\frac{2\pi}{N}(k+1)}f_X\(\frac{2\pi}{N}(k+1)\) \).
\ee
As, given the shape of the spectral density,  we may face problems in 0 and $2\pi$, lets us rewrite 
equation (\ref{eq:acov_sd_sum}) as
\begin{multline}
\gamma_X(h) = \int_{0}^{\frac{2\pi}{N}} e^{i \lambda h} f_X(\lambda) d\lambda +
\int_{\frac{2\pi}{N}(N-1)}^{2\pi} e^{i \lambda h} f_X(\lambda) d\lambda \\
+ \frac{\pi}{N} \sum_{k=1}^{N-2} \( e^{i\frac{2\pi}{N}kh}f_X\(\frac{2\pi}{N}k\) + e^{i\frac{2\pi}{N}(k+1)h}f_X\(\frac{2\pi}{N}(k+1)\) \),
\end{multline}

\begin{multline}
\label{eq:acov_fft}
\gamma_X(h) = 2 K( \alpha,\beta) \sum_{k=0}^\infty \frac{(-1)^k \(\frac{2\pi}{N}h\)^{2k}}{(2k)!(2k +2\alpha+1)}
-\frac{2\pi}{N} \mathcal{R}e \left\{ e^{i\frac{2\pi}{N}h} f_X\(\frac{2\pi}{N}\) \right\} \\
+ \frac{2\pi}{N} \sum_{k=1}^{N-1} e^{i\frac{2\pi}{N}kh}f_X\(\frac{2\pi}{N}k\),
\end{multline}
where the spectral density is deduced from equations (\ref{eq:slkdn2}) and (\ref{eq:beta_F}) :
\be
f_X(\lambda) = \frac{\sigma^2}{2\pi}
\frac{|1 - e^{-i\lambda}|^{2\alpha}}
{\left|1 - \frac{\Gamma(1+\alpha+\beta)}{\Gamma(1+\alpha)\Gamma(1+\beta)}e^{-i\lambda}F(1,-\alpha;1+\beta,e^{-i\lambda})\right|^2},
\ee
with $F(.,.;.,.)$ is the hypergeometric function, and $\sigma^2$ is the variance of $\epsilon_t$,
and
\be
K(\alpha,\beta) = \frac{\sigma^2}{2\pi} \(\frac{2\pi}{N} \)^{2\alpha+1}
\frac{1}{\left| 1 - \frac{\Gamma(\alpha+\beta)}{\Gamma(1+\alpha)\Gamma(\beta)} \right|^2}
\ee
The last term of equation (\ref{eq:acov_fft}) can be easily calculated by the use of the inverse Fast Fourier Transform.

\section{\label{app:acf}Auto-correlation function}

$\E_1[\varphi^k]$ is given by
\begin{equation}
\E_1[\varphi^k] = \frac{1}{2} A_k +  \sum_{n=1}^\infty a_n B_{n,k} + \sum_{n=1}^\infty b_n C_{n,k},
\end{equation}
where

\begin{equation}
A_k =
\begin{cases}
\ds \frac{2}{k+1} &  \textrm{if $k$ is even},\\
0                 &  \textrm{if $k$ is odd},
\end{cases}
\end{equation}

\begin{equation}
B_{n,k} = 
\begin{cases}
\ds(-1)^n \frac{2k}{(n\pi)^2} - \frac{k(k-1)}{(n\pi)^2} B_{n,k-2} &  \textrm{if $k$ is even}, \\
\ds -\frac{k(k-1)}{(n\pi)^2} B_{n,k-2}                            &  \textrm{if $k$ is odd},
\end{cases}
\end{equation}
with $B_{n,0} = 0$ and $B_{n,1} = 0$,

\begin{equation}
C_{n,k} = 
\begin{cases}
\ds -\frac{k(k-1)}{(n\pi)^2} C_{n,k-2}                             & \textrm{if $k$ is even}, \\
\ds (-1)^{n-1} \frac{2}{n\pi} - \frac{k(k-1)}{(n\pi)^2} C_{n,k-2}  & \textrm{if $k$ is odd},
\end{cases}
\end{equation}
with $C_{n,0} = 0$ and $C_{n,1} = \ds (-1)^{n-1}\frac{2}{n\pi}$.

$\E_2[a^k]$ is simply given by
\begin{equation}
\E_2[\varphi^k] = \frac{\Gamma(k+2)}{\Gamma(k + 3 -d)}\Gamma(3-d).
\end{equation}

\section{Implementation of the econometric procedure and convergence of estimators}
\label{App:Conv}

For finite size samples, the distance $\(\hat \eta_{T,L} - \eta_L(\theta)\)^t W_L^{-1} 
\(\hat \eta_{T,L} - \eta_L(\theta)\)$ can exhibit several local minima and, in practice, 
it actually does. Therefore, it turns out to be necessary to use a minimization algorithm 
that is able to deal with such a problem, preventing from being trapped in a local minimum, 
and then to find the global minimum. Genetic algorithms provide relevant solutions for such 
situations and they have been retained to solve our problem.

The idea underlying genetic algorithms is based on the mimicry of the natural selection process 
and genetic principles. The genetic algorithm starts with a population of trial vectors -- called 
{\it genes} -- containing the parameter $\theta$ to optimize and unfolds as follows:
\begin{itemize}
\item  The first step consists in the {\it replication} of the initial trial vectors according to 
their fitness, that is the genes whose distance is the smallest have the highest probability to 
reproduce. Thus, on the average, the new population  has a smaller distance than the initial one, 
but its diversification is also weaker since the fittest genes obviously appear twice or more in 
the new population.
\item The second step is the {\it crossover} which leads to combine the different parameters from 
several vectors drawn from the new population in order to mix their characteristics.
\item  The third and last step is the {\it mutation}, where some genes undergo random changes, 
{\it i.e.}, some parameters of the vectors born of the crossover are randomly modified. This step 
is essential to maintain the diversity of the population which in turn ensures the exploration of 
the whole optimization space. 
\end{itemize} 

The vectors obtained after this third step are then used as initial population and the process is 
reiterated in order to get a new generation of genes and so on. The convergence of this algorithm 
to the global minimum of the problem is ensured by the fundamental theorem of genetic algorithms 
\cite{goldberg89}. An example of particularly efficient genetic algorithms is the Differential 
Evolutionary Genetic Algorithm by \can{price97} or the \can{dorsey95} algorithm.

As the the genetic algorithm is particularly time consuming, we have turned to the \can{Nelder1965} 
simplex method\footnote{see also \can{lagarias98} for a recent discussion of the convergence of the method.}, 
that is a multidimensional unconstrained nonlinear minimization algorithm. This method presents however a 
serious disadvantage in our case : from the starting point chosen to initialize the procedure, it finds 
the nearest local minimizer of the function. In the case we are interested in, we know there are many 
local minima, thus an inadequate choice of the initial value leads to a local minimum instead 
of the global minimum we are looking for. 

In order to bypass this problem, we have developed an iterative procedure hereafter called the ``step by step'' method, 
based on the Nelder Mead method, that has appeared very fast and efficient when 
the coefficients $(a_n, b_n)$ in (\ref{eq:f1q}) decay at least as fast as $1/n^2$. It consists in restricting the optimization 
to $q=1$, in a first step. In such a case, the optimization can be performed by the Nelder Mead method. 
It provides a first estimate $\(\hat a_{1(1)},  \hat b_{1(1)}, \hat \alpha_{(1)},\hat w_{(1)}, \hat \sigma_{(1)}, \hat d_{(1)}\)$. 
In a second step, we set $q=2$, and start the Nelder-Mead algorithm with the value 
$\(\hat a_{1(1)}, 0, \hat b_{1(1)}, 0\right.$, $\left.\hat \alpha_{(1)}, \hat w_{(1)}, \hat \sigma_{(1)}, \hat d_{(1)}\)$ which yields 
a new estimate $\(\hat a_{1(2)}, \hat a_{2(2)}, \hat b_{1(2)}, \hat b_{2(2)},\hat \alpha_{(2)}, \hat w_{(2)}, \hat \sigma_{(2)}, \hat d_{(2)}\)$, 
and so on until the actual value of $q$ is reached. 

\begin{landscape}
\begin{figure}
\begin{center}
\includegraphics[width=11cm]{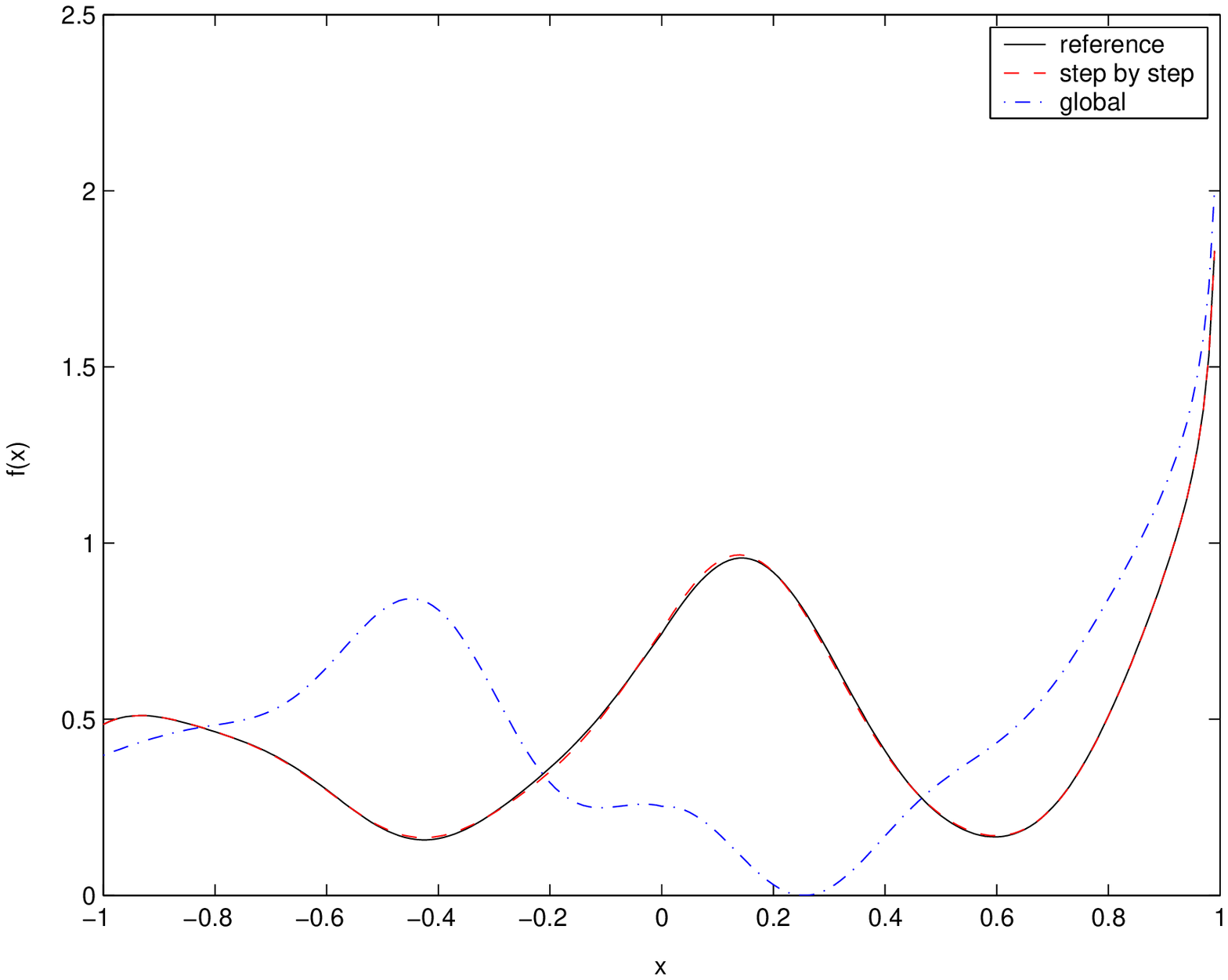} 
\includegraphics[width=11cm]{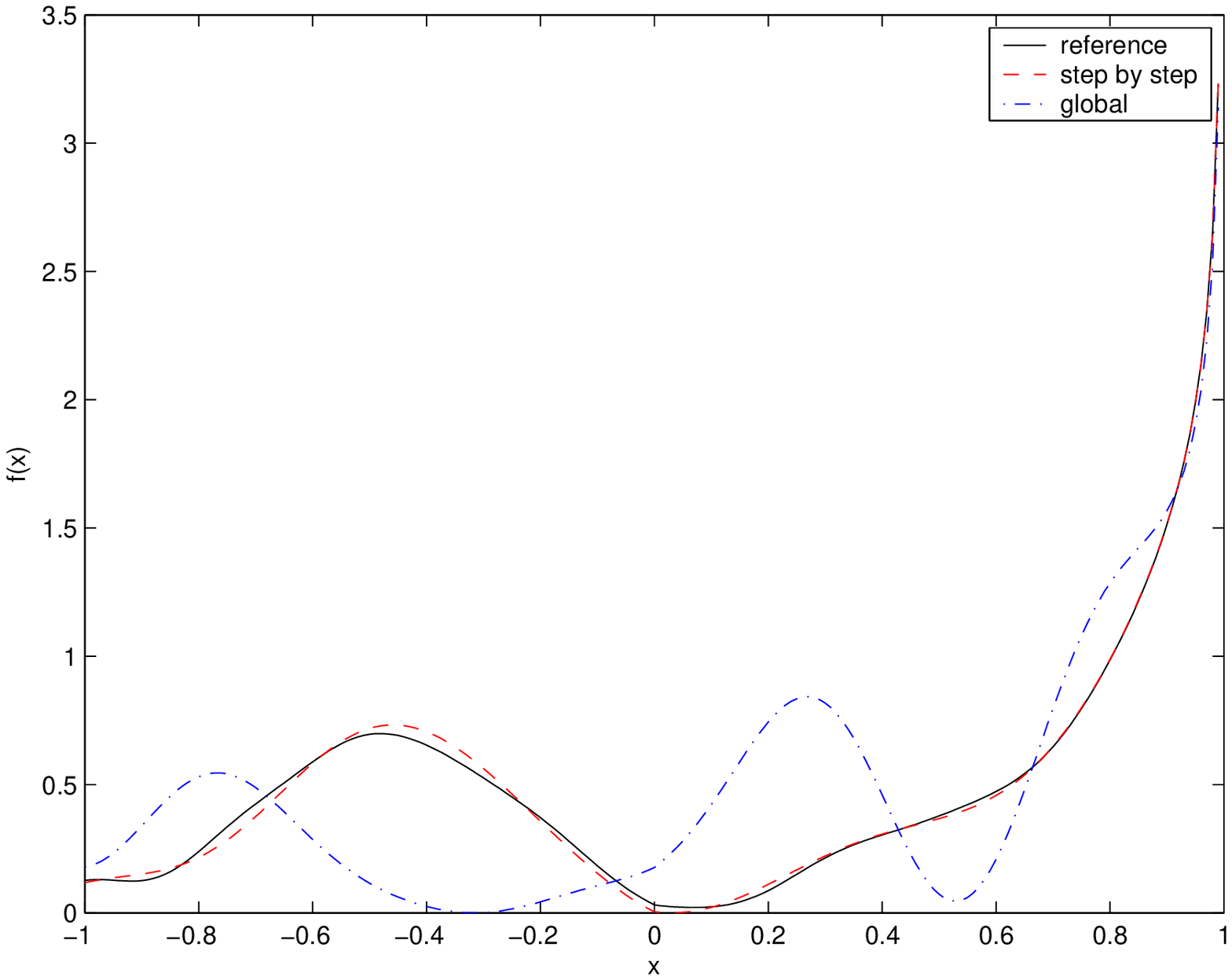}
\end{center}
\caption{\label{fig:E_Xp_densities_n2_OK}Comparison of the goodness of the two Nelder Mead
minimization algorithms. The estimated density is always obtained for $q=5$. On the left panel, the reference 
density is calculated with $q=5$ and on the right one with $q=10$.} 
\end{figure}
\end{landscape}

The figure~\ref{fig:E_Xp_densities_n2_OK} illustrates the convergence of this procedure in the case of a 
numerical experiment that unfolds as follows : 
\begin{enumerate}
\item we generate a reference density with a chosen $q$,
\item we apply the two procedures using the Nelder-Mead method,
\item we compare the accuracy of the two estimated densities to the reference density,
\item step (1) and (2) are iterated one thousand times.
\end{enumerate}
The two graphs show the efficiency of the step by step method. On the left panel, the reference density 
has been drawn for $q=5$ and we estimated the densities for $q=5$ too. The best estimated density is irrevocably 
the one obtained by the iterative procedure. On the right panel, we account for the fact that the density 
should have an infinite number of parameters (see equation (\ref{eq:Fourier_f1})) or, at least, that 
the right order $q$ in (\ref{eq:f1q}) is generally unknown. that is why we have generated a reference 
density with $q=10$ and performed the estimation for $q=5$ only to investigate the effect of the truncation on 
the accuracy of the two approaches. The result obtained for a randomly chosen simulation is displayed 
on the right panel of figure \ref{fig:E_Xp_densities_n2_OK}. One more time, the iterative method gives better results 
than global approach.

\vspace{1cm}
\centerline{\it [Insert figure~\ref{fig:E_Xp_densities_n2_OK} about here]}
\vspace{1cm}

While we have not been able to prove the convergence of the ``step by step'' procedure, our numerical simulations show that it provides estimates that are always close to the true parameter values (within the uncertainty predicted by proposition~\ref{prop1}). In addition, these estimates are almost always more accurate than the estimates obtained by use of the genetic algorithm, due to the very slow convergence of this algorithm.

\clearpage

\begin{table}
\centering
\begin{tabular} {llllrr}
\hline \hline
Symbol & Company                       & Market \quad  & Sector     &  \quad Cap. \\
\hline
BKS    & Barnes \& Noble Inc.          & NYSE   & dist. (c)  &   1.62 \\ 
VLO    & Valero Energy corp (new)      & NYSE   & energy     &   1.70 \\ 
DHI    & DR Horton inc.                & NYSE   & dist. (c)  &   1.83 \\ 
LEN    & Lennar corp CL a common       & NYSE   & dist. (c)  &   1.85 \\ 
TCB    & TCF Financial Corp.           & NYSE   & financial  &   1.92 \\ 
NYB    & New York Bancorp inc.         & NYSE   & financial  &   2.17 \\ 
FNF    & Fidelity Natl Financial inc.  & NYSE   & financial  &   2.27 \\ 
MCHP   & Microchip Technology inc.     & Nasdaq & tech.      &   2.99 \\ 
WPO    & Washington post co clb        & NYSE   & com.       &   3.71 \\ 
GILD   & Gilead Science inc.           & Nasdaq & dist. (nc) &   3.90 \\ 
\\
GM     & General Motors co.            & NYSE   & dist. (c)  &  31.81 \\ 
PG     & Procter \& Gamble co.         & NYSE   & dist. (nc) &  84.30 \\ 
IBM    & Intel Business Machines corp. & NYSE   & tech.      & 108.01 \\ 
CSCO   & Cisco Systems inc.            & Nasdaq & com.       & 112.26 \\ 
MRK    & Merck \& co inc.              & NYSE   & dist. (nc) & 112.73 \\ 
KO     & Coca-Cola co.                 & NYSE   & dist. (nc) & 118.97 \\ 
AIG    & American intl group inc.      & NYSE   & financial  & 125.58 \\ 
INTC   & Intel corp.                   & Nasdaq & tech.      & 140.77 \\ 
WMT    & Wall-Mart Stores inc.         & NYSE   & dist. (c)  & 144.61 \\ 
C      & Citigroup                     & NYSE   & financial  & 148.79 \\ 
PFE    & Pfizer inc.                   & NYSE   & dist. (nc) & 198.20 \\ 
XOM    & Exxon mobile corporation      & NYSE   & energy     & 199.19 \\ 
MSFT   & Microsoft corp.               & Nasdaq & tech.      & 239.81 \\ 
GE     & General Electrics co.         & NYSE   & industrial & 290.44 \\ 
\hline \hline
\end{tabular}
\caption{\label{tab:data}Average capitalization (in billion dollars) of every assets 
over the whole period (01/01/1994 to 12/31/2003) and their characteristics.}
\end{table}
\clearpage

\clearpage

\begin{table}
\centering
\begin{tabular} {lccccc}
\hline \hline
Asset & $d$ & $d_{gph}$ & $d_{hurst}$  \\ 
\hline
BKS  &	0.19 \;(0.04) & 0.36 \;(0.02) & 0.44 \\
VLO  &	0.23 \;(0.04) & 0.32 \;(0.02) & 0.44 \\
DHI  &	0.34 \;(0.04) & 0.34 \;(0.02) & 0.50 \\
LEN  &	0.31 \;(0.05) & 0.34 \;(0.02) & 0.53 \\
TCB  & 	0.31 \;(0.04) & 0.35 \;(0.02) & 0.43 \\
NYB  &	0.00 \;(0.06) & 0.33 \;(0.02) & 0.37 \\
FNF  &	0.25 \;(0.04) & 0.34 \;(0.02) & 0.28 \\
MCHP &	0.19 \;(0.04) & 0.45 \;(0.02) & 0.42 \\
WPO  &	0.45 \;(0.04) & 0.28 \;(0.02) & 0.46 \\
GILD &	0.37 \;(0.06) & 0.44 \;(0.02) & 0.42 \\
\\
GM   & -0.30 \;(0.04) & 0.30 \;(0.02) & 0.30 \\
PG   &	0.22 \;(0.05) & 0.41 \;(0.02) & 0.53 \\
IBM  &	0.21 \;(0.05) & 0.36 \;(0.02) & 0.44 \\
CSCO &	0.33 \;(0.06) & 0.47 \;(0.02) & 0.51 \\
MRK  &	0.22 \;(0.05) & 0.37 \;(0.02) & 0.49 \\
KO   &  0.16 \;(0.05) & 0.41 \;(0.02) & 0.55 \\
AIG  &	0.10 \;(0.03) & 0.44 \;(0.02) & 0.48 \\
INTC &	0.23 \;(0.06) & 0.38 \;(0.02) & 0.53 \\
WMT  &	0.31 \;(0.04) & 0.42 \;(0.02) & 0.46 \\
C    & -0.02 \;(0.04) & 0.40 \;(0.02) & 0.51 \\
PFE  &	0.16 \;(0.05) & 0.40 \;(0.02) & 0.55 \\
XOM  & -0.09 \;(0.04) & 0.41 \;(0.02) & 0.49 \\
MSFT &	0.21 \;(0.03) & 0.43 \;(0.02) & 0.53 \\
GE   &	0.04 \;(0.06) & 0.44 \;(0.02) & 0.50 \\
\hline \hline                                                     
\end{tabular}
\caption{\label{tab:d_all_assets}Estimation of the long memory parameter, $d$, by our method and different 
semi-parameter methods (Geweke and Porter-Hudak (GPH) and Hurst) for all the assets,
over the whole period (from 01/01/1994 to 12/31/2003). For the GPH estimator we also give the standard deviation.}
\end{table}
\clearpage

\begin{table}
\centering
\begin{tabular} {lcccc}
\hline \hline
     &                  & Whole        &   Before     &     After \\
     &                  & period       & bubble burst &  bubble burst \\
\hline
CSCO & d ($\bar\sigma$) &  0.33 (0.06) &  0.27 (0.04) &  0.18 (0.04)  \\ 
     & w                & (0.12)       & (0.57)       & (0.55) \\
     & $\alpha$         & (0.12)       & (0.80)       & (0.55) \\
     
\\ 
MRK  & d ($\bar\sigma$) &  0.22 (0.05) &  0.30 (0.04) &  0.08 (0.04) \\ 
     & w                & (0.43)       & (0.43)       & (0.45)\\ 
     & $\alpha$         & (0.31)       & (0.66)       & (0.39)\\
\\
KO   & d ($\bar\sigma$) &  0.16 (0.05) &  0.19 (0.04) &  0.20 (0.05) \\ 
     & w                & (0.46)       & (0.41)       & (0.47) \\ 
     & $\alpha$         & (0.62)       & (0.66)       & (0.41) \\
\\
AIG  & d ($\bar\sigma$) &  0.10 (0.03) &  0.26 (0.05) & -0.32 (0.04) \\ 
     & w                & (0.64)       & (0.61)       & (0.12) \\ 
     & $\alpha$         & (0.64)       & (0.45)       & (0.85) \\
\\
MSFT & d ($\bar\sigma$) &  0.21 (0.03) &  0.34 (0.04) &  0.00 (2.$10^5$) \\ 
     & w                & (0.66)       & (0.70)       & (0.05) \\
     & $\alpha$         & (0.54)       & (0.50)       & (0.30) \\
\\
\hline \hline
\end{tabular}
\caption{\label{tab:bubble_effects}Estimation of the long memory parameter $d$ (and its satandard deviation), 
the parameter $\alpha$ and the weight $w$ by using the 
auto-correlations method for some large capitalizations, over the whole period (from 01/01/1994 to 12/31/2003).}
\end{table}
\clearpage

\clearpage

\clearpage

\clearpage

\clearpage

\clearpage

\clearpage

\clearpage

\clearpage

\clearpage

\clearpage

\clearpage

\clearpage

\clearpage

\clearpage

\clearpage

\clearpage

\clearpage

\clearpage

\clearpage

\clearpage

\end{document}